\documentclass[10pt,aps,prc,twocolumn,showpacs,floatfix,nofootinbib,tightenlines,superscriptaddress]{revtex4-1}

\usepackage{amsmath,amssymb,graphicx,bm}

\graphicspath{{figures/}}

\newcommand{\beqn}{\begin{equation}}
\newcommand{\eeqn}{\end{equation}}
\newcommand{\bea}{\begin{eqnarray}}
\newcommand{\eea}{\end{eqnarray}}

\newcommand{\abinit}{{\it ab initio }}

\newcommand{\Trel}{T_{\rm rel}}

\newcommand{\fm}{\, \text{fm}}
\newcommand{\fmi}{\, \text{fm}^{-1}}
\newcommand{\mev}{\, \text{MeV}}
\newcommand{\eft}{$\chi$EFT}

\newcommand{\nmax}{\ensuremath{N_{\rm max}}}
\newcommand{\hw}{\ensuremath{\hbar\Omega}}
\newcommand{\atw}{\ensuremath{N_{\rm A2max}}}
\newcommand{\ath}{\ensuremath{N_{\rm A3max}}}
\newcommand{\lam}{\ensuremath{\lambda}}
\newcommand{\NNLO}{N$^2$LO}
\newcommand{\NNNLO}{N$^3$LO}

\newcommand{\he}{$^4$He}
\newcommand{\lip}{$^6$Li}
\newcommand{\li}{$^7$Li}
\newcommand{\bs}{$^7$Be}
\newcommand{\be}{$^8$Be}
\newcommand{\bo}{$^{10}$B}
\newcommand{\ca}{$^{12}$C}

\newcommand{\LamUV}{\Lambda_{UV}}
\newcommand{\kinf}{\ensuremath{k_{\infty}}}

\newcommand{\Einf}{\ensuremath{E_{\infty}}}

\usepackage{color}

\begin{document}

% uncomment \draft to have PACS numbers appear
% \draft

\title{P-shell nuclei using Similarity Renormalization Group \\
evolved three-nucleon interactions}

\author{E.D.\ Jurgenson}
\email{jurgenson2@llnl.gov}
\affiliation{Lawrence Livermore National Laboratory, P.O. Box
808, L-414, Livermore, CA\ 94551, USA}

\author{P.\ Maris}
\email{pmaris@iastate.edu}
\affiliation{Department of Physics and Astronomy, Iowa State University, 
Ames, IA\ 50011, USA}

\author{R.J.\ Furnstahl}
\email{furnstahl.1@osu.edu}
\affiliation{Department of Physics, The Ohio State University, 
Columbus, OH\ 43210, USA}

\author{P.\ Navr\'atil}
\email{navratil@triumf.ca}
\affiliation{TRIUMF, 4004 Westbrook Mall, Vancouver, BC, V6T 2A3, Canada}

\author{W.E.\ Ormand}
\email{ormand1@llnl.gov}
\affiliation{Lawrence Livermore National Laboratory, P.O. Box
808, L-414, Livermore, CA\ 94551, USA}

\author{J.P.\ Vary}
\email{jvary@iastate.edu}
\affiliation{Department of Physics and Astronomy, Iowa State University, 
Ames, IA\ 50011, USA}

%%%%%%%%%%%%%%%%%%%%%%%%%%%%%%%%%%%%%%%%%%%%%%%%%%%%%%%%%%%%%%%%%%%%
%%%%%%%%%%%%%%%%%%%%%%%%%%%% Abstract %%%%%%%%%%%%%%%%%%%%%%%%%%%%%%

\begin{abstract}
The Similarity Renormalization Group (SRG) is used to soften
interactions for \abinit nuclear structure calculations by decoupling
low- and high-energy Hamiltonian matrix elements.  The substantial
contribution of both initial and SRG-induced three-nucleon forces
requires their consistent evolution in a three-particle basis space
before applying them to larger nuclei. While in principle the evolved
Hamiltonians are unitarily equivalent, in practice the need for basis
truncation introduces deviations, which must be monitored.  Here we
present benchmark no-core full configuration calculations with
SRG-evolved interactions in p-shell nuclei over a wide range of
softening.  These calculations are used to assess convergence
properties, extrapolation techniques, and the dependence of energies,
including four-body contributions, on the SRG resolution scale.
\end{abstract}

\date{\today}


\maketitle

%%%%%%%%%%%%%%%%%%%%%%%%% Introduction %%%%%%%%%%%%%%%%%%%%%%%%%%%%%
\section{Introduction
\label{sec:introduction}}

Configuration interaction methods have been used in recent years to
make increasingly accurate large scale \abinit calculations of nuclear
structure and reactions (e.g., see
Refs.~\cite{Maris:2008ax,Navratil:2009ut,Maris:2009bx,Maris:2011as,Navratil:2011zs,Barrett:2013nh})
Improved algorithms and better use of increasing computational
resources are critical for these successes.  However, the reach of
these methods may also be extended by applying renormalization group
(RG) transformations to the input Hamiltonian.  Renormalization
techniques soften the free-space interactions by reducing the coupling
between high and low momenta, leading to improved convergence with the
size of the basis for a fixed number of interacting nucleons.  The
Similarity Renormalization Group
(SRG)~\cite{Glazek:1993rc,Wegner:1994} is an attractive approach for
this purpose due to its relatively simple implementation, general
flexibility, and the feasibility of consistently evolving many-body
operators~\cite{Bogner:2009bt,Furnstahl:2012fn}.

Previous studies of the SRG in nuclear physics established its
usefulness for few-body systems by demonstrating improved convergence
with two-nucleon (NN) interactions
alone~\cite{Bogner:2006pc,Bogner:2007jb,Jurgenson:2007td}. In
Ref.~\cite{Bogner:2007rx}, a detailed study of SRG convergence with NN
forces in the p-shell was made.  The present work extends this study
to include initial and induced three-nucleon (NNN) forces. We build
upon the technology to evolve NNN forces introduced in
Ref.~\cite{Jurgenson:2009qs}, which was applied in
Ref.~\cite{Jurgenson:2010wy} to \he\ and \lip\ to explore the effects
of full two-plus-three-body evolved interactions in light nuclei. Roth
et al.\ have subsequently used the Importance Truncated No-Core
Shell Model (IT-NCSM) with the
SRG~\cite{Roth:2009cw} to significantly push the limits in $A$ and
model space size~\cite{Roth:2011ar,Roth:2011vt}. We focus here on a
wider set of p-shell nuclei and on wider ranges of softening that,
with our extrapolation methods, yield nearly converged results without
implementing importance truncation.

The SRG flow equations generate a continuous series of Hamiltonians
that is analogous to the running of the strong coupling constant in
quantum chromodynamics, having both scale (or resolution) and scheme
dependence~\cite{Furnstahl:2012fn}.  The scheme dependence arises from
the choice of initial nuclear Hamiltonian and the choice of the
operator generating the flow (see below).  While the SRG offers a
useful framework for future comparisons of such choices for both NN
and NNN interactions and exploring the flow to universal
forms~\cite{Bogner:2009bt,Hebeler:2012pr}, in this work, we restrict
our attention to just one choice. In particular, we use the chiral
effective field theory (EFT) potential at \NNNLO\ with 500\,MeV cutoff
from Ref.~\cite{Entem:2003ft} together with an NNN potential at
\NNLO~\cite{Epelbaum:2002vt} in the local form of
Ref.~\cite{Navratil:2007zn}.  This is also the Hamiltonian used in
Refs.~\cite{Jurgenson:2009qs,Jurgenson:2010wy,Roth:2011ar,Navratil:2007we}.

The scale dependence arising from the RG flow is manifested as a
decreasing decoupling scale that marks the energy difference for which
matrix elements between off-diagonal energy states become highly
suppressed.  Formally, all of the evolved Hamiltonians have equivalent
physics content to that of the initial Hamiltonian, so it would seem
to be advantageous to evolve to very low scales to optimize the
convergence of many-body calculations. But in practice, the initial
and running Hamiltonians are expanded in a finite basis (harmonic
oscillators here) and many-body forces are truncated at some level.
Therefore, it is necessary to monitor and characterize the evolution
of many-body forces and the residual running of calculated
observables, which can vary with the size of the nuclear system. In
this paper, we present benchmark calculations using SRG-evolved
interactions in p-shell nuclei. We use them to explore the
characteristics and practical limits of SRG evolution for these
systems by assessing convergence properties, extrapolation techniques,
and the stability of predicted observables.

In Section~\ref{sec:back}, we briefly review the formalism used in
this study and summarize some observations from previous work. A more
complete discussion is provided in~\cite{Jurgenson:2010wy}. The
convergence properties of the evolved NN and NNN potentials in various
nuclei are explored in Section~\ref{sec:convergence} and in
Section~\ref{sec:evolution} we examine the evolution with $\lambda$
and the patterns of induced many-body
forces. Section~\ref{sec:conclusions} summarizes our conclusions and
provides an outlook for future calculations.

%%%%%%%%%%%%%%%%%%%%%%%%% background %%%%%%%%%%%%%%%%%%%%%%%%%%%%%
\section{Background
\label{sec:back}}

\subsection{SRG evolution}

The SRG for low-energy nuclear physics generates a continuous series of
Hamiltonians $H_\lambda$ from an initial free-space  Hamiltonian
$H_{\lambda=\infty}$ by unitary transformations, 
\beqn
H_{\lambda} = U_{\lambda} H_{\lambda=\infty}U_{\lambda}^{\dagger} \;,
\label{eq:Hs}
\eeqn
which is carried out by solving a set of flow equations,
\beqn
 \frac{dH_{\lambda}}{d\lambda} = -\frac{4}{\lambda^5} [[G,H_{\lambda}],H_{\lambda}]  \;.
 \label{eq:flow}
\eeqn
With an appropriate choice of the hermitian operator $G$, the  Hamiltonian is
evolved to band-diagonal form with respect to 
energy~\cite{Wegner:1994,Kehrein:2006}.  In most nuclear applications to date,
the relative kinetic energy $\Trel$ has been used because of the favorable
convergence properties of the evolved Hamiltonians and for its convenience in 
constructing basis expansions~\cite{Bogner:2009bt,Furnstahl:2012fn}.   We use it
here exclusively but note that other choices may be advantageous in reducing the
growth of many-body forces.  This is being investigated
separately~\cite{Anderson:2008mu,li:2011sr}. The flow parameter \lam\ keeps
track of the sequence of Hamiltonians ($s$ or $\alpha$, with $s = \alpha =
1/\lambda^4$,   are also used
elsewhere~\cite{Bogner:2006pc,Bogner:2007jb,Roth:2011ar}). For $G=\Trel$, \lam\
has dimensions of momentum and runs from $\infty$ toward zero with increasing
softening.

Evolution is performed in the Jacobi-coordinate harmonic oscillator
(HO) basis used for the No-Core Shell Model
(NCSM)~\cite{Navratil:2009ut,Barrett:2013nh}.  This is a
translationally invariant, anti-symmetric basis for each $A$-body
sector, in which a complete set of states in the model space is
defined by the maximum excitation of \nmax\hw\ above the minimum
energy configuration, where $\Omega$ is the harmonic oscillator
parameter.  This basis is variational in \nmax; that is, the energy
converges asymptotically from above as more basis states are included.
The SRG preserves this variational characteristic through smooth
unitary evolution, in contrast to Okubo-Lee-Suzuki based
renormalizations~\cite{Navratil:2009ut,Barrett:2013nh}, which are
unitary transformations specific to the model space.

We start by evolving $H_\lambda$ using Eq.~\eqref{eq:flow} in the
$A=2$ subsystem, completely fixing the evolved two-body matrix
elements. Next, by evolving $H_\lambda$ in the $A=3$ subsystem we
determine the combined two-plus-three-body matrix elements. We isolate
the three-body matrix elements by subtracting the evolved two-body
elements within the $A=3$ basis~\cite{Jurgenson:2008jp}. Having
obtained the separate NN and NNN matrix elements we can apply them as
inputs to any \abinit nuclear structure problem.  We are also free to
include an initial three-nucleon force in the starting Hamiltonian
without changing the procedure.

While any initial interaction can be used as input to the SRG
evolution, here we use the chiral EFT NN potential from the 500\,MeV
N$^3$LO interaction of Ref.~\cite{Entem:2003ft} exclusively. As an
initial NNN potential, we use the chiral N$^2$LO
potential~\cite{Epelbaum:2002vt} in the local form of
Ref.~\cite{Navratil:2007zn}. The low-energy constants $c_D = -0.2$ and
$c_E = -0.205$ are the result of a fit to the average of triton and
$^3$He binding energies and to triton beta decay as described in
Ref.~\cite{Gazit:2008ma}.

\begin{table}[tb-]
\caption{Definitions and values of basis-size truncations used on the initial
Hamiltonian and on the model space for the CI calculations. The value
of \ath\ is uniformly reduced for three-body partial-wave J$\pi$T channels 
with higher J to a minimum of 20.
\label{tab:truncs} }
\begin{ruledtabular}
\begin{tabular}{rp{2.4in}c}
  \atw\  &  Maximum sum of two-body oscillator quanta for initial two-body matrix element evolution & 300 \\
  \ath\  &  Maximum sum of three-body oscillator quanta 
  for initial three-body matrix element evolution & 40  \\
 \nmax\  &  Maximum sum of $A$-body oscillator quanta above the minimum
 for the $A$-body system & 2--8 \\
\end{tabular}
\end{ruledtabular}
\end{table}

Hamiltonians obtained via free-space SRG evolution are independent of
the basis choice if the basis is sufficiently complete. That is, a
Hamiltonian evolved to a given \lam\ reproduces the results of a
Hamiltonian evolved to the same \lam\ in a different basis.  But in
practice there are truncations, both in $A$-body forces and in basis
size, that are relevant to controlling the quality and consistency of
SRG-evolved interactions.

In the present work, induced four-body (and higher) forces are not
included, so calculations for $A \geq 4$ will be only approximately
unitary. The many-body interaction matrix elements induced by the
evolution appear in a decreasing hierarchy in few-body
nuclei~\cite{Jurgenson:2009qs,Jurgenson:2010wy}.  One of our goals is
to determine if that hierarchy is maintained for p-shell nuclei or if
the induced many-body contributions become unnaturally large for
certain systems and/or values of $\lambda$.

Because of computational constraints, we are forced to apply separate
truncations to the $A = 2$ and $A = 3$ sectors of the initial
Hamiltonian, which we denote \atw\ and \ath, respectively (see
Table~\ref{tab:truncs}). These cutoffs in the basis size must be large
enough to fully accommodate the ultraviolet (UV) contributions (or
high-momentum components) from the initial NN plus NNN
Hamiltonian. The ultraviolet cutoff in an oscillator basis scales like
$\sqrt{N\hw}$, where $N$ represents the maximum number of
single-particle oscillator quanta in the basis, so there will be an
\hw\ below which the initial and, therefore, evolved Hamiltonian
projections onto the oscillator basis are incomplete.  When we use
such an \hw\ that is too low, we are effectively working with a
different Hamiltonian.  As a consequence, the calculations of
observables in the many-body basis with the too-low \hw\ will not
converge (or extrapolate) to the same results found at larger \hw.
This is not a problem for the NN interaction, for which \atw\ is
sufficiently large for the chiral EFT Hamiltonian for all
\hw\ considered, but becomes a factor for the NNN force, as
illustrated below.

Hamiltonians are derived and evolved in the Jacobi basis for $A=2$ and
3 and then translated to a Slater determinant basis for full
configuration interaction (CI) calculations of larger systems.  The
particular CI procedure used here, including the
extrapolation to infinite basis size and associated uncertainty
estimates, is referred to as no-core full
configuration or NCFC~\cite{Maris:2008ax}. 
Other CI calculations  in the literature using SRG-evolved interactions and extrapolation are called NCSM (e.g., Refs.~\cite{Jurgenson:2009qs,Jurgenson:2010wy}). 
While the original NCSM featured a finite matrix truncation and an effective
Hamiltonian renormalized to that finite space~\cite{Barrett:2013nh},
these SRG-based NCSM and NCFC procedures are equivalent except for
variations in the extrapolation  and uncertainty quantification
procedures. 
 
In these CI calculations,
the size of the largest feasible model space is highly constrained by
the total number of two- and three-body matrix elements in the full
space.  Fortunately, the MFDn
code~\cite{Sternberg:2008:ACI:1413370.1413386,DBLP:journals/procedia/MarisSVNY10,DBLP:conf/europar/AktulgaYNMV12}
that carries out the Lanczos matrix diagonalization algorithm is
highly optimized for parallel computing.  The calculations were
performed on the Intel Xeon cluster Sierra at LLNL, using up to about
15\,TB of memory across 7,200 cores on the Cray XE6 Hopper at NERSC, 
using up to about
100\,TB of memory across 76,320 cores; and on the Cray XK6 Jaguar at
ORNL, using over 500\,TB of memory across 261,120 cores.  MFDn has
been demonstrated to scale well on these platforms for these types of
runs~\cite{Maris:2013}.

\subsection{Guide to the calculations}

\begin{table}[tb-]
\caption{\label{tab:guide}Guide to the calculations.}
\begin{ruledtabular}
\begin{tabular}{rp{2.0in}}
  NN-only  & No initial NNN interaction   
   and do not keep NNN-induced interaction.   \\
  NN + NNN-induced  & No initial NNN interaction  
      but keep the SRG-induced NNN interaction arising from the
      NN interaction alone.   \\
  NN + NNN  & Include an initial NNN interaction   
   	\emph{and} keep the SRG-induced NNN interaction
	arising from the combination of NN and NNN interactions.   \\
\end{tabular}
\end{ruledtabular}
\end{table}

\begin{figure}[thb]
\includegraphics*[width=3in]{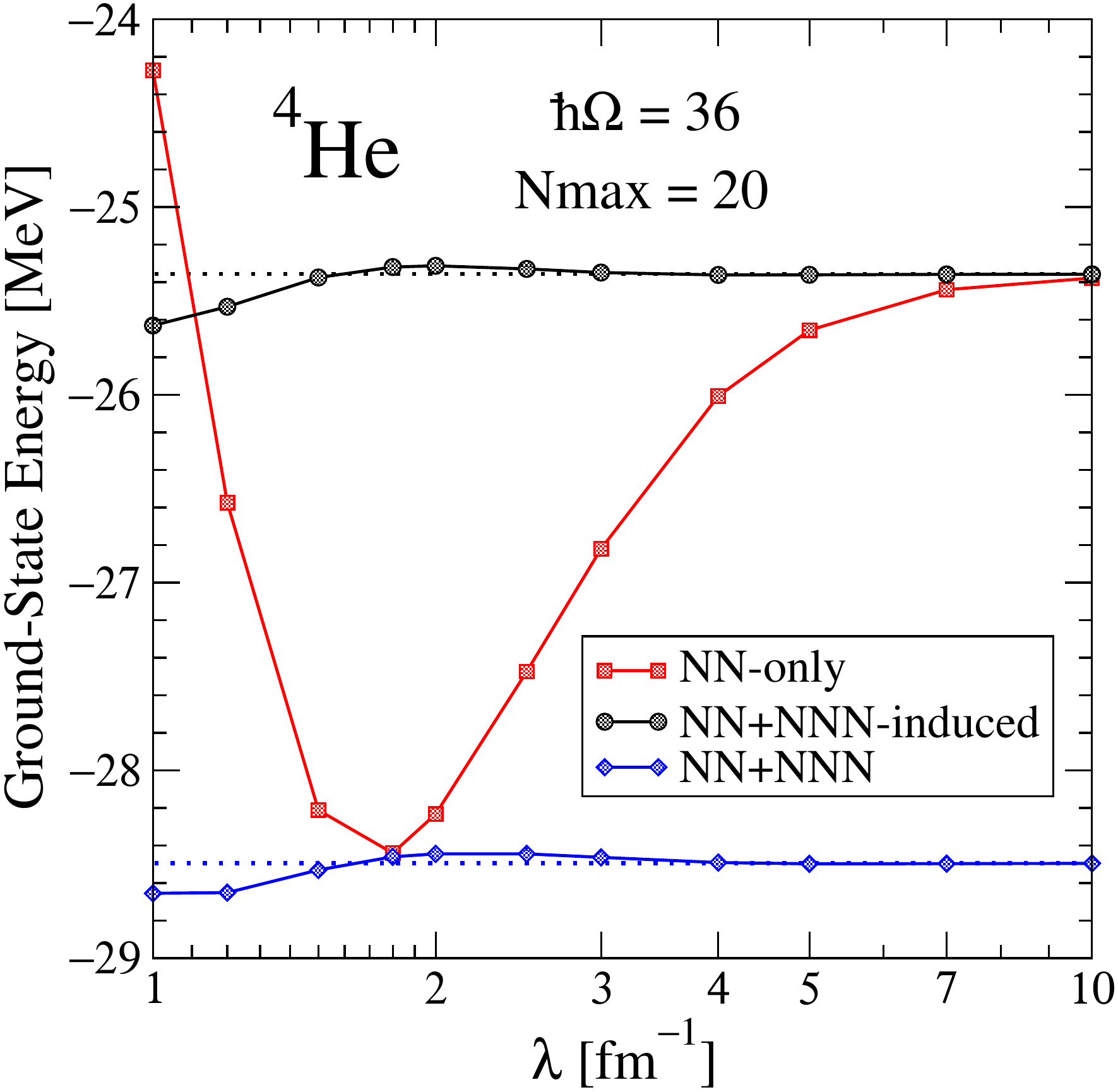}
\caption{(color online) Ground-state energy of $^4$He 
as a function of $\lambda$ for the three calculations
in Table~\ref{tab:guide}.  The results at all $\lambda$ are converged
at the 2\,keV level or better.  The dotted line is the
unevolved result.
\label{fig:He4_running}}
\end{figure}

To set the stage for our examination of p-shell nuclei, we show in
Fig.~\ref{fig:He4_running} the ground-state energy of $^4$He as a
function of $\lambda$~\cite{Jurgenson:2009qs,Jurgenson:2010wy}. Here,
and throughout the paper, we will compare three types of calculations,
which are summarized in Table~\ref{tab:guide}. The NN-only
calculations include two-body matrix elements that are phase-shift
equivalent to the initial NN interaction. When this part of the
Hamiltonian is used alone to study $A>2$ systems, the results are not
unitarily equivalent at different $\lambda$ because the SRG has
reorganized the degrees of freedom to reduce coupling of high- and
low-energies. Formally, in the process of maintaining unitary
equivalence in all sectors the SRG induces new contributions to
many-body matrix elements, but in the NN-only result these induced
interactions are omitted. We get a characteristic pattern (see the
NN-only results in Fig.~\ref{fig:He4_running}) where the converged
result varies with evolution, starting at some (underbound) level for
the initial Hamiltonian, then falling and rising again with subsequent
evolution (decreasing \lam).

In the NN+NNN-induced calculations, the Hamiltonian includes the
evolved NN matrix elements as well as all three-body matrix elements
induced by the SRG starting from only an NN interaction (i.e., no
initial NNN).  In an $A=3$ system this will be unitarily equivalent to
the initial NN-only Hamiltonian, so the energy spectrum will be the
same, up to numerical truncation errors (e.g., because of an
insufficient \ath). Finally, the NN+NNN calculations include an
initial three-body interaction as well as the induced three-body
matrix elements that now arise from the combined evolution of the
NN+NNN interactions. For $A=3$, it should be unitarily equivalent to
the initial NN+NNN Hamiltonian. For $A \geq 4$, there will be induced
four-body (and higher-body) interactions that are not included in any
of the present calculations.  Their omission causes differences in the
predicted energy spectra as a function of $\lambda$.

The computed ground-state energies for $^4$He in
Fig.~\ref{fig:He4_running} are well converged at all $\lambda$, so the
interpretation is clear. For the NN-only calculations, deviations from
unitary equivalence are evident just below $\lambda = 10\fmi$, where
binding is increasing by a maximum of about three MeV (10\% of the
total binding energy), peaking just below $\lambda = 2\fmi$ and then
decreasing rapidly and overshooting the original ground-state energy
by $\lambda = 1\fmi$.  A similar pattern for NN-only was shown in
Ref.~\cite{Bogner:2007rx} for several p-shell nuclei.

The NN+NNN-induced calculation shows a dramatic reduction in the
variation of the energy for $\lambda > 1.5\fmi$, with only a small
decrease in the binding energy peaking near $\lambda = 2\fmi$.  The
deviations near $\lambda = 1\fmi$, which imply net induced four-body
interaction contributions, are only about 300\,keV, or still an order
of magnitude reduced from the largest NN-only variations.  The same
pattern for the (implied) induced four-body interaction is seen when
an initial NNN interaction is included, with just a slight change in
the pattern at $\lambda = 1\fmi$.  When we compare to the larger
nuclei in the present work, we will not be able to examine the full
range of $\lambda$ used in Fig.~\ref{fig:He4_running} because
convergence is only sufficient for reliable extrapolation with small
errors for \lam\ up to about $2\fmi$.

\subsection{Extrapolation methods
\label{subsec:extrapolation}}

For well-evolved Hamiltonians in lighter nuclei (e.g., $^3$H or $^4$He
at $\lambda \leq 1.5\fmi$), our predictions for ground-state energies
are well converged at computationally accessible values of \nmax.
However, for larger nuclei and greater \lam\ values we will need to
extrapolate calculated energies to $\nmax = \infty$.  To do so, we
primarily use empirical extrapolation procedures based on those
described in Refs.~\cite{Bogner:2007rx,Jurgenson:2010wy,Maris:2008ax}
but also compare to a new procedure from Ref.~\cite{Furnstahl:2012qg}.
 
The empirical model used for ground-state energies is
\beqn
  E_{\alpha i} = E_\infty + A_\alpha\, e^{-b_\alpha N_{\alpha i}}
  \;,
  \label{eq:Ealphai}
\eeqn
where $A_\alpha$ and $b_\alpha$ are (\hw\ dependent) constants,
  $N_{\alpha i}$ are the \nmax\ values, and $\alpha$ labels the
  \hw\ value.  The goal is to determine the common parameter
$E_\infty$, which is the estimate for the ground-state energy
extrapolated to $\nmax = \infty$.  This can be cast as a
one-dimensional constrained minimization problem with the function
\beqn
  g(E_\infty) = \sum_{\alpha,i} 
    ( \log(E_{\alpha i} - E_\infty) - a_\alpha - b_\alpha N_{\alpha i} )^2
    /\sigma_{\alpha i}^2 
    \;,
    \label{eq:residual}
\eeqn
where the $\{a_\alpha\}$ and $\{b_\alpha\}$ are determined directly
within the function $g$ by invoking a constrained linear least-squares
minimization routine. The constraint is the bound $E_\infty \leqslant
\min(\{E_{\alpha i}\})$, where $E_\infty < 0$ and ``min'' means ``most
negative''. (One can also allow for weights depending on \nmax\ and/or
\hw.)

In the present investigation, we have applied this extrapolation model
for individual values of \hw, determining error estimates as in
``Extrapolation B'' from Ref.~\cite{Maris:2008ax}, but also including
several values of \hw\ in a constrained fit over a range where they are
considered reliable.  We emphasize that while this model has been
generally successful when applied in NCFC calculations with
SRG-evolved interactions, its validation is empirical rather than
theoretical.

An alternative EFT-motivated approach to extrapolation is based on
explicitly considering the ultraviolet (UV) and infrared (IR) cutoffs
imposed by a truncated harmonic oscillator
basis~\cite{Coon:2012ab,Furnstahl:2012qg}.  This has led to a
theoretically motivated IR correction formula and an empirical UV
correction formula~\cite{Furnstahl:2012qg} in which the basic
extrapolation variables are the effective hard-wall size $L$ and the
analogous cut-off in momentum, $\LamUV$.  In terms of the oscillator
length $b\equiv\sqrt{\hbar/(m\Omega)}$, rough estimates of these
variables are $\LamUV \approx \sqrt{2(N+3/2)}\hbar/b$ and $L \approx
\sqrt{2(N+3/2)} b$, where $N = \nmax + 1$ for p-shell
nuclei~\cite{Coon:2012ab,Furnstahl:2012qg}.  A formula combining both
corrections (i.e., they are treated independently) takes the
form~\cite{Furnstahl:2012qg}
\beqn
 E(\LamUV,L)
   \approx \Einf + B_0 e^{-2\LamUV^2/B_1^2} + B_2 e^{-2\kinf L}
   \;.
   \label{Ecombi}
\eeqn
Note that this formula contains exponentials with arguments
proportional to both $N$ (from $\LamUV^2$) and $\sqrt{N}$ (from $L$),
in contrast to Eq.~\eqref{eq:Ealphai}.

Following Ref.~\cite{Furnstahl:2012qg}, we apply Eq.~\eqref{Ecombi}
with $\Einf$, $B_0$, $B_1$, $B_2$, and $\kinf$ treated as fit
parameters that are determined from a simultaneous optimization to
data at all \hw, including the intermediate region where both IR
and UV corrections are significant.  It may be advantageous in general
to isolate the IR or UV corrections by using only large \hw\ or small
\hw\ results, respectively.  However, most of the present calculations
were made with \hw\ values close to the energy minimum, which means
comparable UV and IR contributions~\cite{Furnstahl:2012qg}.  (The
exception is for very low $\lambda$, where UV convergence is reached
for all \hw\ considered.)  We also exploit a recent observation that
the expressions for $L$ and $\LamUV$ give more accurate energy
corrections if we take $N \rightarrow N+2$, which is particularly
effective when $\nmax$ is small~\cite{More:2013aa}.  Thus we will use
$N = (\nmax + 1) + 2 = \nmax + 3$ for the calculations in
Section~\ref{subsec:comparisons}.  Equation~\eqref{Ecombi} has been
successfully applied to NN-only calculations from
Ref.~\cite{Bogner:2007rx}, but here we test it for the first time with
three-body forces included.

%%%%%%%%%%%%%%%%%%%%%%%%% Convergence Plots %%%%%%%%%%%%%%%%%%%%%%%%%%%%%
\section{Convergence
\label{sec:convergence}}

For fixed \nmax, both the UV and IR momentum cutoffs scale with
$\sqrt{\hw}$, which means that there is a trade-off: increasing
\hw\ increases the ability to accommodate high-momentum components
while decreasing the ability to accomodate long-distance physics. The
result is a familiar variational minimum with respect to \hw.  For
NN-only calculations, it was observed in Ref.~\cite{Bogner:2007rx}
that with decreasing SRG $\lambda$ at fixed \nmax, the minimum
systematically shifts to lower \hw\ and convergence becomes much more
rapid.  Here we examine if these observations are modified by the
presence of a three-nucleon force (3NF) in NN+NNN-induced and NN+NNN
calculations.

\subsection{Size of three-body evolution basis
\label{subsec:A3nmax}}

\begin{figure}[thb-]
\includegraphics*[width=3in]{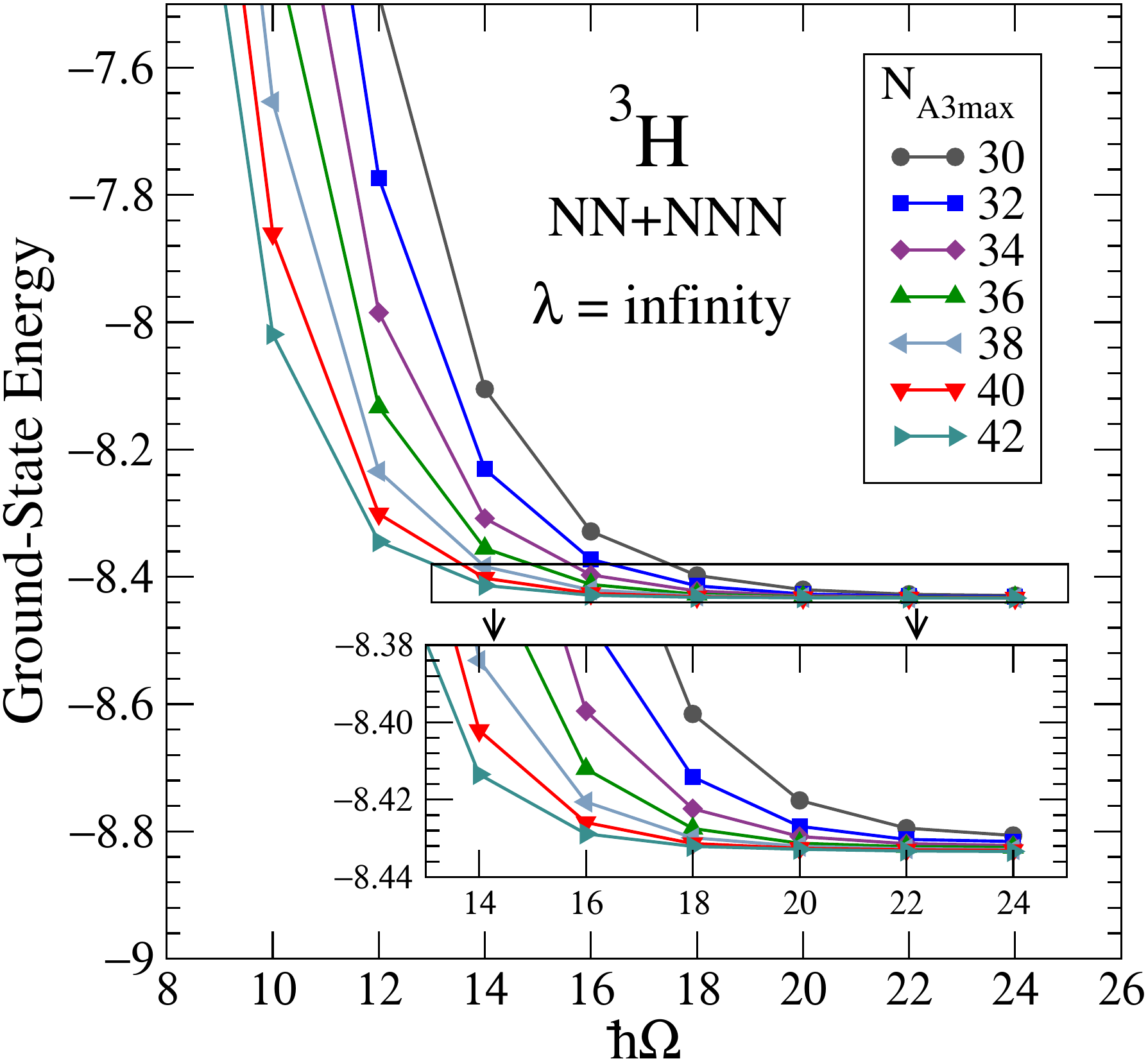}
\caption{(color online) Ground-state energy of the triton for 
the unevolved chiral EFT Hamiltonian
in different three-body basis sizes
(\ath) with a large, fixed two-body
basis ($\atw = 300$).
\label{fig:convergence_H3_A3nmax}}
\end{figure}

As noted earlier, the size \ath\ of the $A=3$ basis we use to evolve
the Hamiltonian before embedding in larger systems is limited by
computational constraints. In Fig.~\ref{fig:convergence_H3_A3nmax},
the impact on the calculated ground-state of the triton is shown for
the unevolved interaction using different values of \ath\ with
\hw\ ranging from 10 to 24\,MeV.  At each \hw, the signal that
\ath\ is sufficiently large is convergence of the ground-state energy,
which is evident for $\hw \ge 18$.  In contrast, the systematic
underbinding at lower \hw\ values with the \ath\ from
Table~\ref{tab:truncs} will be preserved when the Hamiltonian is
evolved; in effect a different initial Hamiltonian will be used. The
spread of points at fixed \hw\ and particularly the deviations at
$\ath=40$ from the fully converged energy imply that low oscillator
parameters (i.e., below $\hw=18\mev$), will be unreliable for energy
calculations in larger nuclei (because we cannot predict the degree of
underbinding, as shown in Fig.~\ref{fig:extrapolation_Be8_A3nmax}
discussed below).  In contrast, the size of the $A=2$ basis used here
is sufficient for convergence within 1\,keV in the full range of
\hw\ considered.

\begin{figure}[thb-]
\includegraphics*[width=2.7in]{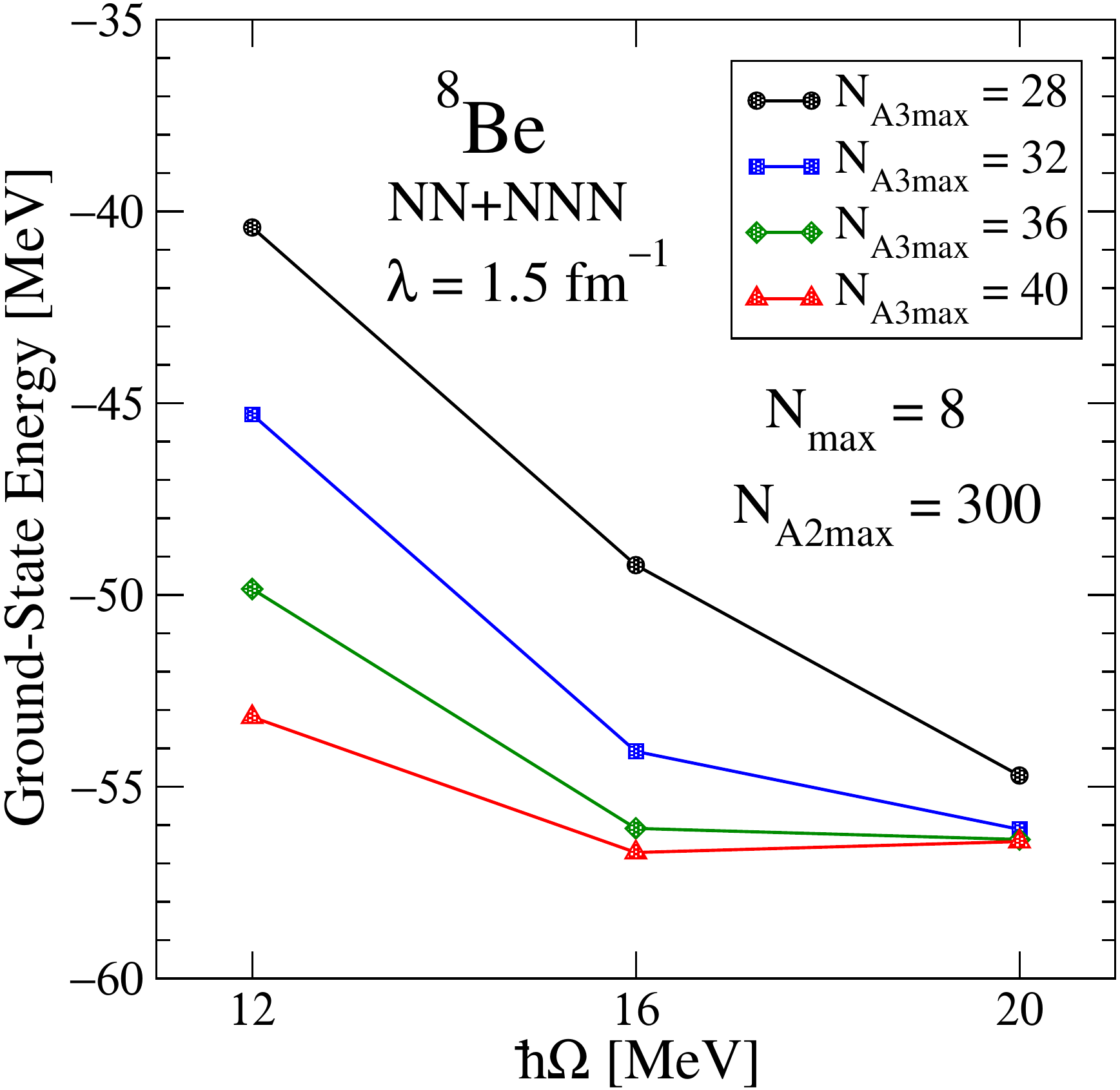}
\caption{(color online) Ground-state energy of \be\ for a fixed many-body basis
size of $\nmax=8$ for Hamiltonians evolved to $\lambda = 1.5\fmi$ in different
three-body basis sizes (\ath) with a fixed, large two-body basis ($\atw = 300$).
\label{fig:convergence_Be8_A3nmax}}
\end{figure}

\begin{figure}[thb-]
\includegraphics*[width=2.7in]{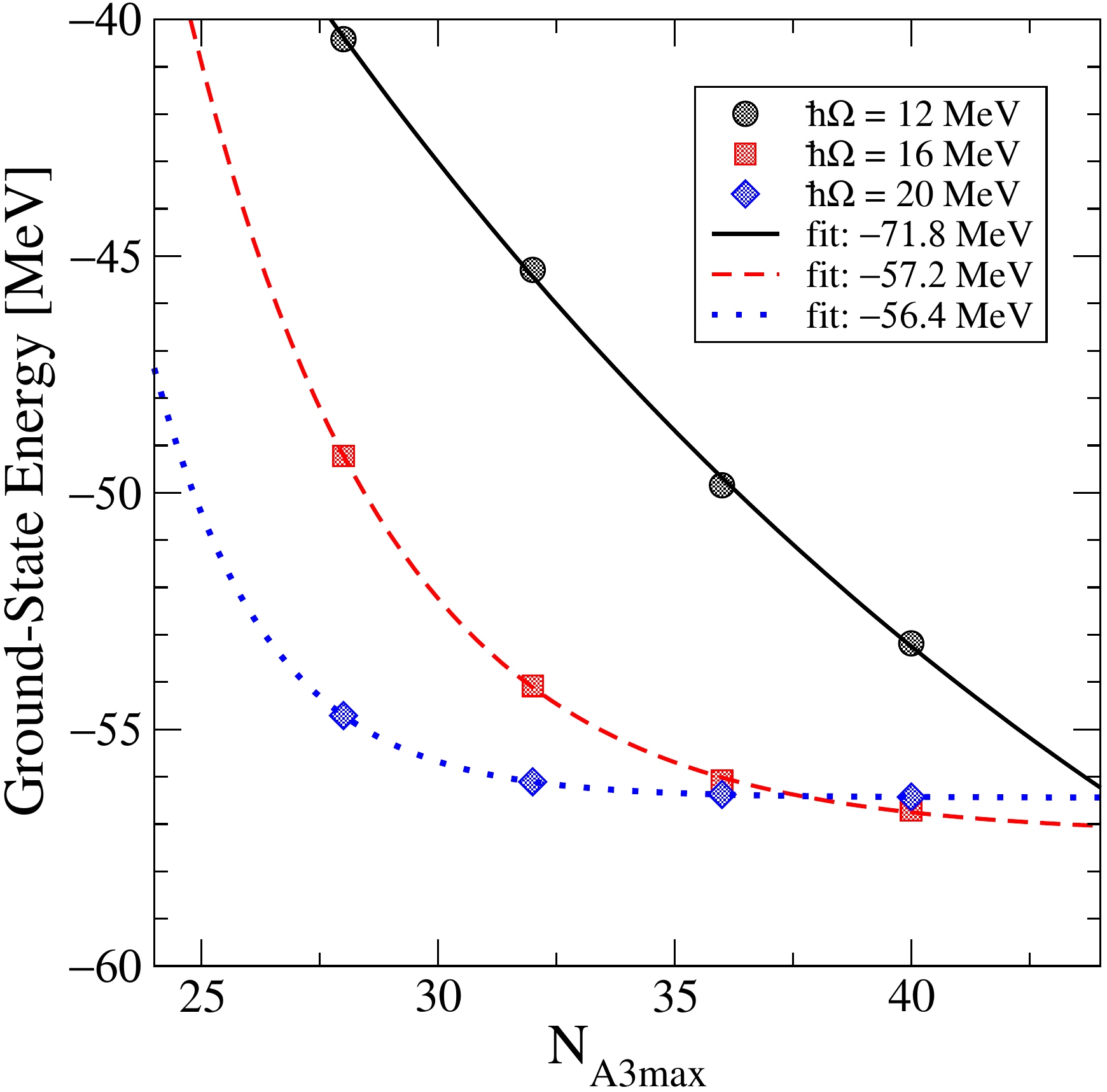}
\caption{(color online) Attempted extrapolation in \ath\ at fixed $\nmax=8$ of
\be\ ground-state energies from Fig.~\ref{fig:convergence_Be8_A3nmax} at several
values of \hw.
\label{fig:extrapolation_Be8_A3nmax}}
\end{figure}

The implications of a too-small \ath\ at smaller \hw\ for SRG
evolution are illustrated in Fig.~\ref{fig:convergence_Be8_A3nmax},
which shows fixed $\nmax=8$ (and fixed $\atw=300$) calculations of the
\be\ ground state of the same initial NNN interaction evolved in
different three-body basis sizes, for a range of \hw.  All results are
for $\lambda = 1.5\fmi$. It is evident that $\hw=12\mev$ is far from
converged even at the largest value of \ath\ available.  Note that we
do not expect the different \hw\ calculations in this figure to have
the same energy as $\ath\rightarrow\infty$, because $\nmax=8$ is still
unconverged in the many-body system. However, for each individual
\hw\ we need convergence for the largest \ath, as observed for
$\hw=20\mev$.  We also note that an exponential model for the
convergence in \ath\ does not work, as shown in
Fig.~\ref{fig:extrapolation_Be8_A3nmax}.  For $\hw=20\mev$ we observe
good convergence at $\ath=40$.  However, the quality of the
exponential fits in Fig.~\ref{fig:extrapolation_Be8_A3nmax}
deteriorates as \hw\ decreases below $20\mev$.  We conclude that simple
exponential extrapolation in \ath\ is not an option for $\hw=16\mev$
and below.  In particular, while $\hw=16\mev$ is close to converged
for \be, this may be less true for larger nuclei, so we will (perhaps
conservatively) only consider $\hw \ge 18\mev$ to be reliable in the
following.

\subsection{Convergence with model space size \nmax}

\begin{figure*}[p-]
\includegraphics*[width=5.7in]{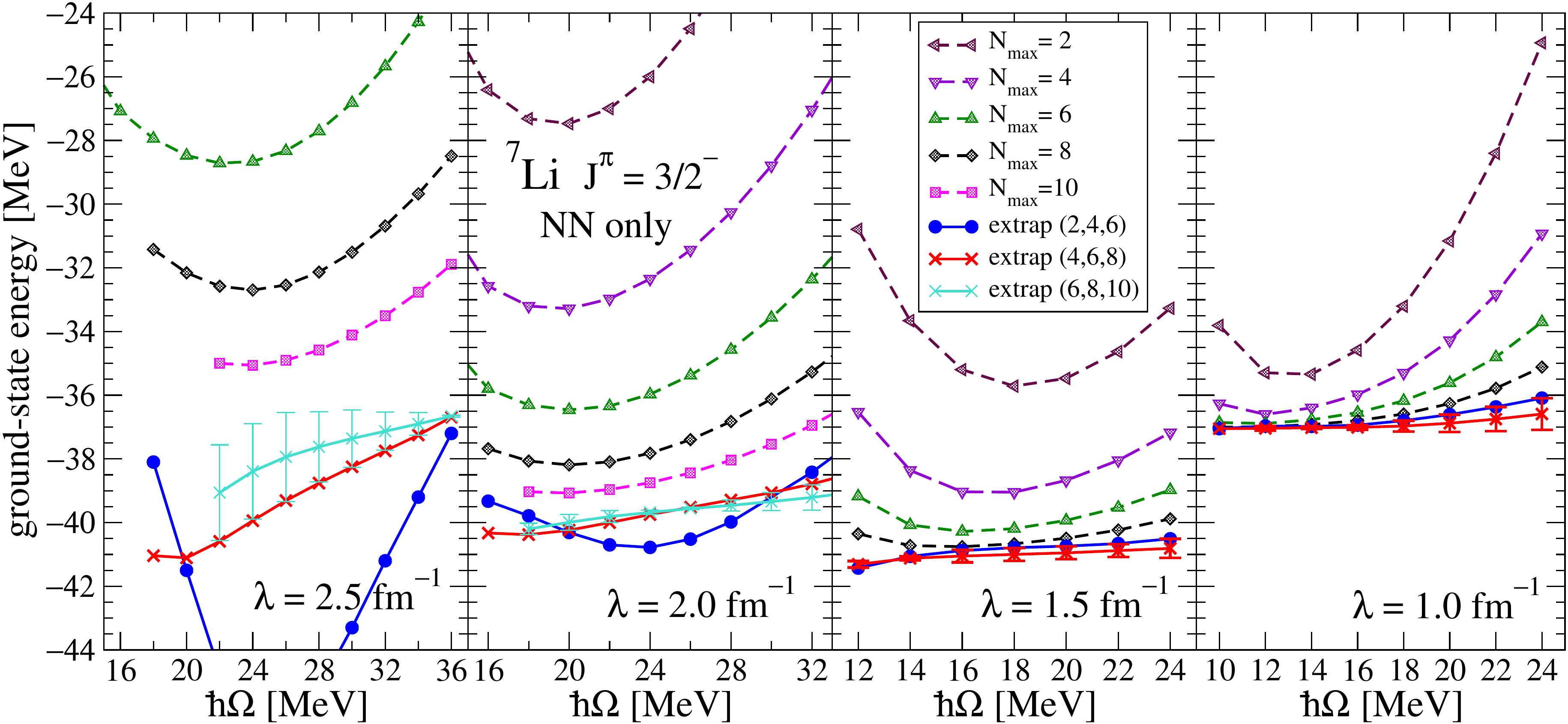}
\vspace*{-.1in}
\caption{(color online) Ground-state energy of $^{7}$Li 
for NN-only evolved Hamiltonians at $\lambda = 2.5$, $2.0$, $1.5$, and $1.0\fmi$, plus extrapolations based on ``Extrapolation B'' from Ref.~\cite{Maris:2008ax}.  See text for further details.}
\label{fig:Li7convergenceNNonly}

\vspace*{.1in}

\includegraphics*[width=5.7in]{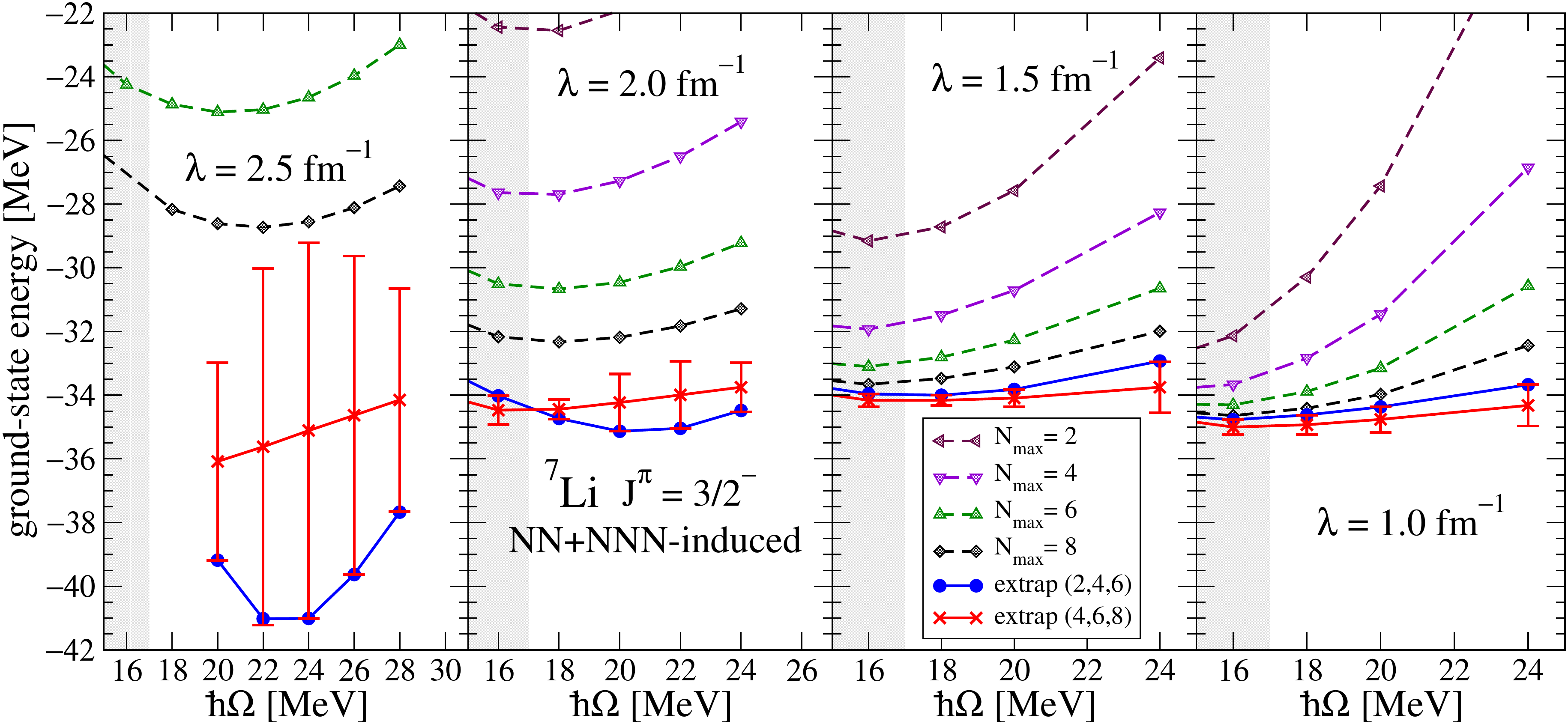}
\vspace*{-.1in}
\caption{(color online) Ground-state energy of $^{7}$Li  for NN+NNN-induced
evolved Hamiltonians at $\lambda = 2.5$, $2.0$, $1.5$, and $1.0\fmi$, plus extrapolations based on ``Extrapolation B'' from Ref.~\cite{Maris:2008ax}.  See text for further details.}
\label{fig:Li7convergenceNNNind}

\vspace*{.1in}

\includegraphics*[width=5.7in]{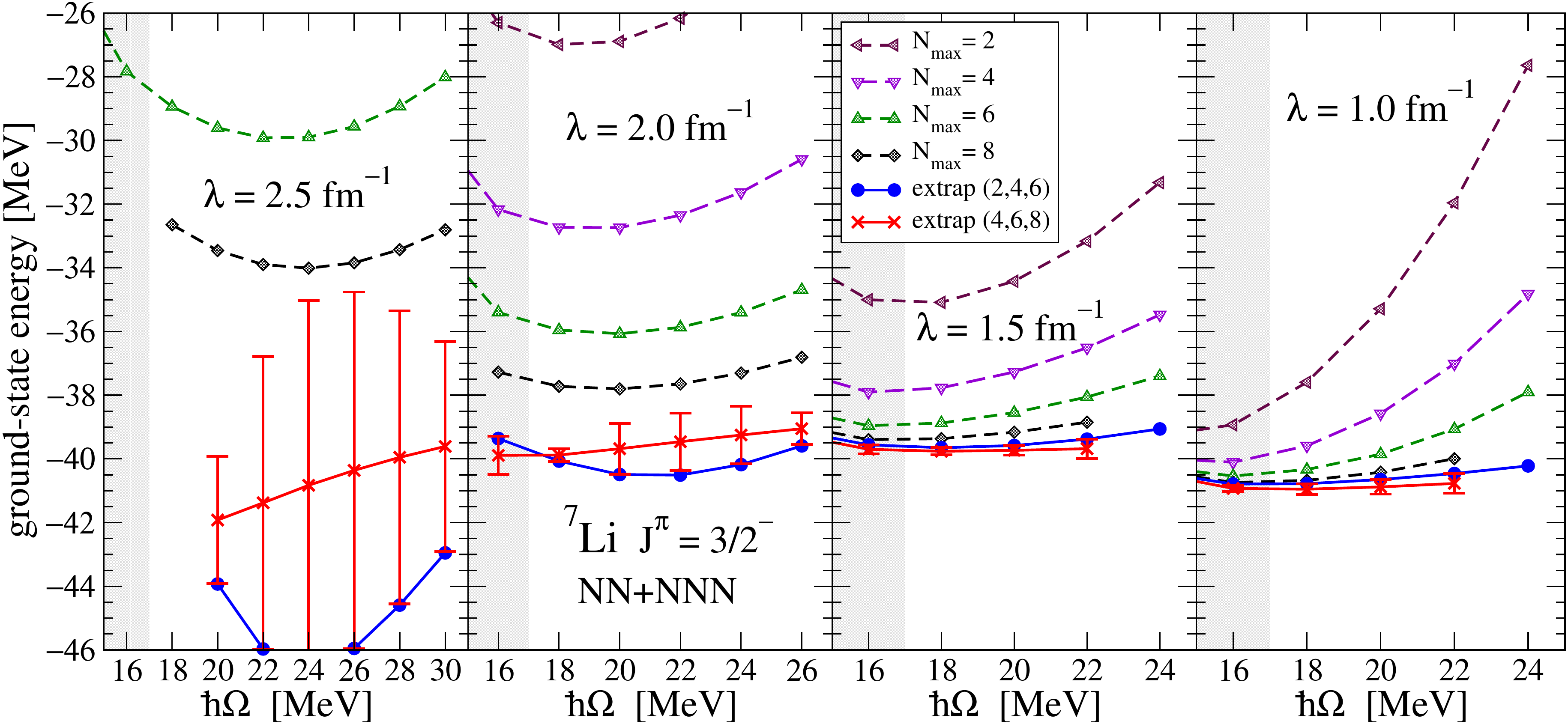}
\vspace*{-.1in}
\caption{(color online) Ground-state energy of $^{7}$Li 
for NN+NNN evolved Hamiltonians at $\lambda = 2.5$, $2.0$, $1.5$, and $1.0\fmi$, plus extrapolations based on ``Extrapolation B'' from Ref.~\cite{Maris:2008ax}.  See text for further details.}
\label{fig:Li7convergenceNNN}

\end{figure*}

\begin{figure*}[p-]
\includegraphics*[width=5.7in]{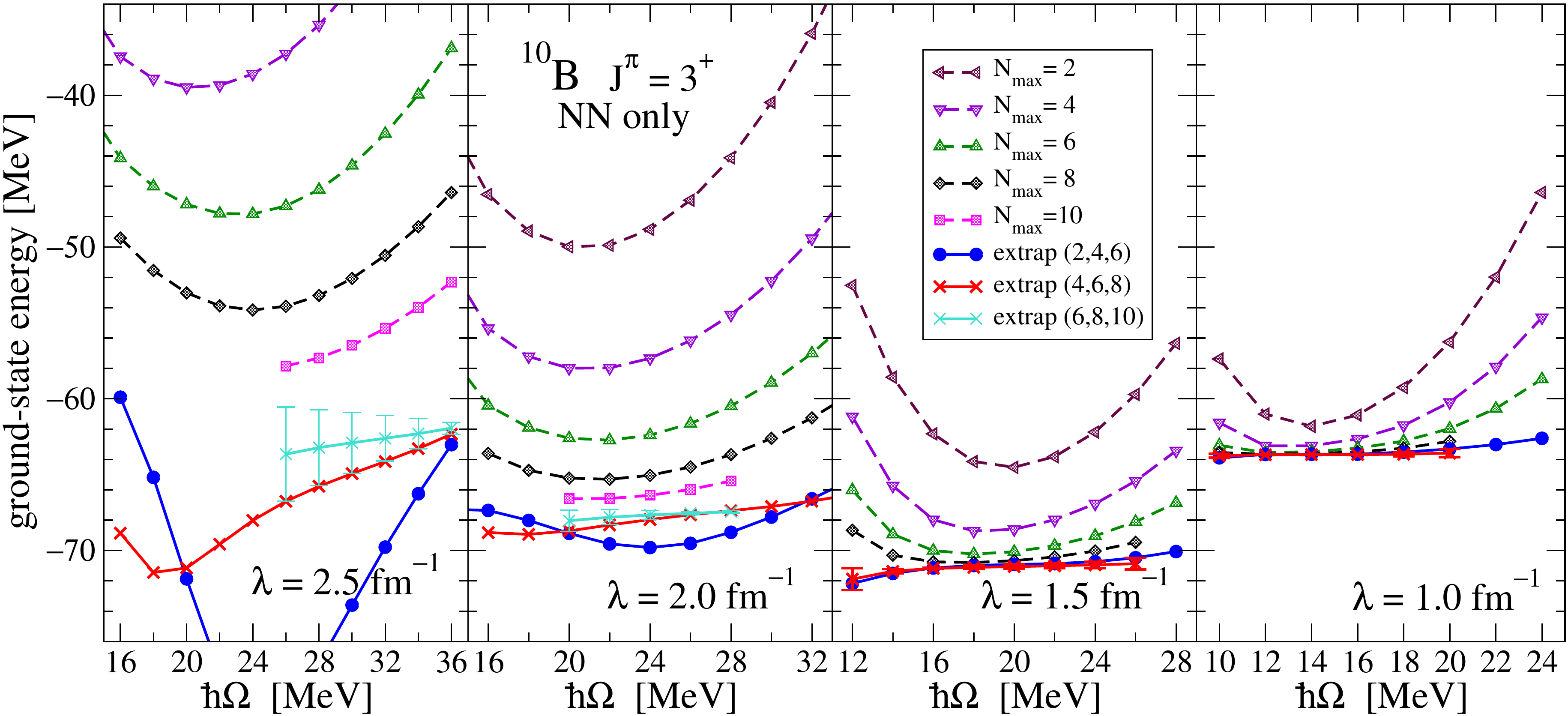}
\vspace*{-.1in}
\caption{(color online) Ground-state energy of $^{10}$B 
for NN-only evolved Hamiltonians at $\lambda = 2.5$, $2.0$, $1.5$, and $1.0\fmi$, plus extrapolations based on ``Extrapolation B'' from Ref.~\cite{Maris:2008ax}.  See text for further details.}
\label{fig:B10convergenceNNonly}

\vspace*{.1in}

\includegraphics*[width=5.7in]{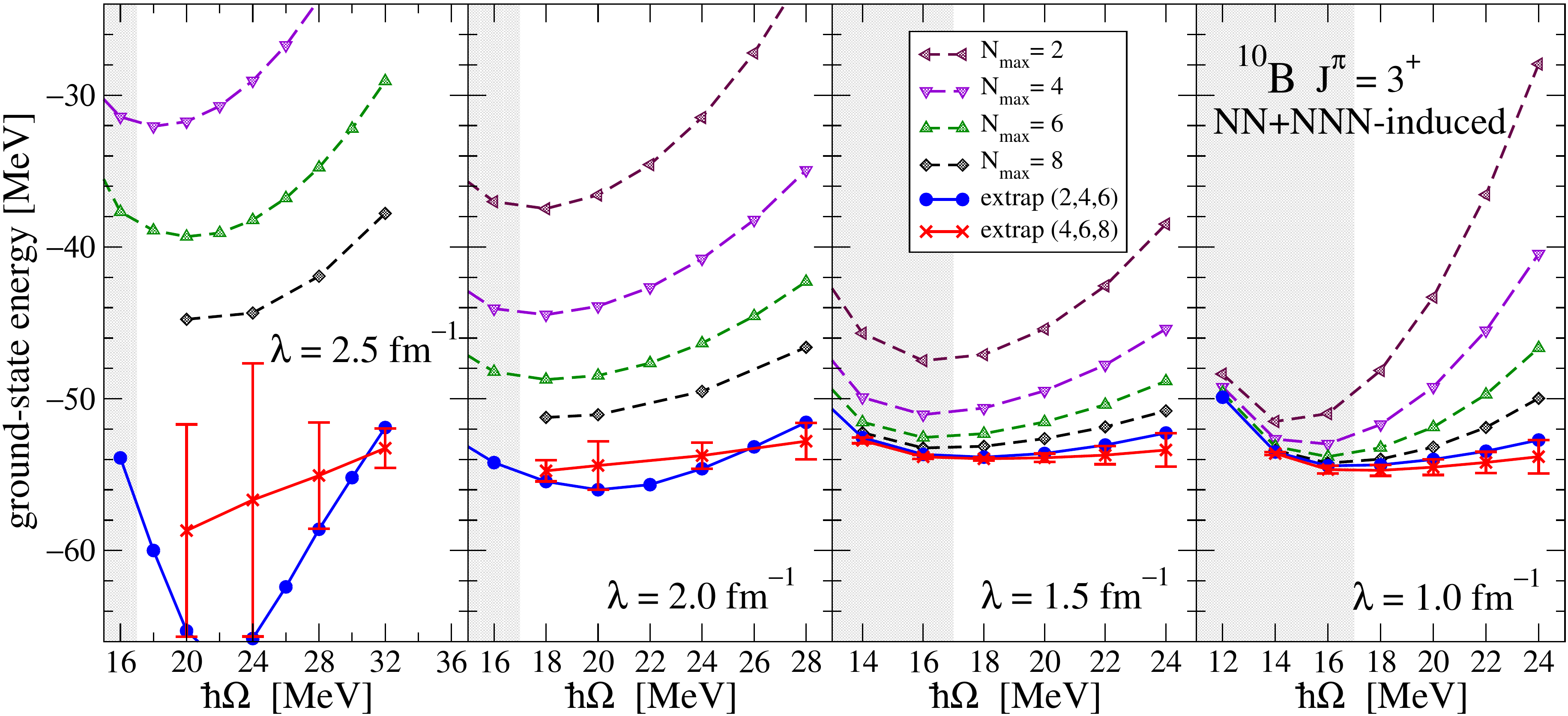}
\vspace*{-.1in}
\caption{(color online) Ground-state energy of $^{10}$B 
for NN+NNN-induced
Hamiltonians evolved to $\lambda = 2.5$, $2.0$, $1.5$, and $1.0\fmi$, plus extrapolations based on ``Extrapolation B'' from Ref.~\cite{Maris:2008ax}.  See text for further details.}
\label{fig:B10convergenceNNNind}

\vspace*{.1in}

\includegraphics*[width=5.7in]{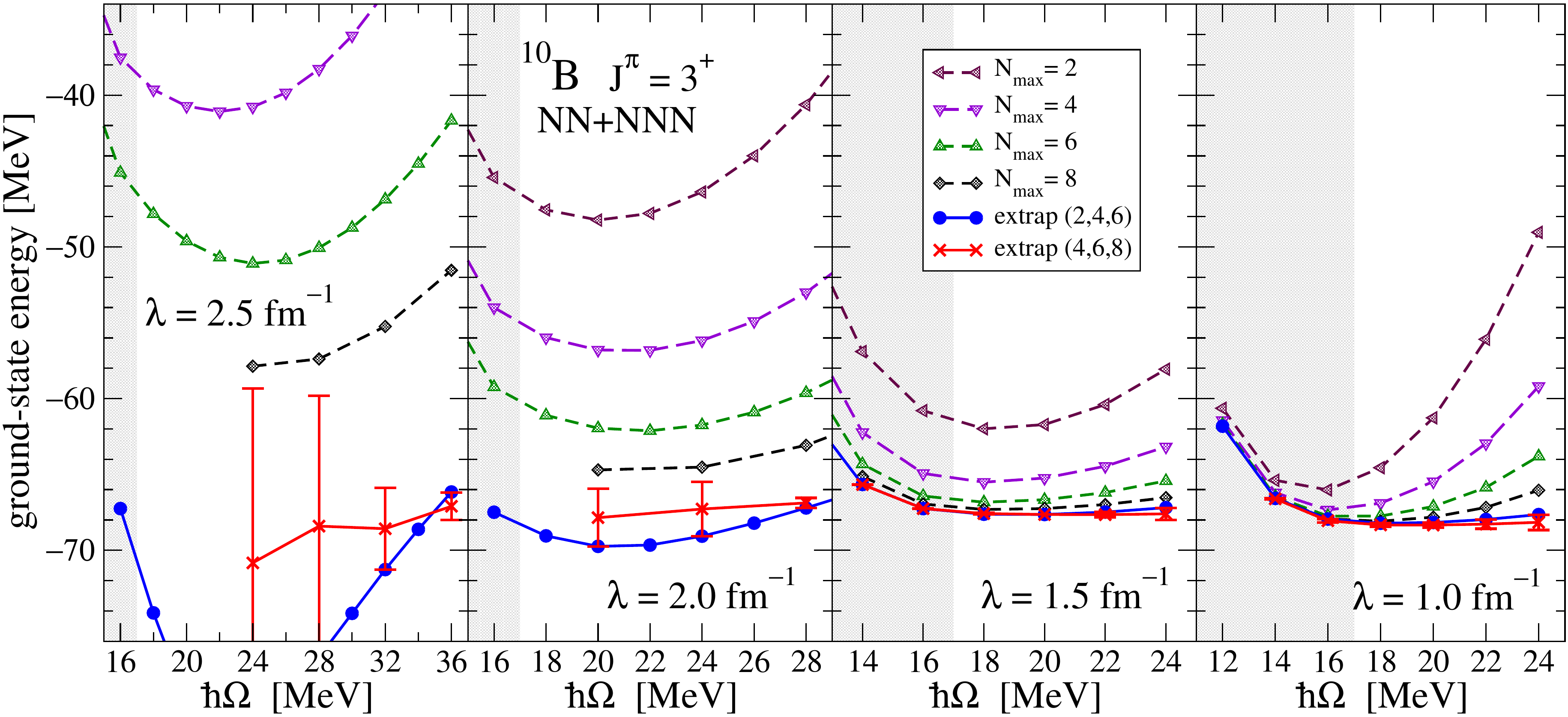}
\vspace*{-.1in}
\caption{(color online) Ground-state energy of $^{10}$B 
for NN+NNN evolved Hamiltonians at $\lambda = 2.5$, $2.0$, $1.5$, and $1.0\fmi$, plus extrapolations based on ``Extrapolation B'' from Ref.~\cite{Maris:2008ax}.  See text for further details.}
\label{fig:B10convergenceNNN}

\end{figure*}

\begin{figure*}[p-]
\includegraphics*[width=5.3in]{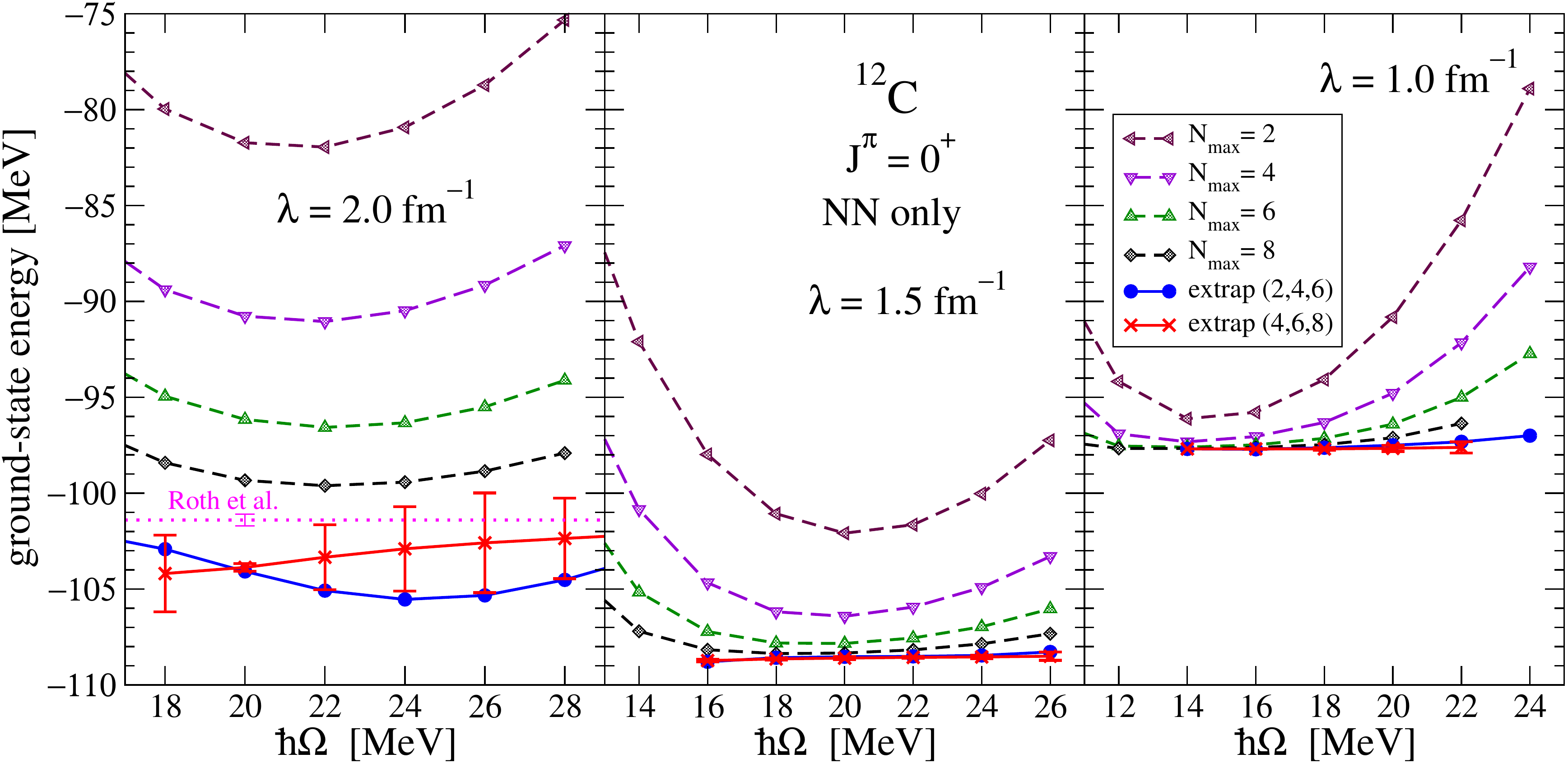}
\vspace*{-.1in}
\caption{(color online) Ground-state energy of $^{12}$C 
for NN-only evolved Hamiltonians at $\lambda = 2.0$, $1.5$, and $1.0\fmi$, plus extrapolations based on ``Extrapolation B'' from Ref.~\cite{Maris:2008ax}.  See text for further details.}
\label{fig:C12convergenceNNonly}

\vspace*{.1in}

\includegraphics*[width=5.3in]{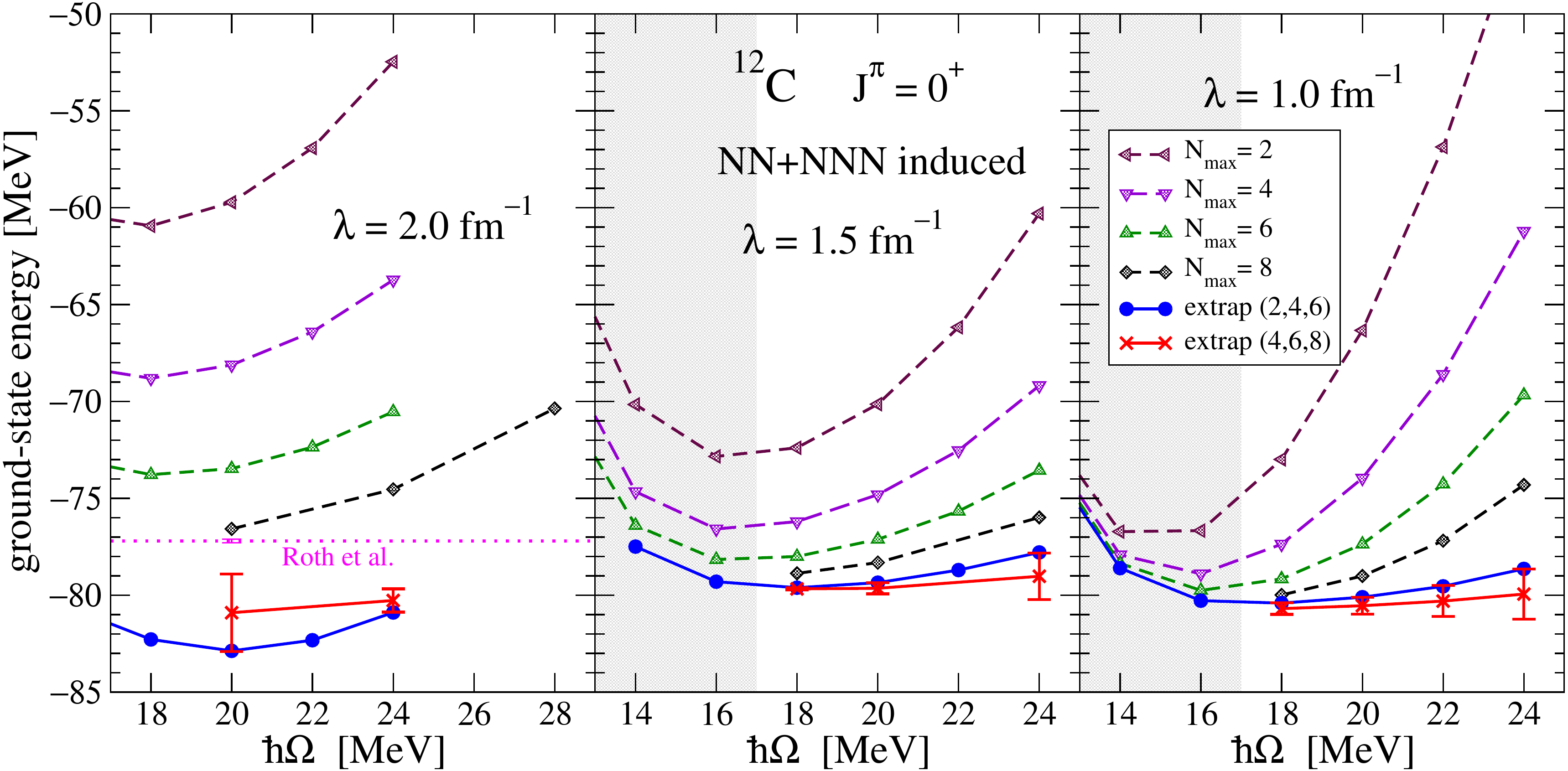}
\vspace*{-.1in}
\caption{(color online) Ground-state energy of $^{12}$C 
for NN+NNN-induced
Hamiltonians evolved to $\lambda = 2.0$, $1.5$, and $1.0\fmi$, plus extrapolations based on ``Extrapolation B'' from Ref.~\cite{Maris:2008ax}.  See text for further details.}
\label{fig:C12convergenceNNNind}

\vspace*{.1in}

\includegraphics*[width=5.3in]{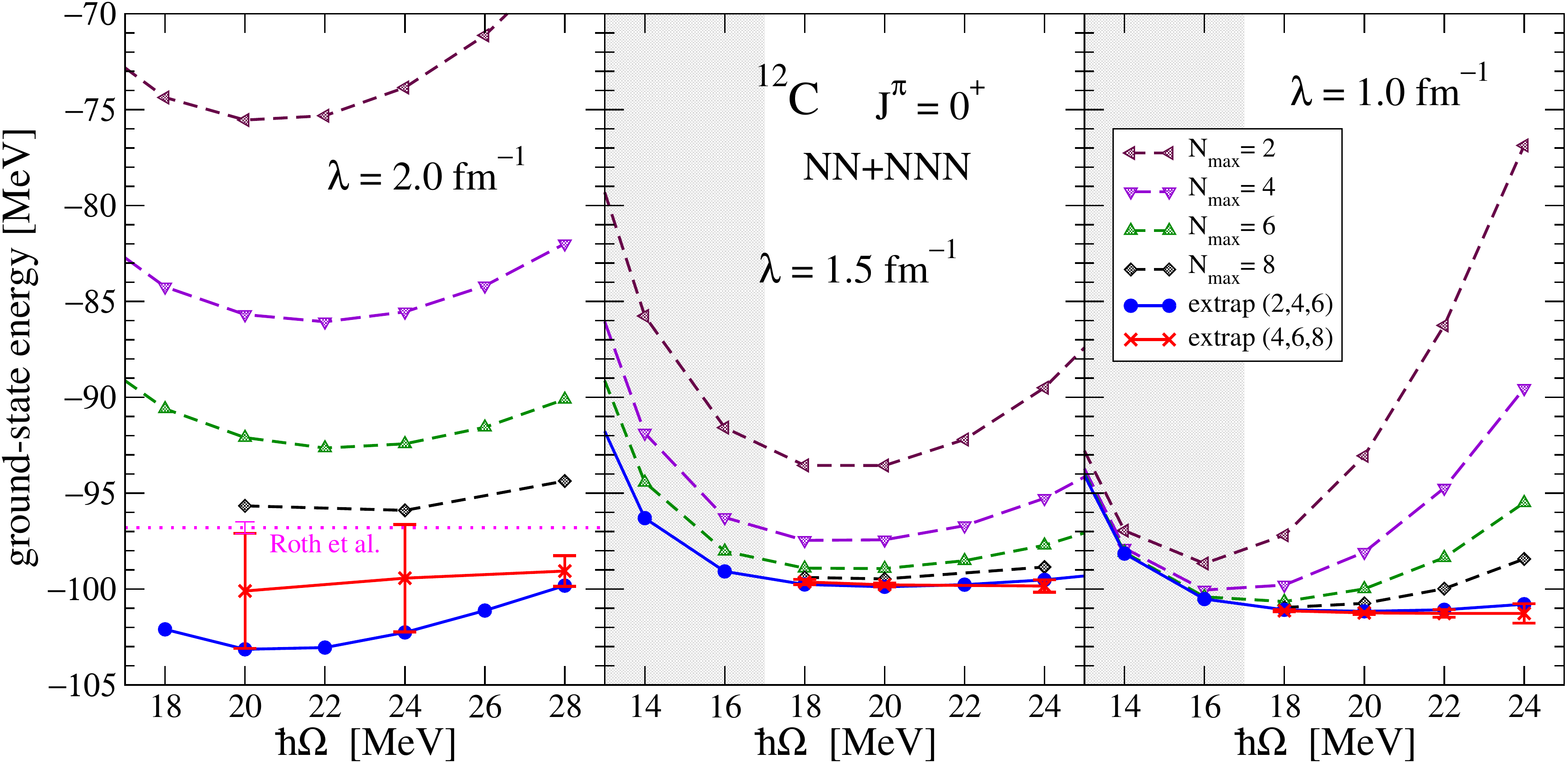}
\vspace*{-.1in}
\caption{(color online) Ground-state energy of $^{12}$C 
for NN+NNN evolved Hamiltonians at $\lambda = 2.0$, $1.5$, and $1.0\fmi$, plus extrapolations based on ``Extrapolation B'' from Ref.~\cite{Maris:2008ax}.  See text for further details.}
\label{fig:C12convergenceNNN}

\end{figure*}

In Figs.~\ref{fig:Li7convergenceNNonly} through
\ref{fig:C12convergenceNNN}, we show detailed results for the
ground-state energy for each of the calculations of
Table~\ref{tab:guide} as a function of \hw\ for three representative
nuclei (\li, \bo, and \ca) and basis sizes \nmax\ from 2 to 8 (and 10
for some NN-only cases).  Panels from left to right show results for
decreasing SRG $\lambda$.  In all panels the symbols connected by
dashed lines denote the \abinit calculated points at a single \nmax,
while the solid lines are exponential extrapolations at each
individual \hw\ based on ``Extrapolation B'' from
Ref.~\cite{Maris:2008ax}.  Extrapolations are given for
$\nmax=2\mbox{--}6$ and $\nmax=4\mbox{--}8$, with error bars in the
latter determined by the difference from the central prediction of the
former.  For \li\ and \bo\ NN-only calculations, there are also
extrapolations for $\nmax=6\mbox{--}10$ with error bars based on
$\nmax=4\mbox{--}8$; see Fig.~5 of Ref.~\cite{Maris:2013} for \ca\ for
results up to $\nmax=10$ with the same NN-only interactions, but
without the Coulomb interaction.  These figures display the
systematics of NCFC convergence of SRG-evolved interactions with and
without three-body forces.

Here, we make various summary observations based on these figures
(with further discussion of extrapolations in the next section):
\begin{itemize}

\item For all calculations, the rate of convergence with basis size
  \nmax\ is greatly accelerated with decreasing $\lambda$ for
  \hw\ near the variational minimum.  The addition of NNN interactions
  does not affect the rate substantially.  The dependence on \hw\ at
  low $\lambda$ is very flat for higher \nmax.

\item In almost all cases, the location of the variational
  minimum in \hw\ for a given \nmax\ shifts to smaller \hw\ as
  \lam\ decreases. This is expected because the minimum is where the
  ultraviolet (UV) and infrared (IR) corrections are roughly
  equal~\cite{Furnstahl:2012qg}.  (See
  Fig.~\ref{fig:extrap_IRUV_7Li_contributions} and the accompanying
  discussion below.)  The contribution from high momentum components
  of the interactions to low-energy states decreases as
  \lam\ decreases, so lower \lam\ Hamiltonians are less sensitive to
  the UV reach in momentum of a truncated HO basis.  In other words, the
  more evolved Hamiltonian is less sensitive to the UV cutoff
  intrinsic to the HO basis and thus the interplay between UV cutoff
  and IR distortion is shifted. This results in a systematic shift of
  the optimal \hw\ to lower values as the Hamiltonian is evolved.

\item The location of the minima for NN+NNN-induced calculations are
  systematically lower in \hw\ than for the corresponding NN+NNN
  interaction calculations.  This is consistent with the induced NNN
  interaction being softer than the initial NNN interaction.  The
  significance of high momentum components in the initial NNN
  interaction were already evident above where we saw that they are
  not as well converged in the $A=3$ space used for evolution and
  require larger \hw.  Furthermore, with NN+NNN-induced interactions,
  the nuclei are systematically less bound than with the NN+NNN
  interaction, so the wavefunctions will have longer-range exponential
  tails; as a consequence, the corresponding wavefunction is better
  represented with a lower \hw\ for NN+NNN-induced calculations than
  for NN+NNN calculations.

\item Results for $\lambda = 1.5\fmi$ and $1.0\fmi$ are sufficiently
  converged that extrapolations in the non-gray regions (gray
  shadowing signifies the unreliable region of \hw\ discussed above)
  are the same for different \hw\ within the (small) error bars.  Near
  the minima in this $\lambda$ range extrapolation is largely
  superfluous.  In cases including initial or induced NNN
  interactions, the minimum is sometimes in the gray region.  However,
  in such cases the convergence for larger \hw\ is well advanced and
  the individual \hw\ extrapolations are consistent with each other.

\item Extrapolated results for $\lambda = 2.5\fmi$ (and to some degree
  for $\lambda = 2.0\fmi$) appear to depend systematically on \hw.
  Although it has generally been considered most reliable to
  extrapolate using energies for \hw\ near the variational minimum,
  selected results in $\nmax=10$ spaces (e.g., for \bo\ NN-only)
  suggest that the extrapolations from 4,6,8 are overbound near the
  minimum for larger $\lambda$ (and 2,4,6 even more so), so that
  estimates from larger \hw\ are more robust.  This may be
  related to the fact that the location of the variational minimum is
  shifting to larger values of \hw\ as \nmax\ increases.  In general,
  our $\lambda \geq 2.2\fmi$ extrapolations for these
  nuclei have error bars too large to allow useful quantitative
  conclusions about \lam\ dependence.

\item In Figs.~\ref{fig:C12convergenceNNonly},
  \ref{fig:C12convergenceNNNind}, and \ref{fig:C12convergenceNNN}, the
  $\lambda = 2\fmi$ results for \ca\ at $\hw=20\mev$ include a
  horizontal line marking the best extrapolated value (with error bar)
  from the IT-NCSM calculations of Roth et al~\cite{Roth:2011ar}. The
  $\nmax \le 6$ values are relatively consistent with the extrapolated
  IT-NCSM results, but beginning with the extrapolated NCFC results that
  include the $\nmax=8$ points, we appear to predict somewhat more
  binding than the IT-NCSM extrapolated result, though our uncertainties
  are large enough that a definitive conclusion is elusive.  Note,
  however, that the prescription for the \ath\ truncation is slightly
  different in Ref.~\cite{Roth:2011ar} than what we use here.  Both in
  Ref.~\cite{Roth:2011ar} and in our calculations use $\ath=40$ for the
  leading, $J=\frac{1}{2}$, 3NF contributions, but we use different
  cutoffs for the higher-$J$ terms.  A detailed analysis of the effect
  of importance truncation 
  for this particular case is in progress~\cite{Calci:2013}.

\item It is evident by comparing extrapolated values for NN-only to
  the NN+NNN calculations that the \lam\ dependence of the
  extrapolated energies is significantly reduced by including 3NFs.
  This is summarized in the figures discussed in
  Section~\ref{sec:evolution}.

\end{itemize}

\noindent
Overall, the convergence patterns with \nmax\ and \hw\ previously
observed as a consequence of NN-only SRG
evolution~\cite{Bogner:2007rx} are still present when 3NF
contributions are included.

\subsection{Comparison of extrapolation methods
\label{subsec:comparisons}}

The extrapolations in Figs.~\ref{fig:Li7convergenceNNonly} through
\ref{fig:C12convergenceNNN} are a series of individual extrapolations
using Eq.~(\ref{eq:Ealphai}), each for a fixed value of \hw.  As noted
earlier, the plotted error bars for each \hw\ are determined by
comparison of each individual $\nmax=4\mbox{--}8$ extrapolation to
that obtained from $\nmax=2\mbox{--}6$.  The resulting set of
predictions can then be analyzed to obtain a predicted energy and
overall error bar~\cite{Maris:2008ax}.  We consider two criteria
for determining a best ``Extrapolation B'' result for \Einf\ and the
corresponding error bar:
\begin{itemize}
\item the result at the \hw\ value for which the amount of
  extrapolation is minimal (i.e. the point where 
  $E(\nmax)-\Einf$ is minimal);
\item the result at the \hw\ value for which the numerical error
  estimate is minimal.
\end{itemize}
In both cases, we restrict the best ``Extrapolation B'' to \hw\ values
at or above the variational minimum at the highest \nmax\ employed in
the extrapolation.  For the error estimates we use the average of the
error bars in a region of 8\,MeV around this best \hw\ value.  This
initial error estimate is enlarged as necessary in order to get
consistent results, such that the central values are within the error
estimate in the entire 8\,MeV range.

Note that below $\lam = 1.5\fmi$ with 3NFs, we cannot apply either of
these criteria, because we are only using results for $\hw \ge
18\mev$, and the variational minimum is typically at or below
$\hw=18\mev$ for the lowest \lam\ values; however, for these values of
\lam\ the results are quite close to convergence, and we base our
error estimate on the results for $18 \le \hw \le 24\mev$.  Above
$\lam = 1.8\fmi$, the two criteria give (slightly) different results,
but generally with overlapping extrapolation error estimates.  

The convergence pattern and extrapolations of the NN-only data up
through $\nmax=10$ (see Figs.~\ref{fig:Li7convergenceNNonly} and
\ref{fig:B10convergenceNNonly}, as well as Fig.~5 of
Ref.~\cite{Maris:2013}), suggest that for $\lam = 2.5\fmi$ the
\hw\ value that minimizes the numerical error estimate, $\hw \simeq
36\mev$, is more reliable for the extrapolation (at least at this
\lam\ value) than the \hw\ value that minimizes $E(\nmax)-\Einf$.
(With the nonlocal NN interaction JISP16,
for which this extrapolation has been used
extensively~\cite{Maris:2008ax,Maris:2009bx,Cockrell:2012vd} these two
criteria rarely lead to significant differences.)

An alternative approach is to use a constrained optimization that uses
all (or a specified subset) of \hw\ data, requiring the same
extrapolated energy $\Einf$ in Eq.~\eqref{eq:Ealphai} for every \hw,
such as ``Extrapolation A'' of Ref.~\cite{Maris:2008ax}.  Here, we use
basically the same procedure for the constrained optimization, using
four subsequent \nmax\ values and five \hw\ values for each
constrained fit, with the same weights and error estimate as in
Ref.~\cite{Maris:2008ax}.  (For the current calculations five
\hw\ values span an 8\,MeV range of \hw, whereas in
Ref.~\cite{Maris:2008ax} it spans a 10\,MeV range.)  Again, with 3NFs
this procedure cannot be applied below $\lam = 1.5\fmi$, but the
results are close to convergence for these cases, and therefore less
sensitive to the details of the extrapolation.  For $1.5\fmi \ge \lam
\ge 2\fmi$ the results from ``Extrapolation A'' are consistent with
those from ``Extrapolation B'', but for $\lam > 2 \fmi$ the procedure
of Ref.~\cite{Maris:2008ax} for ``Extrapolation A'', modified as
described above, leads to results that show a systematic deficiency
at $\nmax=10$.

\begin{figure}[tbh-]
\includegraphics*[width=3in]{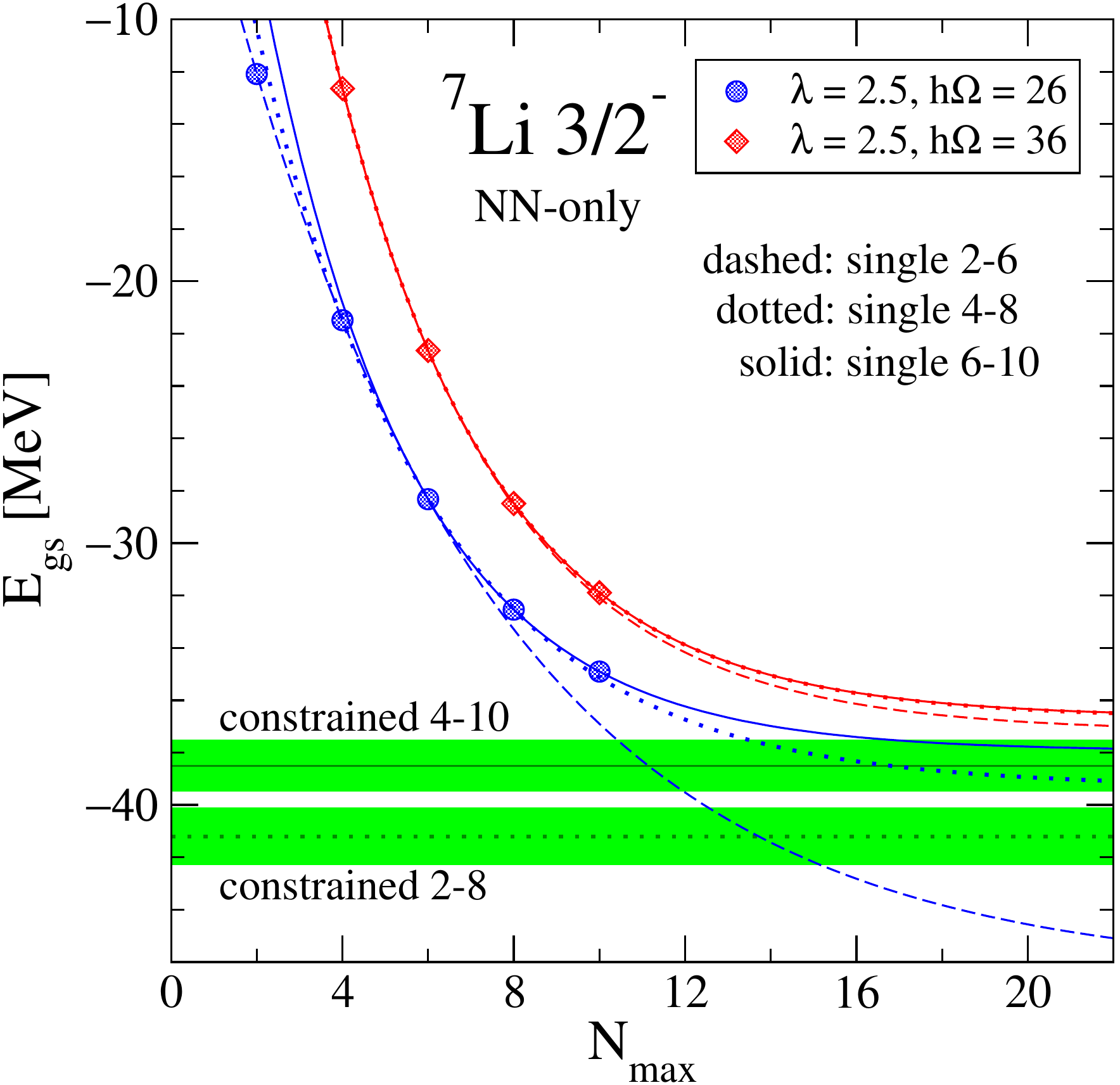}
\caption{(color online) Ground-state energy extrapolations of \li\ as
  a function of \nmax\ with an N$^3$LO NN
  interaction~\cite{Entem:2003ft} evolved to $\lam=2.5\fmi$.  The
  symbols are the calculated points.  The curves show single
  extrapolations using Eq.~(\ref{eq:Ealphai}) with $\nmax=2\mbox{--}6$
  (dashed), $4\mbox{--}8$ (dotted) and $6\mbox{--}10$ (solid) at (blue)
  $\hw=26\mev$ which minimize the amount of extrapolation and at (red)
  $\hw=36\mev$ which minimize the numerical error estimate.  The
  horizontal dotted and solid lines, with the band indicating the
  associated error bars, are the result from a constrained fit
  following the procedure of Ref.~\cite{Maris:2008ax} for five
  \hw\ values from $22$ to $30\mev$.
  \label{fig:extrap_example}}
\end{figure}
Figure \ref{fig:extrap_example} shows a comparison of the single and
the constrained extrapolation schemes for $\lam=2.5\fmi$.  In this
example we use \li\ with NN-only calculations up to $\nmax=10$.  The
variational minimum is at $24\mev$, so we use five \hw\ values from
$22\mev$ to $30\mev$, and four \nmax\ values for each \hw\ for the
each of the constrained fits.  Figure \ref{fig:extrap_example} shows
clearly that this procedure with the $\nmax=2$ to $8$ results leads to
an overestimate of the binding energy and an underestimate of the
extrapolation uncertainty.  In Fig.~\ref{fig:extrap_example}, we can
also see why this procedure leads to erroneous results: near the
variational minimum the convergence is not a simple exponential for
this value of \lam, as is evident from the single \hw\ fits (blue
curves).

\begin{figure}[tbh-]
\includegraphics*[width=3in]{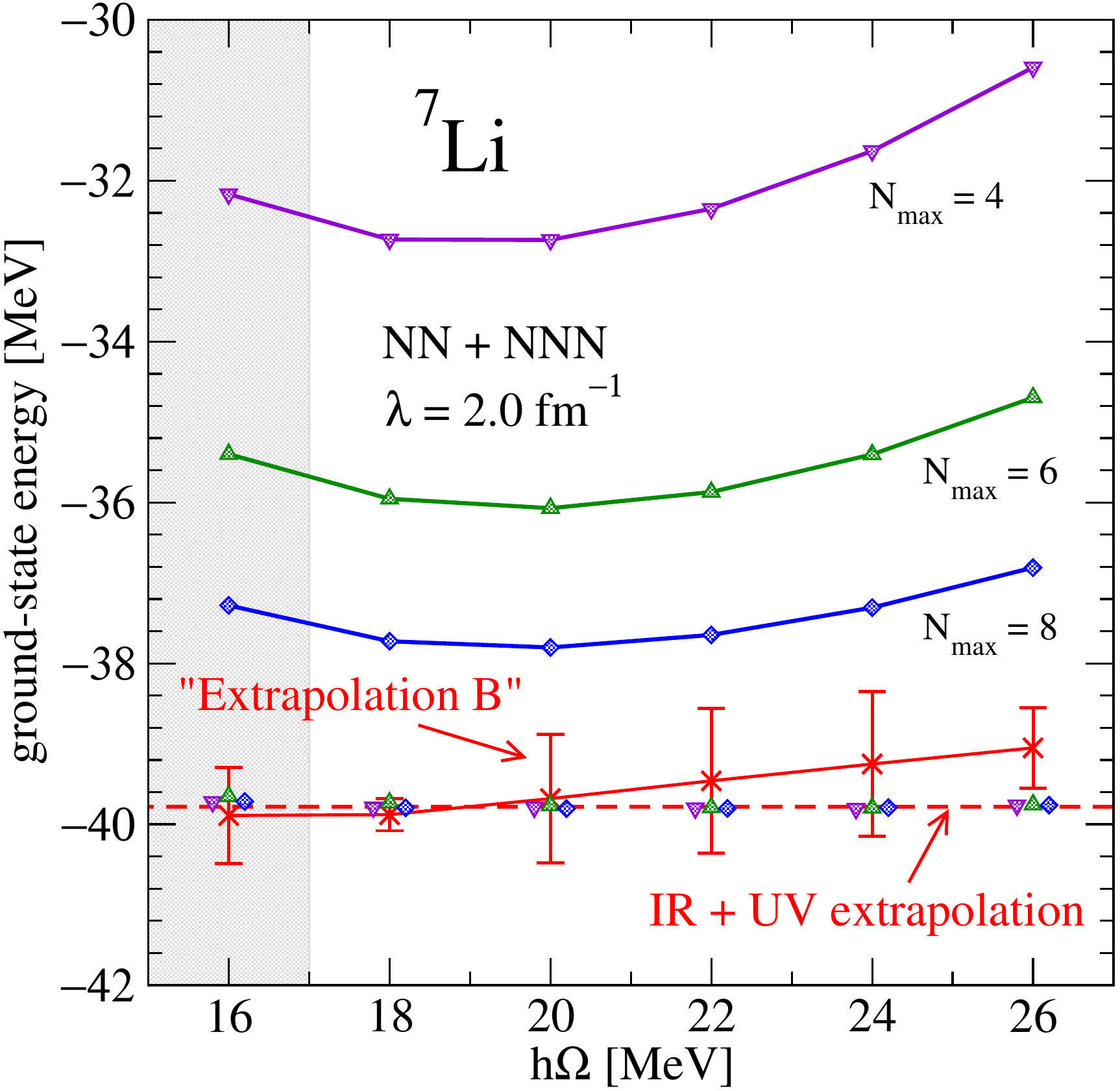}
\caption{(color online) Ground-state energy of \li\ for NN+NNN evolved
  Hamiltonians at $\lambda = 2.0\fmi$, plus extrapolations based on
  ``Extrapolation B'' from Ref.~\cite{Maris:2008ax} (solid line and
  points with error bars) and on the combined IR/UV correction formula
  Eq.~\eqref{Ecombi} that yields $\Einf$ (dashed line) and individual
  corrections for each \hw\ and \nmax\ combination (points near the
  dashed line) based on the single set of best-fit parameters.
  \label{fig:extrap_IRUV_7Li}}
\end{figure}

\begin{figure}[tbh-]
\includegraphics*[width=3in]{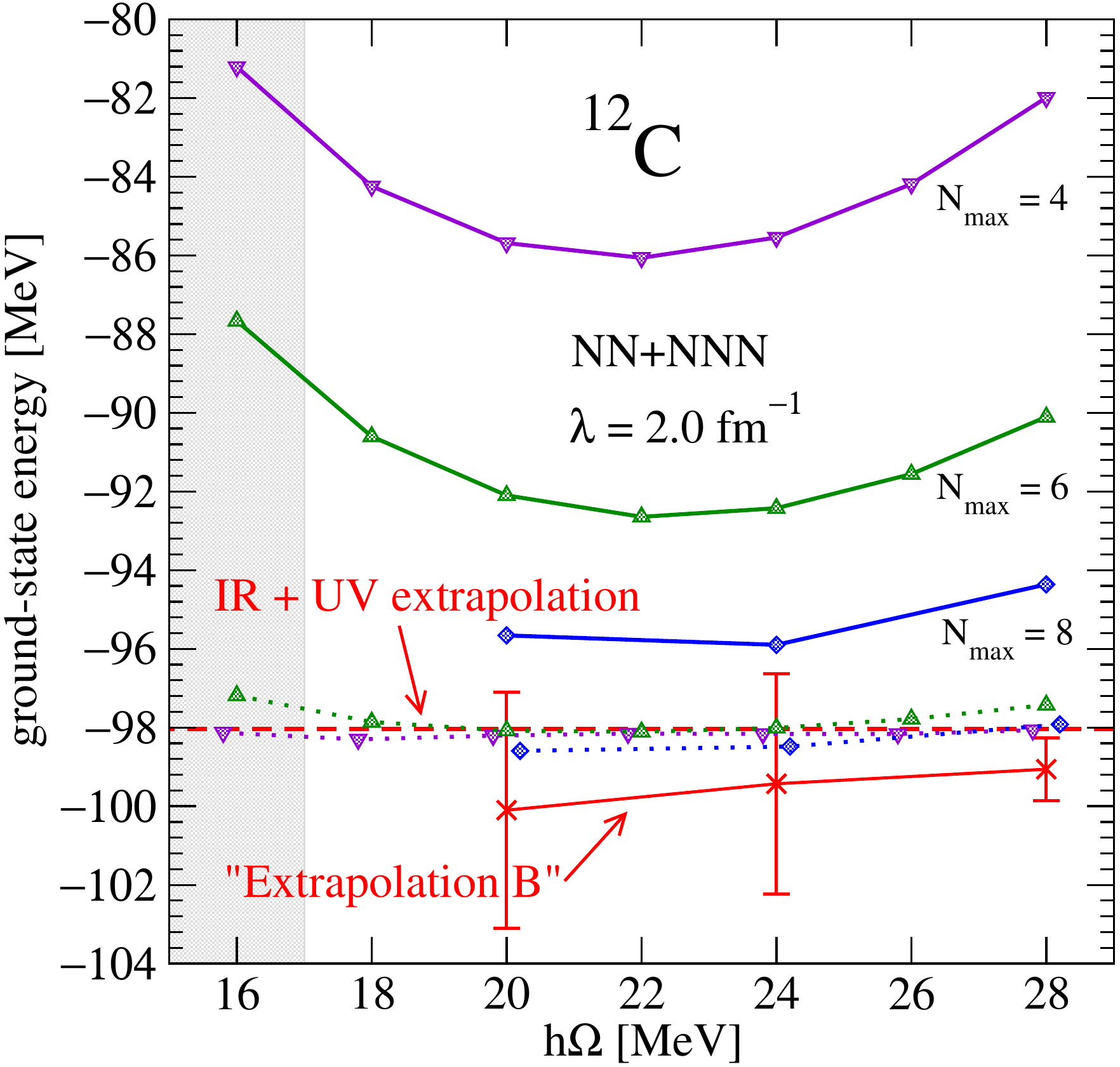}
\caption{(color online) Ground-state energy of \ca\ for the NN+NNN
  evolved Hamiltonians at $\lambda = 2.0\fmi$, plus extrapolations
  based on ``Extrapolation B'' from Ref.~\cite{Maris:2008ax} (solid
  line and points with error bars) and on the combined IR/UV
  correction formula Eq.~\eqref{Ecombi} that yields $\Einf$ (dashed
  line) and individual corrections for each \hw\ and
  \nmax\ combination (points near the dashed line) based on the single
  set of best-fit parameters.
  \label{fig:extrap_IRUV_12C}}
\end{figure}

The alternative EFT-motivated approach described in
Sec.~\ref{subsec:extrapolation} is complementary to the schemes used
in Figs.~\ref{fig:Li7convergenceNNonly}--\ref{fig:extrap_example}.  A
sample application of Eq.~\eqref{Ecombi}, which includes both UV and
IR corrections to the energy at each \nmax\ and \hw, is shown in
Fig.~\ref{fig:extrap_IRUV_7Li} for the NN+NNN calculation of the
\li\ ground state.  In this example, the fifteen points with $\hw \ge
18\mev$ and \nmax\ from 4 to 8 are inputs to a simultaneous fit of
the five parameters of $E(\LamUV,L)$ with $\LamUV$ and $L$ given as
functions of \hw\ and \nmax\ (see Sec.~\ref{subsec:extrapolation} for
the formulas, recalling that we will use $N = \nmax +3$ for p-shell
nuclei).  The result for $\Einf$ (dashed line) is consistent with the
individual ``Extrapolation B'' results (from
Fig.~\ref{fig:Li7convergenceNNN}) and the overall result is within the
error bars of that scheme.  An error analysis procedure for the IR/UV
correction model is not yet available, which limits its utility for
the present analysis, but the goodness-of-fit can be assessed by the
(very small) spread of corrected points about the dashed line.
Corrected points represent the use of the best fit parameters with
Eq.~(\ref{Ecombi}) (except $\Einf$) to extend each finite basis result
to infinite \nmax\ at fixed \hw.  Note also the predictions for the
$\hw=16\mev$ points, although those data points were not included in
the fit.

A second extrapolation based on the IR/UV correction model is shown in
Fig.~\ref{fig:extrap_IRUV_12C}, where the spread of corrections
indicates a still good but less-than-ideal fit for the NN+NNN
calculation of the \ca\ ground state.  Note that the largest
deviations of the corrected results from the fit $\Einf$ occur for two
of the three highest \nmax\ points.  The implication is that the true
$\Einf$ should be slightly more negative, which is also the conclusion
from comparing with the ``Extrapolation B'' analysis.  Note, however,
that although we used the same total number of points for this IR/UV
extrapolation, we have only three $\nmax=8$ points (see
Fig.~\ref{fig:C12convergenceNNN}, compared to five of the fifteen
points that we used for Fig.~\ref{fig:extrap_IRUV_7Li}.

In general, a good fit requires the UV and IR functional forms to be
adequate models for smaller \nmax\ values (with increasing
\nmax\ there is decreasing sensitivity while the computational cost is
increasing dramatically).  Much remains to be explored for heavier
nuclei but detailed investigation of two-particle models and the
deuteron suggest that the forms in Eq.~\eqref{Ecombi} can be improved
and that the predictions can be sensitive to optimizing the choice of
expressions for $L$ and $\LamUV$~\cite{More:2013aa} (e.g., using
$\nmax+3/2+2$ rather than $\nmax+3/2$ for two-body systems).  We are
also not yet able to take advantage of the theoretical prediction that
$\kinf$ should be related to the nucleon separation energy and the
empirical observation that $B_1$ is found to be numerically close to
\lam.

The two examples considered so far are for $\lam=2\fmi$.  For $\lambda
\leq 1.5\fmi$, the new extrapolation method using data up to $\nmax=8$
gives predictions for the $\nmax=\infty$ energies consistent with the
other extrapolation schemes.  As already noted, there are some
systematic differences for $\lam=2\fmi$, but they are within the
``Extrapolation B'' uncertainties.  For larger $\lam$ these
differences grow, but it is not possible at present to determine which
approach is superior.

\begin{figure}[tbh-]
\includegraphics*[width=3in]{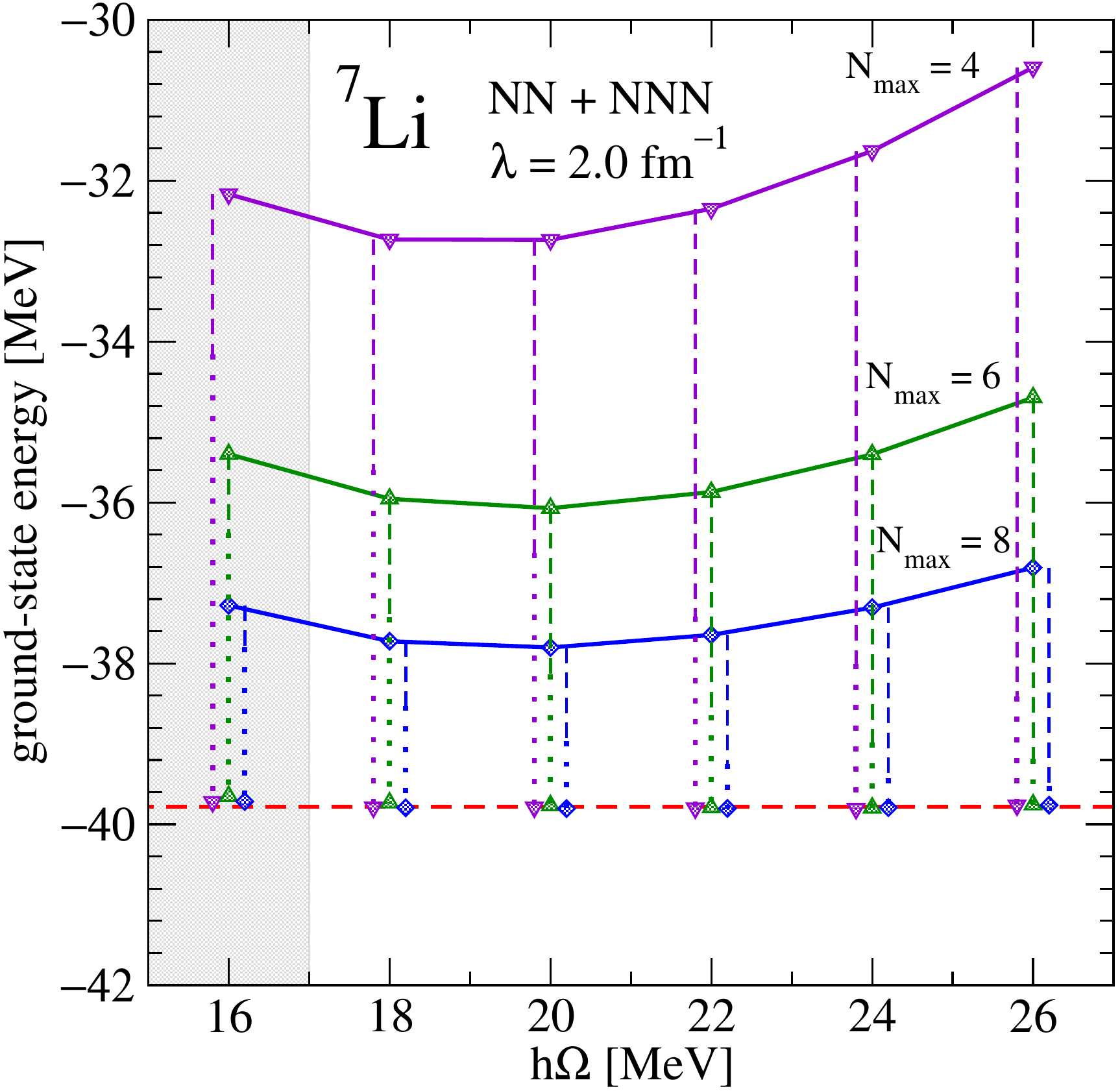}
\caption{(color online) Ground-state energy of \li\
for the NN+NNN evolved Hamiltonians at $\lambda = 2.0\fmi$, with IR
(vertical dashed) and UV (vertical dotted) 
corrections from Eq.~\eqref{Ecombi} that add to predicted
$\Einf$ values (points near the horizontal dashed line, which is
the global $\Einf$).}
\label{fig:extrap_IRUV_7Li_contributions}
\end{figure}

\begin{figure}[tbh-]
\includegraphics*[width=3in]{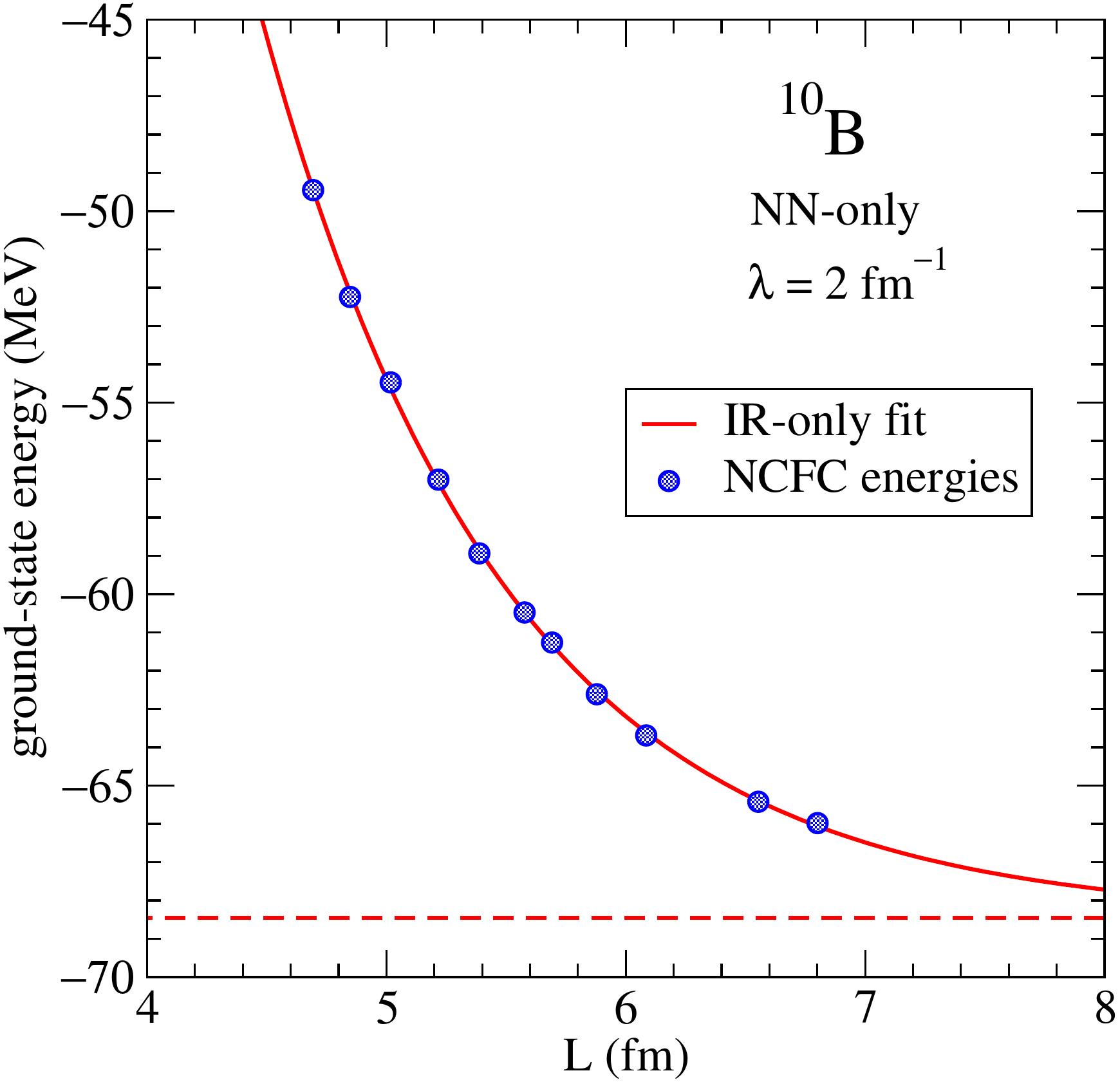}
\caption{(color online) Ground-state energy of \bo\
for the NN-only evolved Hamiltonians at $\lambda = 2.0\fmi$
for $\nmax=4\mbox{--}8$ and $\hw=28\mbox{--}32\mev$ plus $\nmax=10$ and $\hw=26,28\mev$ with
an IR-only fit using Eq.~\eqref{EIRonly}.
The fit value of $\Einf$ is the dashed line.
}
\label{fig:extrap_IRonly_10B}
\end{figure}

It is instructive after making a global fit to decompose each
correction for a given \hw\ and \nmax\ into the individual IR and UV
contributions.  This is done in
Fig.~\ref{fig:extrap_IRUV_7Li_contributions} for the \li\ fit of
Fig.~\ref{fig:extrap_IRUV_7Li}.  This figure verifies our prior claim
that IR and UV corrections are roughly equal at the variational
minima, while the IR(UV) correction rapidly dominates when we move to
the right(left) of a minimum.  If there are enough $(\hw,\nmax)$
points calculated where one of the two corrections is numerically
insignificant, a simpler extrapolation with only three fit parameters
is possible.  
An example of an IR-only fit is given for the NN-only
calculation of \bo\ at $\lam=2\fmi$ is given in
Fig.~\ref{fig:extrap_IRonly_10B}.  
Such IR-only fits are not possible for the calculations here including NNN
because we do not have enough points
sufficiently removed from the minimum in \hw.

The points in Fig.~\ref{fig:extrap_IRonly_10B} were chosen to the right
of the variational minimum (see Fig.~\ref{fig:B10convergenceNNonly}),
where a fit to Eq.~\eqref{Ecombi} implies that the UV correction to
these points is much smaller than the IR correction.  The observation
that the points with different \hw\ and \nmax\ values all lie on the
same curve verifies that $L$ is the correct variable.  
We note similar demonstrations for $^6$He results in
Fig.~4 of Ref.~\cite{Furnstahl:2012qg} with the same NN-only
interaction and in Fig.~11 of Ref.~\cite{Coon:2012ab} using a
different NN interaction.
The fit to
\beqn 
  E(L) \approx \Einf + B_2 e^{-2\kinf L} \;,
   \label{EIRonly}
\eeqn 
for the energies
shown in Fig.~\ref{fig:extrap_IRonly_10B} is very good.  
However,
the prediction for $\Einf$  is about 0.6 to 1\,MeV more
bound than that from the ``Extrapolation B'' analysis.

It would be premature to draw robust conclusions on the relative
efficacy of the extrapolation schemes used here.  In particular,
further comparisons are needed where large \nmax\ results are
available to check small \nmax\ extrapolations.  However, for our
present purposes it is sufficient that the results of the different
schemes are consistent with each other to within the assessed
uncertainties currently available.  For the remainder of this work, we
will use extrapolation procedures based on Eq.~(\ref{eq:Ealphai}).

%%%%%%%%%%%%%%%%%%%%%%%%% Evolution Plots %%%%%%%%%%%%%%%%%%%%%%%%%%%%%
\section{Evolution}
\label{sec:evolution}

\subsection{Running of ground-state energies}

\begin{figure}[tbh-]
\includegraphics*[width=0.9\columnwidth]{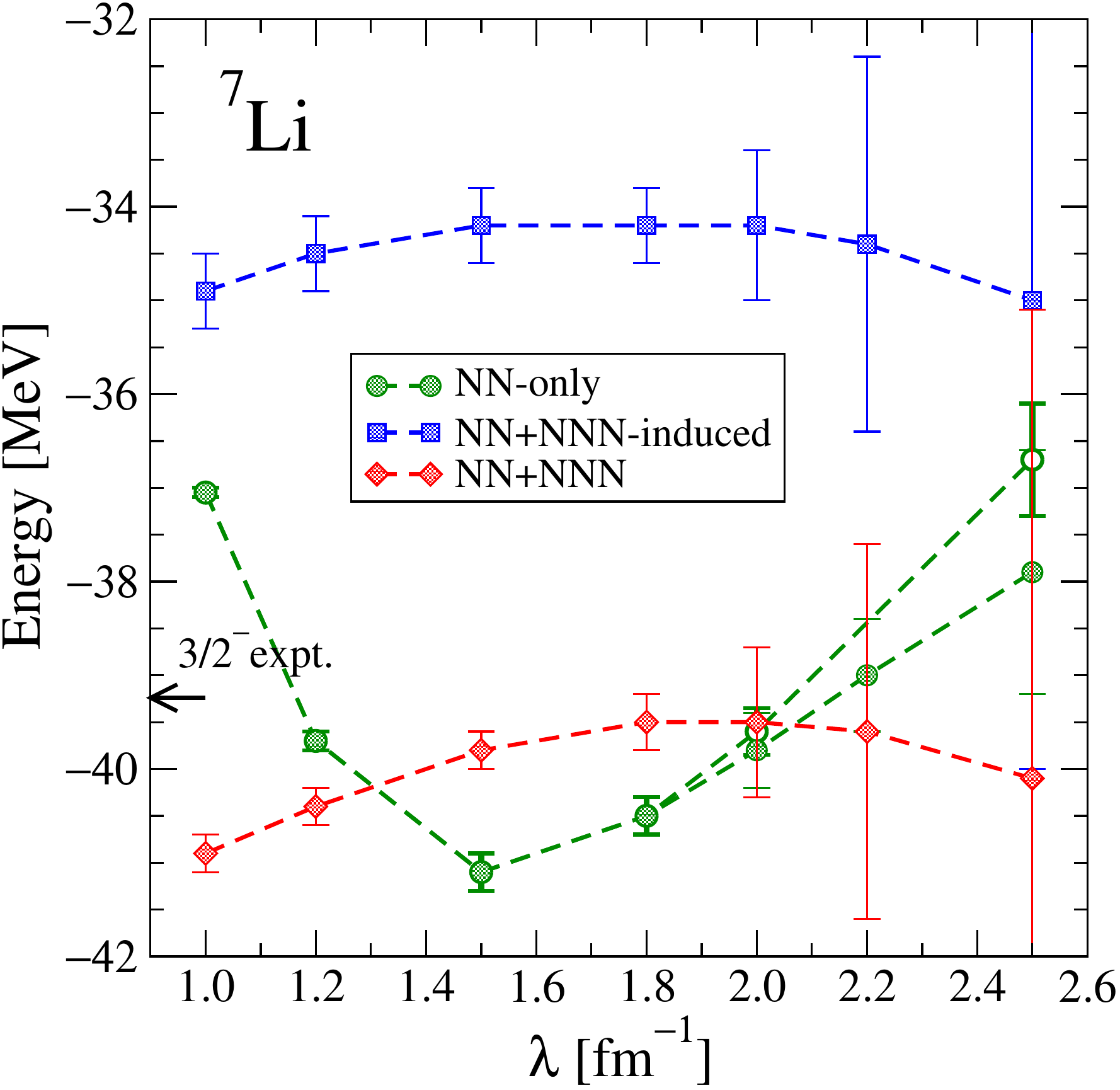}
\caption{(color online) Extrapolated ground-state energy of \li\ as a
  function of \lam\ for each SRG calculation.  The initial interaction
  was N$^3$LO NN~\cite{Entem:2003ft} included up to $\atw = 300$ and
  N$^2$LO NNN~\cite{Epelbaum:2008ga,Gazit:2008ma} up to $\ath = 40$.
  The dashed curves connect data points and error bars obtained using
  the extrapolations described in the text.  The small black arrow
  on the left shows the experimental value.
\label{fig:running_Li7}}
\end{figure}

\begin{figure}[tbh-]
\includegraphics*[width=0.9\columnwidth]{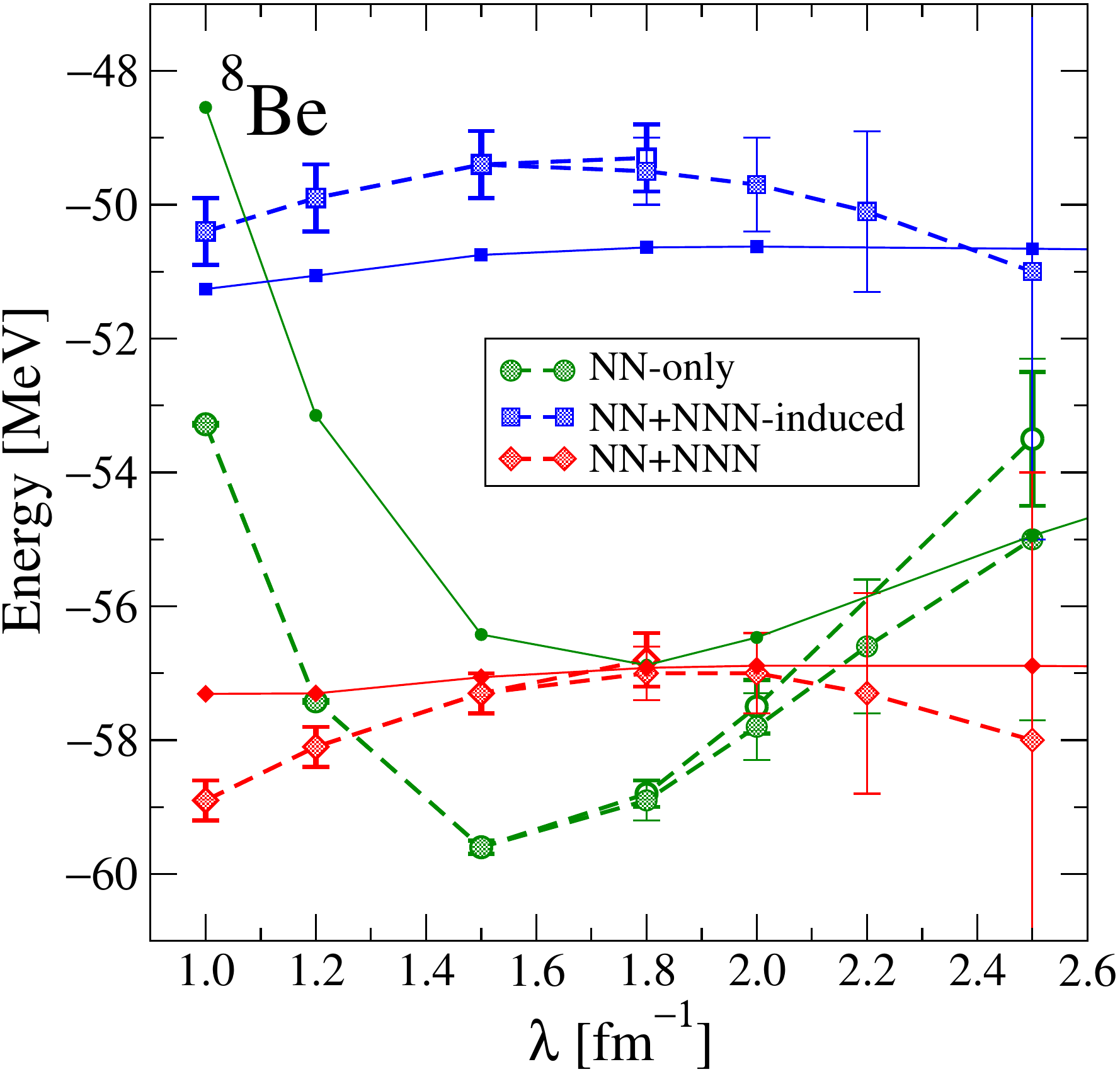}
\caption{(color online) Same as Fig.~\ref{fig:running_Li7} but for the
  \be\ ground state, as well as twice the \he\ ground state energy
  (solid curves).
\label{fig:running_Be8}}
\end{figure}

\begin{figure}[tbh-]
\includegraphics*[width=0.9\columnwidth]{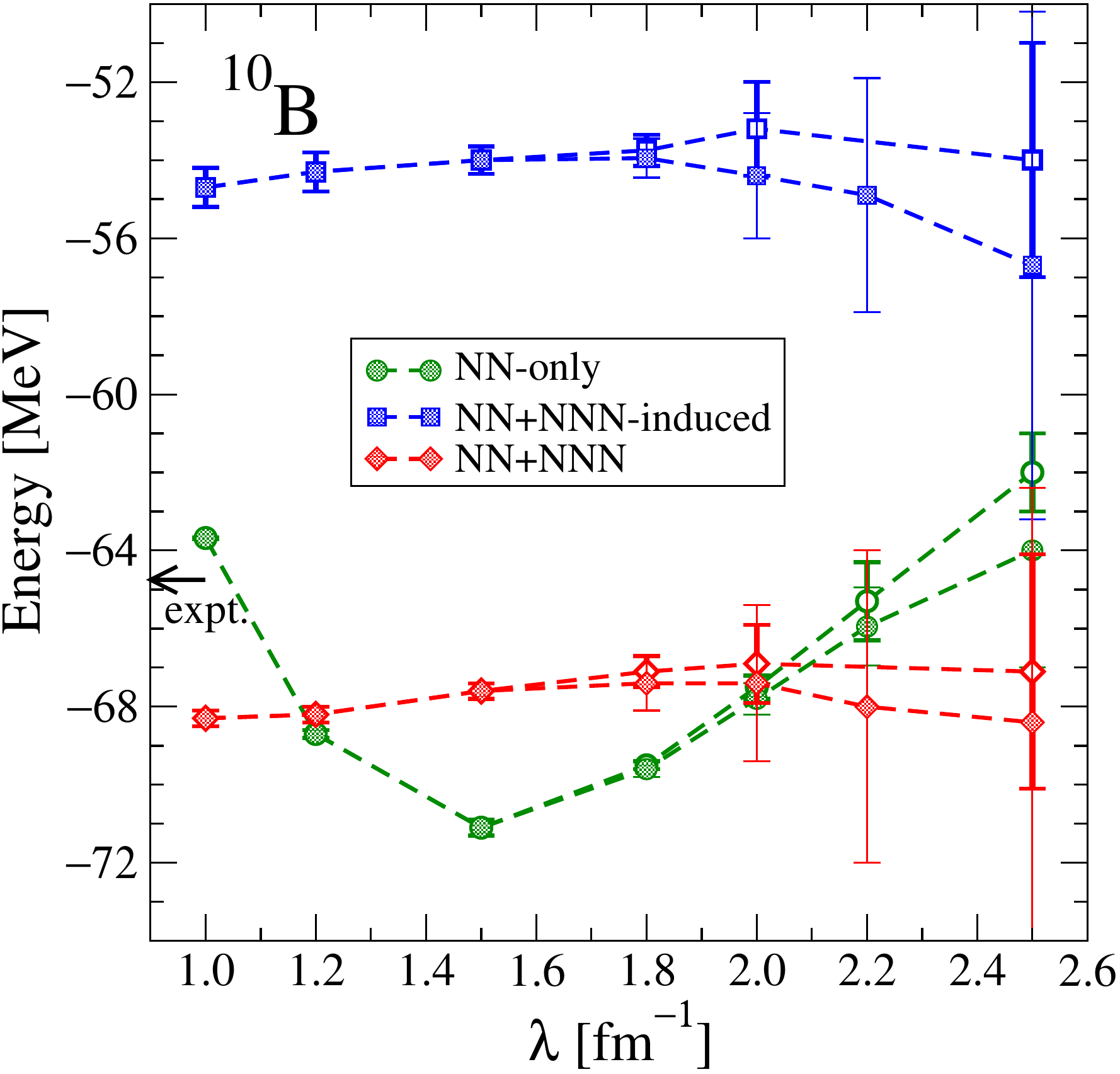}
\caption{(color online) Same as Fig.~\ref{fig:running_Li7} but for the
  \bo\ ground state.  The small black arrow on the left shows the
  experimental value.
\label{fig:running_B10}}
\end{figure}

\begin{figure}[tbh-]
\includegraphics*[width=0.9\columnwidth]{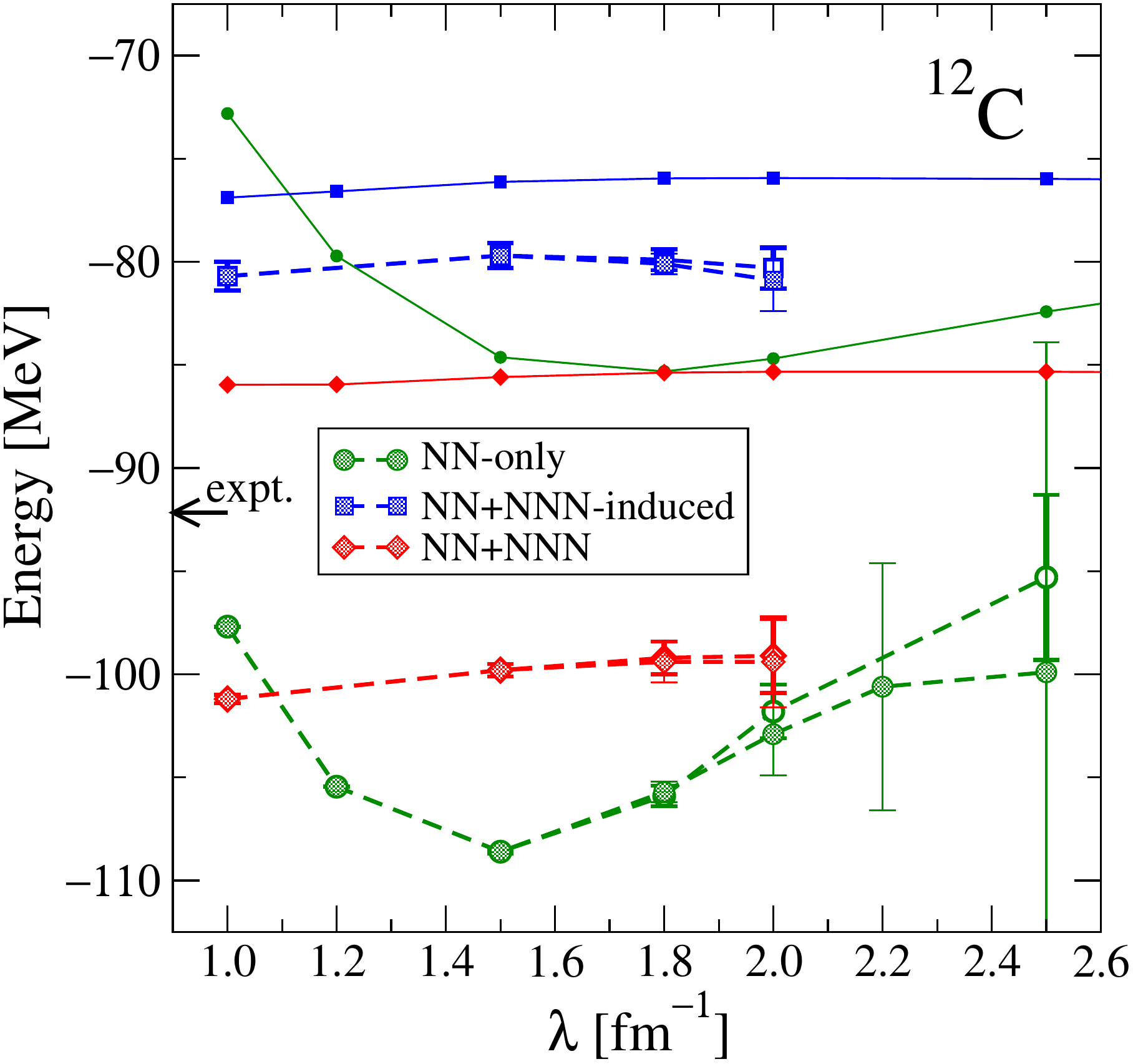}
\caption{(color online) Same as Fig.~\ref{fig:running_Li7} but for the
  \ca\ ground state, as well as three times the \he\ ground state
  energy (solid curves).
\label{fig:running_C12}}
\end{figure}

In Figs.~\ref{fig:running_Li7}--\ref{fig:running_C12}, we show the
dependence on \lam\ of the ground-state energy in \li, \be\, \bo, and
\ca\ using our best estimate for infinite-basis space results based on
``Extrapolation B'' described in the previous section.  For
$\lam < 1.5\fmi$ there is good convergence with small error bars,
although the numerical accuracy of the extrapolation with 3NFs is
limited by the \ath\ cutoff.  For $\lam > 1.5\fmi$ the two criteria
for selecting the optimal \hw\ lead to (slightly) different results,
and their difference grows with \lam; the shaded symbols in
Figs.~\ref{fig:running_Li7}--\ref{fig:running_C12} correspond to using
the \hw\ value that minimizes $E(\nmax)-\Einf$, whereas the open
symbols correspond to using the \hw\ value that minimizes the
numerical error estimate.  For many cases we currently do not have
data available to perform extrapolations at or near the \hw\ value
that would minimize the numerical error estimate.

Nevertheless, we see in these figures that for $\lambda$ between $1.0$
and $2.0\fmi$ the general pattern is the same as that observed for
$^4$He in Fig.~\ref{fig:He4_running}.  We expect that the NN-only
(green circles) start for $\lambda=\infty$ at an (underbound) energy
but the larger \lam's required to verify this are not sufficiently
converged here.  However, the characteristic dip due to omitted
induced NNN forces is clear in each of the nuclei.  Including the
induced NNN matrix elements (blue squares) significantly reduces but
does not eliminate the dependence on \lam.  (Note that, as observed in
Fig.~\ref{fig:He4_running}, the NN-only curve should be equal to the
NN+NNN induced result at large \lam, which is plausible from
Figs.~\ref{fig:running_Li7}--\ref{fig:running_C12} but not verifiable
in the present calculations.)  The trend of the results with induced
NNN interactions is increased binding as $\lambda$ decreases from
$1.8$ to $1.0\fmi$, consistent with $^4$He shown in
Fig.~\ref{fig:He4_running}.  The magnitude of the decrease is about
0.7--1.1\,MeV, without a systematic dependence on the nucleus.  When
initial NNN interactions are included (red diamonds), the qualitative
dependence on \lam\ over this same range is similar, but the magnitude
of the decrease is systematically larger by roughly 1\,MeV or a factor
of two (less in \bo).  The downward shift in ground-state energies
takes them below the experimental values for the four nuclei of
Figs.~\ref{fig:running_Li7}--\ref{fig:running_C12}.  The additional
binding provided by the initial 3NF increases from less than one MeV
per nucleon in \li\ to more than 1.5\,MeV per nucleon in \ca, almost
independent of \lam\ for \lam\ between $1.0$ and $2.0\fmi$.

The shape of an evolution curve (as in
Figs.~\ref{fig:running_Li7}--\ref{fig:running_C12}) is determined by
the interplay of short- and long-range effects in a given $A$-body
sector.  This has been demonstrated explicitly in
Ref.~\cite{Jurgenson:2008jp} for a model but needs to be more
systematically verified in realistic systems.  For a given Hamiltonian
(i.e. fixed initial interactions and fixed truncation of the evolution
equations at the 2-body or at the 3-body level) we observe close
similarity between the various nuclear ground-state evolution curves
presented here.  Focusing on the range in \lam\ where the error bars
are smaller ($\lam = 1.0\mbox{--}1.8 \fmi$), the energy variations for
the NN-only calculations are reduced by a factor of 4--5 for the
NN+NNN calculations (and more for NN+NNN-induced).  Thus the induced
NNN interaction acts to (almost) restore $\lam$-independence in this
region, with the residual variation attributed to four-body (and
higher) forces.

This suggests that the induced NNN interaction, as in the $^4$He case,
is the leading correction to the NN-only results arising from SRG
evolution.  Moreover, the similar shapes of the evolution curves
through the range of $A$ is also consistent with the preservation of
hierarchical induced many-body forces; that is, we expect that the
induced four-body forces will provide a residual contribution smaller
than the induced NNN interaction.  This conclusion is also consistent
with previous analysis of expectation values for components of the
evolved interaction~\cite{Jurgenson:2010wy}.

The net induced 4NF contribution in \he\ was found to be a few hundred
keV at $\lam=1\fmi$ (see Fig.~\ref{fig:He4_running}).  In
Figs.~\ref{fig:running_Li7}--\ref{fig:running_C12} the effect of
omitted induced four-body (and higher) forces at lower \lam\ (judging
solely from the vertical range of each curve) is at least of order 1
to 2\,MeV, depending on whether initial 3NFs are included.  This is
still small enough that for light nuclei it may be possible to exploit
very highly evolved Hamiltonians, especially if a simple approximation
for the 4NF contribution can be found.  Alternatively, incorporating
the induced 4NF contribution explicitly may become computationally
feasible in the near future.  A quantitative understanding of the
magnitude and scaling of (assumed) induced 4NFs with \lam\ and with
$A$ is still lacking.  This has become an important issue in light of
the growing overbinding with larger $A$ observed by Roth et al.\ when
an initial 3NF is included (unless the 3NF cutoff is significantly
reduced with respect to the NN cutoff)~\cite{Roth:2011ar}.

In Fig.~\ref{fig:running_Be8}, we also show the evolution curves for
two $\alpha$-particles, noting that the \lam-dependence is
significantly stronger for \be\ than for two $\alpha$-particles.  With
the NN-only interaction, \be\ is actually bound for $\lam \le 2\fmi$,
but once the induced 3NF is included, it appears to be unbound by
about 1\,MeV.  With the initial 3NFs we find that \be\ is bound for
$\lam \le 1.5\fmi$, but for larger values of \lam\ we cannot draw a
firm conclusion.  Experimentally, \be\ is unbound by about 0.1\,MeV
(i.e. the lowest lying narrow resonance is 0.1\,MeV above the
two-$\alpha$ threshold).
In Fig.~\ref{fig:running_C12}, we also show three times the ground
state energy of \he.  With the NN-only potential, \ca\ is bound by
about 20 to 25\,MeV relative to three $\alpha$ particles with the same
interaction for \lam\ between $1$ and $2\fmi$, but once the induced
3NFs are taken into account, it is bound by only about 4\,MeV.  Once
the initial N2LO chiral 3NFs are incorporated, the binding relative to
three $\alpha$'s increases to about 14\,MeV, and is nearly independent of
\lam.  Experimentally, this energy difference is about 7.5\,MeV.  We
expect that the overbinding by almost a factor of two relative to the
three $\alpha$ threshold with the chiral N2LO 3NFs will have important
consequences for the low-lying spectrum of \ca, in particular for the
Hoyle state.

\subsection{Low-lying excited states}

\begin{figure*}[tbh-]
\includegraphics*[width=0.9\columnwidth]{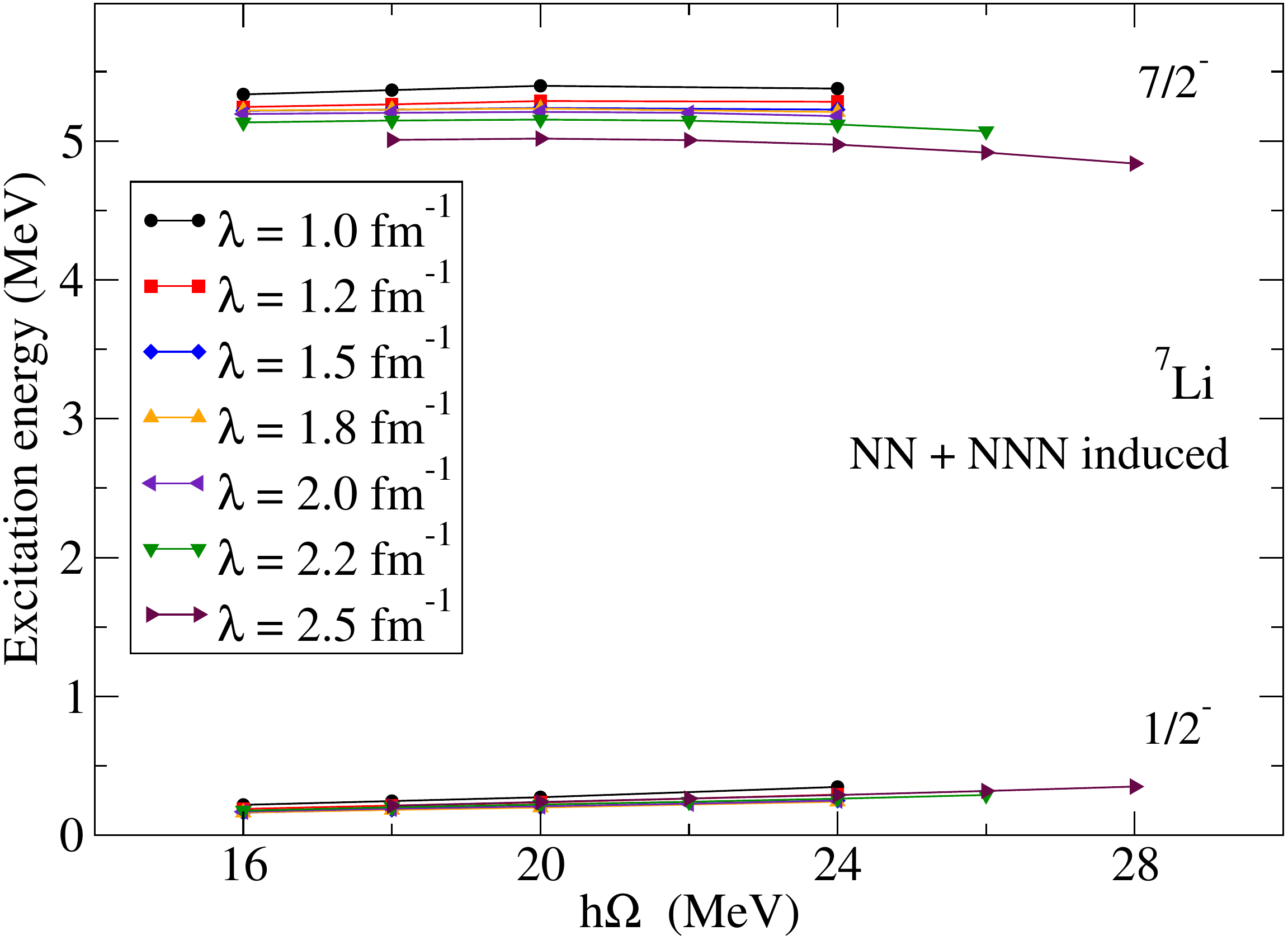}\qquad
\includegraphics*[width=0.9\columnwidth]{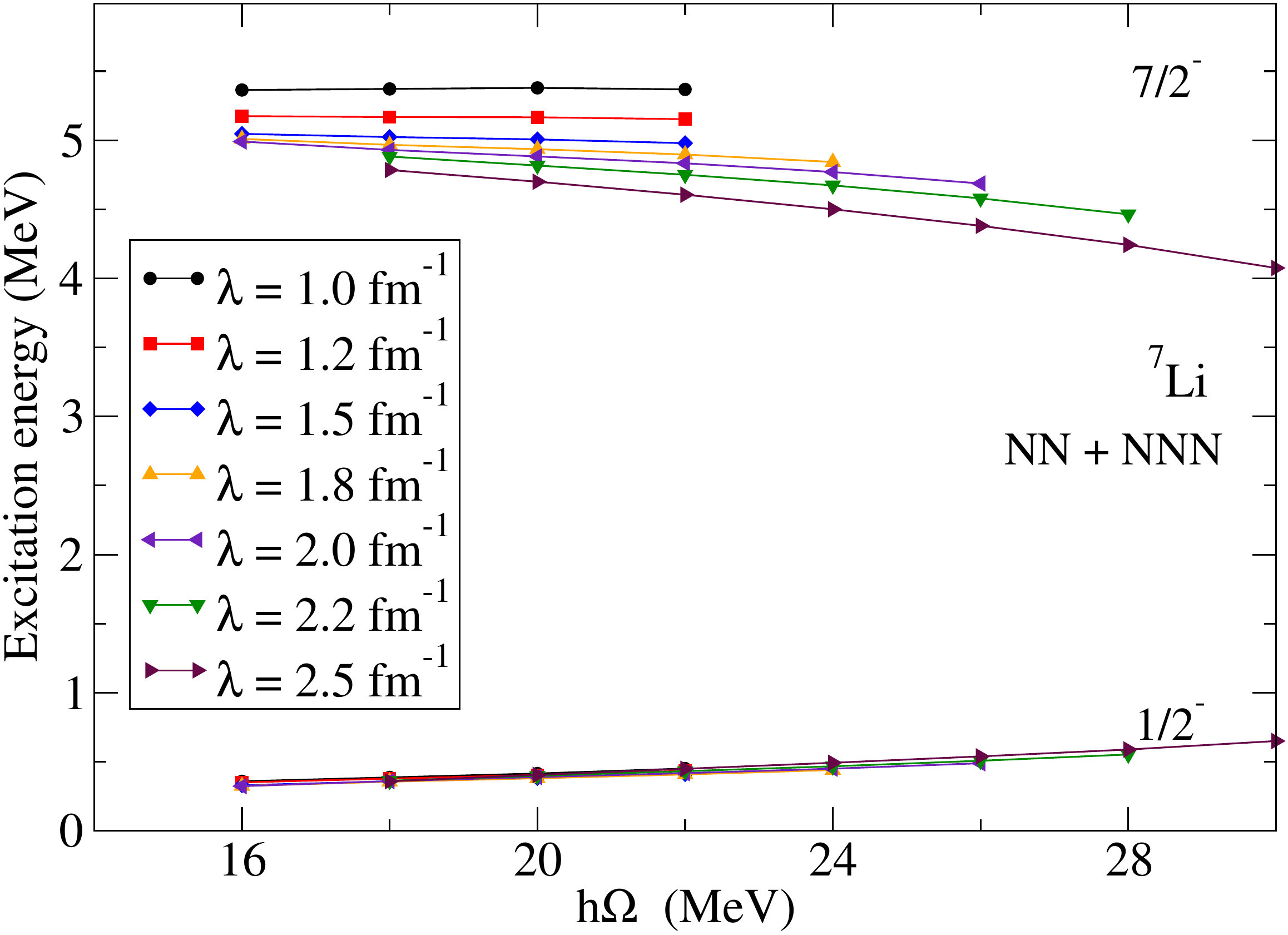}
\caption{(color online) Lowest two excited states \li\ as a function
  of \hw\ for each SRG \lam\ value at $\nmax=8$.  The initial
  interaction was N$^3$LO NN~\cite{Entem:2003ft} included up to
  $\atw = 300$ and N$^2$LO NNN~\cite{Epelbaum:2008ga,Gazit:2008ma} up
  to $\ath = 40$.  The small black arrow on the left shows the
  experimental value.
\label{fig:excitation_Li7}}
\end{figure*}

\begin{figure*}[tbh-]
\includegraphics*[width=0.9\columnwidth]{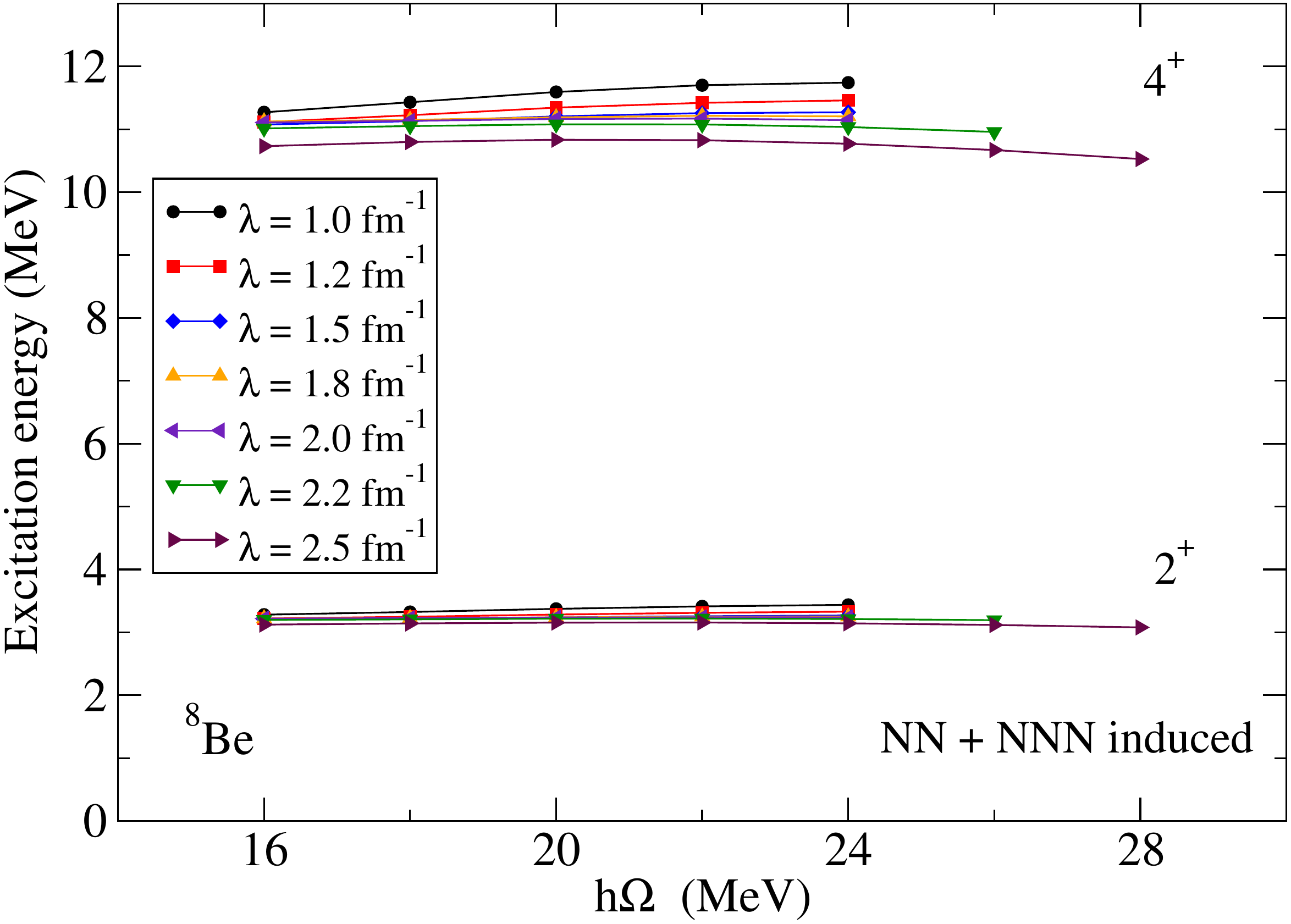}\qquad
\includegraphics*[width=0.9\columnwidth]{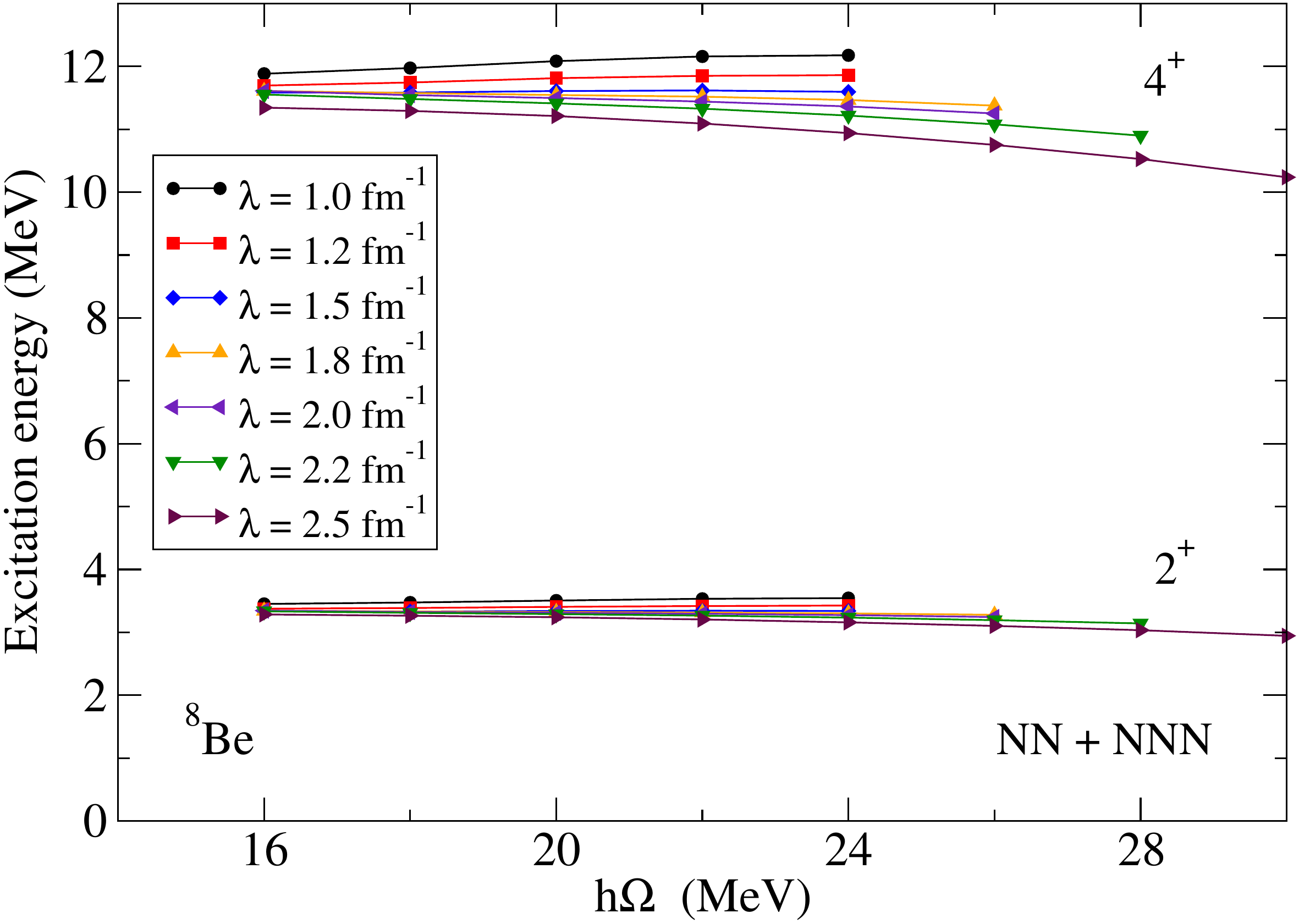}
\caption{(color online) Lowest two excited states \be\ as a function of
\hw\ for each SRG \lam\ value at $\nmax=8$.  
The small black arrow on the left shows the experimental value.
\label{fig:excitation_Be8}}
\end{figure*}

\begin{figure*}[tbh-]
\includegraphics*[width=0.9\columnwidth]{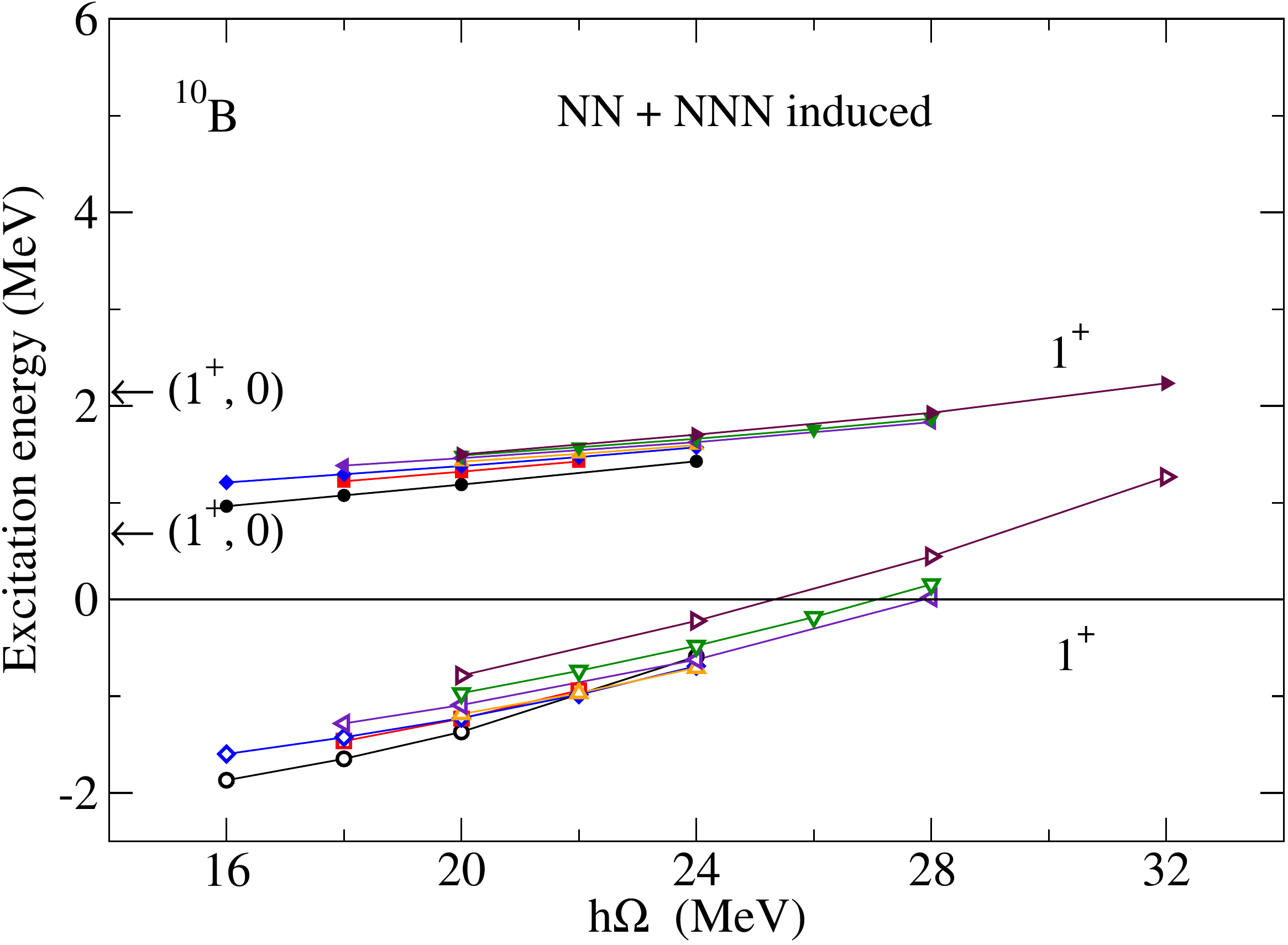}\qquad
\includegraphics*[width=0.9\columnwidth]{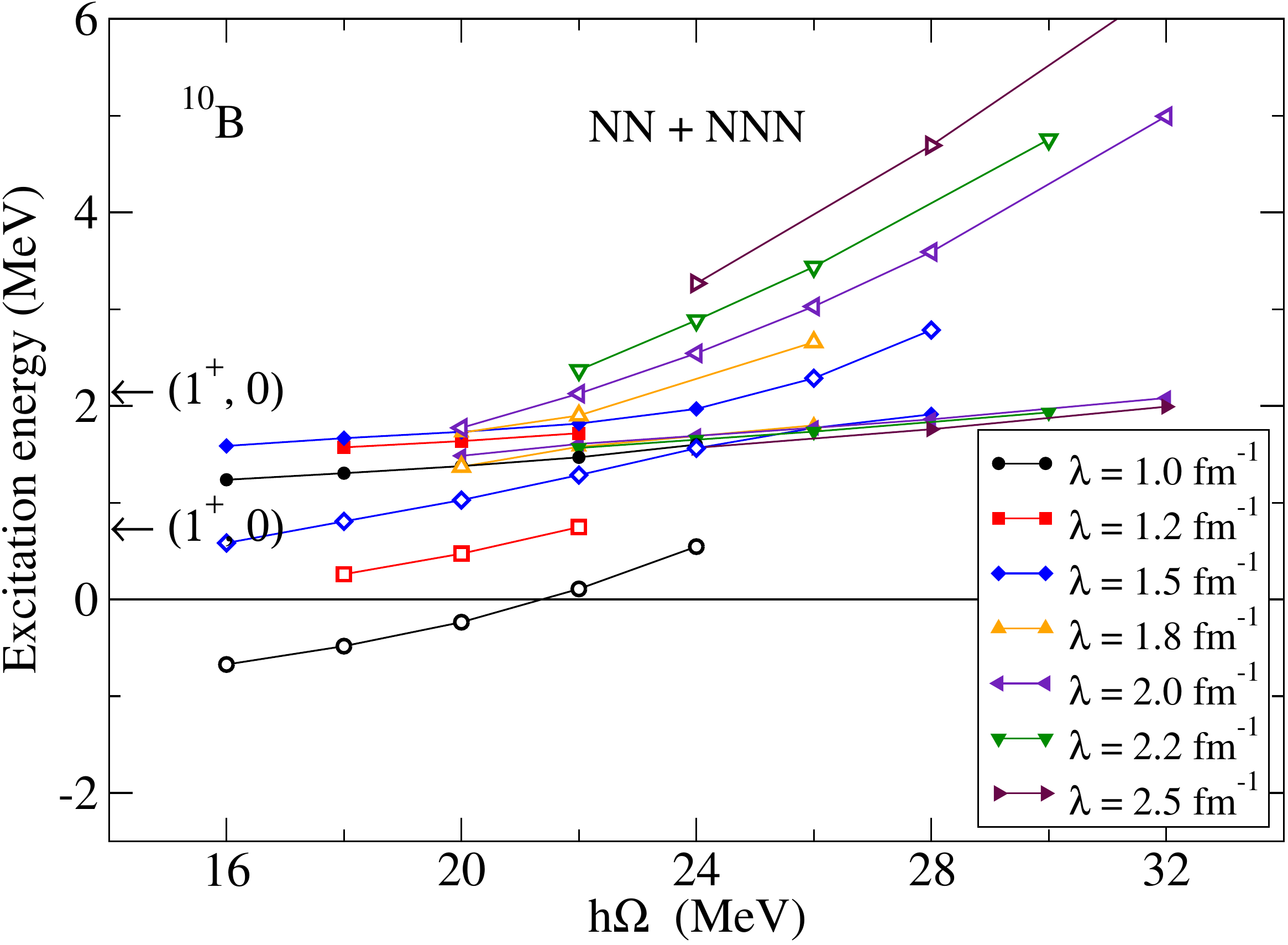}
\includegraphics*[width=0.9\columnwidth]{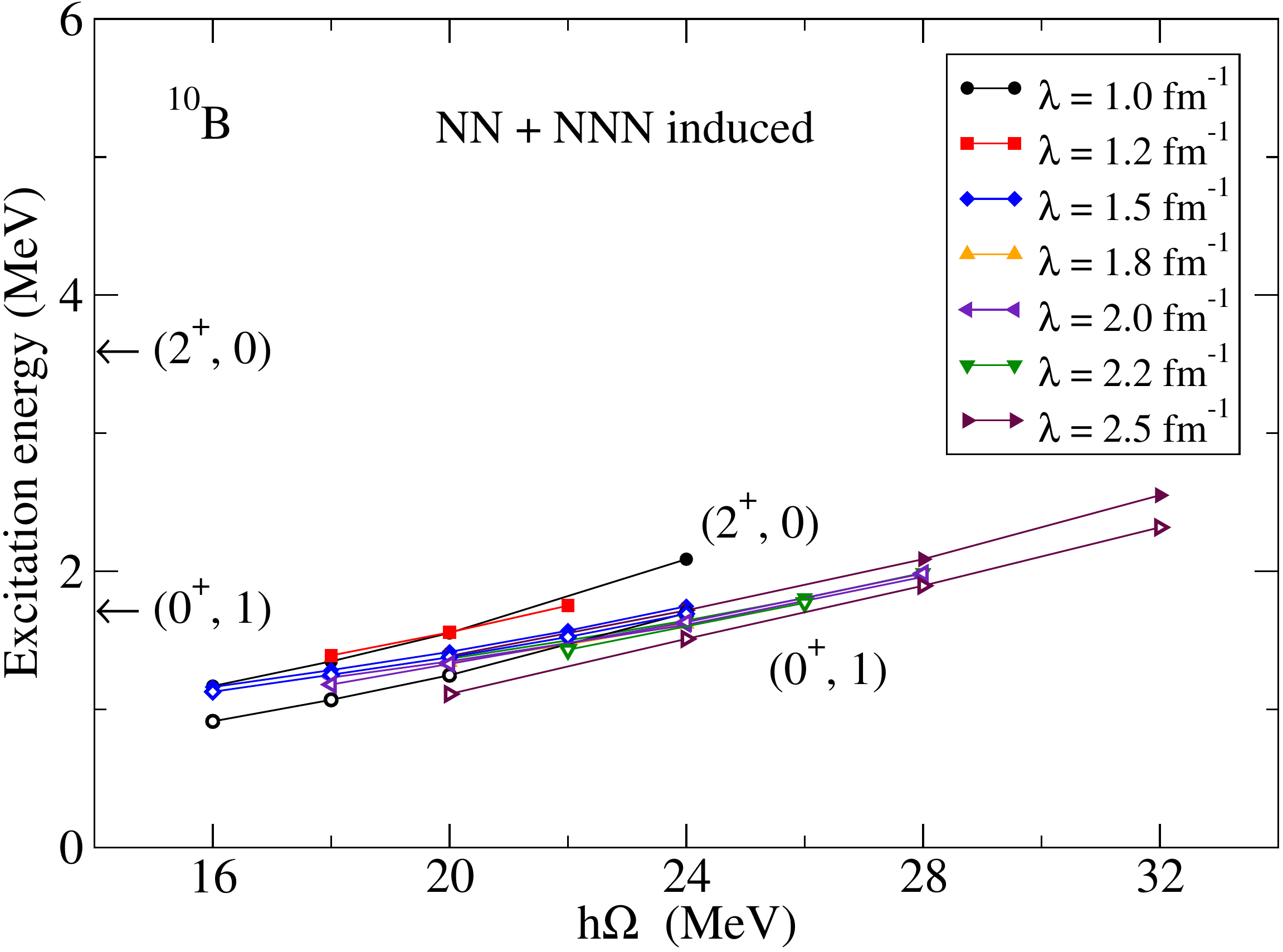}\qquad
\includegraphics*[width=0.9\columnwidth]{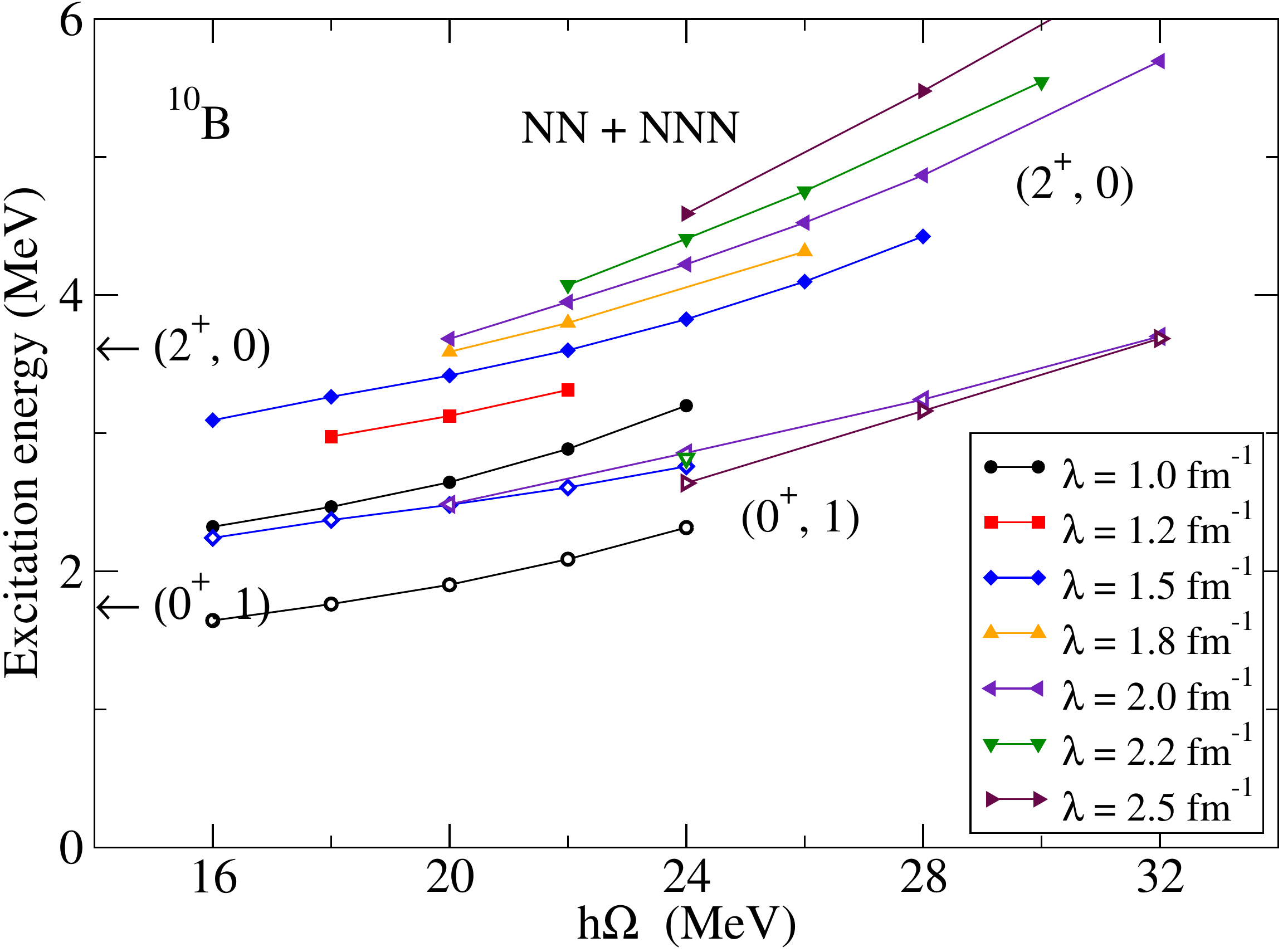}
\caption{(color online) Lowest excited states \bo\ as a function of
\hw\ for each SRG \lam\ value at $\nmax=8$.  
The small black arrows on the left shows the experimental values.
\label{fig:excitation_B10}}
\end{figure*}

In Figs.~\ref{fig:excitation_Li7}--\ref{fig:excitation_B10}, we show
the excitation energies of the lowest excited states in \li, \be, and
\bo\ as function of \hw\ for different values of \lam\ at fixed
$\nmax=8$.  For \li\ and \be, the excitation energy of the first excited
state is almost independent of both \lam\ and \hw, and in good
agreement with experiment.  This independence suggests that these excitation
energies are close to being converged.  The second excited state of
these nuclei shows a slight dependence on both \lam\ and \hw.  In
particular, at larger values of \lam\ the excitation energies show a
variation with \hw\, indicating that these excitation energies are not
yet converged.  However, this variation is significantly less than the
estimated extrapolation error in the absolute (ground state) energies
for these nuclei.  More interesting is that the (albeit small)
dependence on \lam\ is significantly larger with initial 3NFs than
with induced 3NFs only, as is evident by the smaller spread of the
curves in the left panels of Figs.~\ref{fig:excitation_Li7} and
\ref{fig:excitation_Be8}.

In \bo\ the situation is much more complicated.  In
Fig.~\ref{fig:excitation_B10}, we show the excitation energies of the
lowest four excited states of \bo\ relative to the $(3^+,0)$ (which is
the ground state of \bo): two $(1^+,0)$ states, a $(2^+,0)$ state, and
a $(0^+,1)$ state (analog state of the ground state of $^{10}$Be and
$^{10}$C).  It is immediately obvious that there is a much larger
spread in the excitation energies than for \li\ and \be, in particular
with initial 3NFs.

With induced 3NFs only, the lowest state is actually a $(1^+,0)$
state, rather than the $(3^+,0)$.  This state however is rather poorly
converged relatively to the $(3^+,0)$ state, as is evident from the
strong \hw\ dependence of its (mostly negative) excitation energy,
even for very small values of \lam.  Furthermore, the second
$(1^+,0)$, as well as the lowest $(2^+,0)$ and $(0^+,1)$ states are
very close to each other, all with excitation energies between 1 and
2\,MeV, and all with a similar weak \lam\ and moderate \hw\ dependence.
Experimentally, these three states have excitation energies of
2.154\,MeV, 1.740\,MeV, and 3.587\,MeV respectively.  Thus, in the
absence of initial 3NFs, the chiral interactions not only predict the
wrong ground state for \bo\, but also a much too dense spectrum for
the other low-lying states.

With initial 3NFs the spectrum looks quite different.  One of the two
$(1^+,0)$ becomes strongly dependent on \lam, whereas the other
$(1^+,0)$ state remains almost independent of \lam.  In the region of
parameter space where these two states are well-separated, they can be
distinguished by their quadrupole and magnetic moments.  Although the
quadrupole moments are not converged, they are clearly different for
these two states: one has a small positive quadrupole moment of the
order of one~$e\,\fm^2$ or smaller (open symbols in
Fig.~\ref{fig:excitation_B10}), whereas the other has a negative
quadrupole moment around $-2$ to $-3\ e\,\fm^2$ (solid symbols in
Fig.~\ref{fig:excitation_B10}).  The latter of these two states
depends only weakly on \lam\ and \hw, and appears to be reasonably
well converged, whereas the former is very strongly dependent on both
\lam\ and \hw, and is not converged at all.  However, for \lam\ from
$1.5\fmi$ to $2.0\fmi$ these two states show significant mixing.
(Experimentally, they are separated by about 1.4\,MeV.)  For
comparison, the quadrupole moment of the ground state is about +7 to
+8~$e\,\fm^2$ with induced 3NFs only, and about +6 to +7~$e\,\fm^2$
with initial 3NFs, in reasonable agreement with the experimental value
of +8.472(56)~$e\,\fm^2$, given the fact that the quadrupole moments
are not yet converged in these basis spaces.  It will take a major
effort to develop robust extrapolation and error quantification tools
for long-range operators, such as quadrupole operators.  We therefore
defer a detailed consideration of these operators until those tools
are developed.

For small values of \lam, the slowly converging $(1^+, 0)$ state with
a small quadrupole moment actually becomes the lowest state, even with
initial 3NFs.  It is likely that for this state the induced four-body
(and higher-body) interactions, which have been omitted in the present
calculations, are important, though without convergence it is hard to
draw firm conclusions. It is also possible that the strong
\lam\ dependence is caused by the lack of convergence, and that once
convergence (i.e. independence of both \nmax\ and \hw) is reached the
results will be much less dependent on \lam.

The other excited states shown in Fig.~\ref{fig:excitation_B10},
namely the $(0^+,1)$ (open symbols) and the $(2^+,0)$ (solid symbols), show a
strong dependence on both \lam\ and \hw\ with initial 3NFs. Clearly,
these excitation energies are not very well converged, but
nevertheless we can see that the overall effect of the initial 3NFs is
to increase their excitation energy and to separate these two states
from each other, in qualitative agreement with the data.

A striking difference between Figs.~\ref{fig:excitation_Li7} and
\ref{fig:excitation_Be8} on the one hand, and
Fig.~\ref{fig:excitation_B10} on the other, is the strong \lam\ and
\hw\ dependence of the excitation energies in
Fig.~\ref{fig:excitation_B10} (with the possible exception of one of
the two low-lying $(1^+, 0)$ states) compared to the independence of \lam\ and
\hw\ of the excitation energies in Figs.~\ref{fig:excitation_Li7} and
\ref{fig:excitation_Be8}.  One possible explanation for this
observation is that the excited states in
Figs.~\ref{fig:excitation_Li7} and \ref{fig:excitation_Be8} can be
interpreted as rotational excitations of the ground
state~\cite{Caprio:2013yp}.  Thus these states have a very similar
structure, and are therefore likely to exhibit a similar convergence
pattern (i.e. \hw\ dependence) and \lam\ dependence.  We also see this
in \ca, where the lowest $2^+$ and $4^+$ states form a rotational band
with the ground state~\cite{Maris:2013}, but other low-lying states in
\ca\ are much more sensitive to the 3NFs and have a different
convergence pattern~\cite{Calci:2013}.

\begin{table}[thb-]
\caption{\label{tab:magmom} Magnetic moments for \li, \bs, and
  \bo\ with the NN+NNN-induced and NN+NNN interactions.
  }
\begin{tabular}{|c|c||c|c||c|}
\hline
 Nucleus & state & induced  & NNN & expt. \\
\hline
 $^{7}$Li & $\frac{3}{2}^-$  &  3.0(1) &  3.0(1) & 3.2564 \\
 $^{7}$Li & $\frac{1}{2}^-$  & $-0.8(1)$ & $-0.8(1)$ &  \\
 $^{7}$Li & $\frac{7}{2}^-$  &  3.5(2) &  3.2(3) &  \\
\hline
 $^{7}$Be & $\frac{3}{2}^-$  & $-1.15(5)$& $-1.15(6)$ & $-1.399$ \\
 $^{7}$Be & $\frac{1}{2}^-$  &  1.18(3)& 1.22(3) &        \\
 $^{7}$Be & $\frac{7}{2}^-$  & 0.24--0.56& 0.37--1.07 & \\
\hline
 $^{10}$B & $3^+$ &  1.85(1) & 1.83(2)        & 1.8006   \\
 $^{10}$B & $1^+$ &  0.84(2) & 0.78--0.85 & 0.63(12) \\
 $^{10}$B & $1^+$ &  0.35(2) & 0.34--0.41 &          \\
\hline
\end{tabular}
\end{table}

Finally, we also calculated the magnetic moments of \li, \bs, and \bo,
see Table~\ref{tab:magmom}.  The quoted numerical uncertainty in
Table~\ref{tab:magmom} includes both the dependence on the basis space
parameters (i.e. \hw\ and \nmax\ dependence) and the \lam\ dependence.
At $\nmax=8$ the magnetic moments are typically converged to within a
few percent, and the influence of the SRG evolution is less than a few
percent, except for the $\frac{7}{2}^-$ state of \bs\ and the two
$(1^+, 0)$ states in \bo.  A closer look at the different components
contributing to the magnetic moment of the $\frac{7}{2}^-$ state of
\bs\ shows that the contributions from the neutron intrinsic spin and
from the proton angular momentum nearly cancel, leaving the proton
intrinsic spin contribution to dominate the magnetic moment, both with
and without initial 3NFs.  For the two states in \bo, most of the
parameter dependence of the magnetic moments for these two states is
due to (strong) mixing.  In general, adding initial 3NFs to the chiral
N3LO NN interaction does not have a significant effect on these
magnetic moments.

These magnetic moments are calculated in impulse approximation, using
the canonical M1 operator.  Of course, we should use a current
operator that is consistent with the chiral Hamiltonian that we are
using: for the ground state of \li\ and \bs\ we might expect, based on
Refs~\cite{PhysRevC.78.065501,Pastore:2012rp}, a correction due to
meson exchange currents of about 10\%, in the direction that would
bring our results toward agreement with experiment.

%%%%%%%%%%%%%%%%%%%%%%%%%%%%%%%%%%%%%%%%%%%%%%%%%%%%%%%%%%%%%%%%%%%%%%%%%%
\section{Conclusions}
\label{sec:conclusions}

We have presented \abinit NCFC calculations of energies in the p-shell
using SRG-evolved two- and three-nucleon forces.  Several different procedures
were considered to extrapolate energies to infinite
harmonic-oscillator basis size; for the range in the evolution
parameter \lam\ we focus on (from $1$ to $2\fmi$) they give
consistent results within estimated error bars.  Error bars above
$2\fmi$ grow rapidly and limit what we can quantitatively conclude
about \lam\ dependence.  As anticipated from results with lighter
nuclei, inclusion of induced NNN interaction significantly reduced the
$\lambda$ dependence of ground-state energies compared to NN-only
calculations for all nuclei considered.  Furthermore, the rapid
improvement of convergence with decreasing $\lambda$ of ground-state
and low-lying state energies observed for NN-only
calculations~\cite{Bogner:2007rx} carries over when only induced NNN
interactions are included.  With initial NNN interactions,
ground-state convergence is similarly improved but some excited states
show very different behavior, see Fig.~\ref{fig:excitation_B10}.

With NN+NNN-induced interactions (but without initial NNN
interactions), the net change in total ground-state energy for nuclei
in the p-shell was found to be small (and within extrapolation error
bars) between $\lambda = 2.0\fmi$ and $\lambda = 1.5\fmi$, but
systematically decreases (becomes more bound) as \lam\ decreases to
$1.0\fmi$ by about 1\,MeV.  The $A$ dependence is small.  This
additional binding can be attributed to four- and higher-body forces,
which is of natural size (as implied by EFT power counting) despite
the extreme degree of softening.  While this might appear to be a
narrow range in $\lambda$, we emphasize that there is significant
evolution (e.g., note that the natural SRG evolution variable $s =
1/\lambda^4$ increases by a factor of 16 as $\lambda$ decreases from
$2$ to $1\fmi$).  When initial 3NFs are included, a similar pattern is
found, but the decrease is more nucleus dependent and larger by as
much as a factor of 2 (i.e., up to about 2\,MeV).  This increase in
binding is consistent with the difference in running between
NN+NNN-induced and NN+NNN observed for \ca\ in Ref.~\cite{Roth:2011ar}
with the same interactions.

Results with 3NFs for higher $\lambda$ are consistent with
small changes, but the uncertainties after extrapolation are too large
to be definitive. 
In contrast, Roth et al.\ found a steady linear increase in binding
from \lam\ above $2.2\fmi$ down to $1.6\fmi$ for \ca~\cite{Roth:2011ar}.
This is not evident in the systematics of the central values of
our extrapolations, but possible within the estimated error bars.
It would be most helpful to have
accurate energy
calculations for the initial Hamiltonian to fully assess the degree of
running down to $\lambda = 2\fmi$.
For excited states we also find similar quantitative differences between
NN+NNN-induced and NN+NNN calculations and for some particular states there
are qualitative differences.

Further investigations are warranted.  For example, it will be
important to compare our present results, which use harmonic
oscillator evolved SRG 3NFs, to forthcoming calculations using the
same initial interactions but evolved in momentum space using recently
developed SRG technology~\cite{Hebeler:2012pr}.  Further explorations
at low \lam\ will help to map out the quantitative scaling of induced
4NF contributions.  Improved convergence at these low resolutions
motivates searching for simple approximations to account for these 4NF
contributions to the energy and other observables.  Besides additional
CI calculations, application of highly evolved two- and three-nucleon
forces may also be fruitful for coupled cluster~\cite{Hagen:2007hi}
methods, \abinit density functional theory~\cite{Drut:2011te}, and
NCSM/RGM~\cite{Quaglioni:2008sm,Navratil:2010jn,Navratil:2011zs,Navratil:2011sa,Quaglioni:2012bm}
calculations of light nuclear reactions.

Our studies were limited in basis size by the available codes and
computer resources. We anticipate further developments in improved
basis construction and evolution algorithms. We could also study
non-oscillator basis spaces~\cite{Caprio:2012rv,Caprio:2012as} and
apply the importance truncation technique~\cite{Roth:2009cw} 
to increase available basis sizes.  We look forward to a detailed
extrapolation analysis of such results.  In addition, a difficulty
encountered here with an insufficient initial $A=3$ basis at smaller
\hw\ (see Section~\ref{subsec:A3nmax}) might be circumvented by the
momentum-space evolution technology~\cite{Hebeler:2012pr}, so that
matrix elements in the oscillator basis are only calculated
\emph{after} evolution.

We considered only a single initial NN+NNN Hamiltonian in our
analysis. While in past investigations~\cite{Jurgenson:2010wy} other
choices of initial NN Hamiltonians have displayed no qualitative
difference in the effects of the SRG procedure, studies with a range
of Hamiltonians are highly desirable.  It will also be important to
consider consistent operators for other observables.  In this regard,
new extrapolation methods for radii and other long-distance operators
may be particularly valuable~\cite{Furnstahl:2012qg}.

Our results demonstrate a level of precision in induced many-body
force effects that allows further analysis of the impact of additional
\eft\ inputs. In all cases presented here the addition of 3NFs from
the initial \eft\ Hamiltonian overbinds the ground states.  Our
results suggest that the effects of missing induced forces due to
softening transformations are small (at least in larger nuclei)
compared to discrepancies with experiment.  These discrepancies may be
reduced by additional 3NF and 4NF contributions at N$^3$LO in the
\eft.  Suitable matrix elements of these terms will be available for
calculations in the near future and it will be important to
incorporate them.

\begin{acknowledgments}

This work was supported in part by the National Science Foundation
under Grant Nos.~PHY--1002478 and PHY--0904782 and the Department of
Energy under Grant Nos.~DE-FG02-87ER40371, DE-FC02-07ER41457
(SciDAC-2/UNEDF), DE-FC02-09ER41582 (SciDAC-2/UNEDF) and DESC0008485
(SciDAC-3/NUCLEI).  Prepared in part by LLNL under Contract
DE-AC52-07NA27344.  
Support from the Natural Sciences and Engineering Research Council of Canada (NSERC) Grant No. 401945-2011 is acknowledged. TRIUMF receives funding via a contribution through the National Research Council Canada.
A portion of the computational resources were
provided by the National Energy Research Scientific Computing Center
(NERSC), which is supported by the DOE Office of Science, and by an
INCITE award, "Nuclear Structure and Nuclear Reactions", from the DOE
Office of Advanced Scientific Computing.  This research also used
resources of the Oak Ridge Leadership Computing Facility at ORNL,
which is supported by the DOE Office of Science under Contract
DE-AC05-00OR22725.

\end{acknowledgments}

\bibliography{srg_refs}

%merlin.mbs apsrev4-1.bst 2010-07-25 4.21a (PWD, AO, DPC) hacked
%Control: key (0)
%Control: author (8) initials jnrlst
%Control: editor formatted (1) identically to author
%Control: production of article title (-1) disabled
%Control: page (0) single
%Control: year (1) truncated
%Control: production of eprint (0) enabled
\begin{thebibliography}{50}%
\makeatletter
\providecommand \@ifxundefined [1]{%
 \@ifx{#1\undefined}
}%
\providecommand \@ifnum [1]{%
 \ifnum #1\expandafter \@firstoftwo
 \else \expandafter \@secondoftwo
 \fi
}%
\providecommand \@ifx [1]{%
 \ifx #1\expandafter \@firstoftwo
 \else \expandafter \@secondoftwo
 \fi
}%
\providecommand \natexlab [1]{#1}%
\providecommand \enquote  [1]{``#1''}%
\providecommand \bibnamefont  [1]{#1}%
\providecommand \bibfnamefont [1]{#1}%
\providecommand \citenamefont [1]{#1}%
\providecommand \href@noop [0]{\@secondoftwo}%
\providecommand \href [0]{\begingroup \@sanitize@url \@href}%
\providecommand \@href[1]{\@@startlink{#1}\@@href}%
\providecommand \@@href[1]{\endgroup#1\@@endlink}%
\providecommand \@sanitize@url [0]{\catcode `\\12\catcode `\$12\catcode
  `\&12\catcode `\#12\catcode `\^12\catcode `\_12\catcode `\%12\relax}%
\providecommand \@@startlink[1]{}%
\providecommand \@@endlink[0]{}%
\providecommand \url  [0]{\begingroup\@sanitize@url \@url }%
\providecommand \@url [1]{\endgroup\@href {#1}{\urlprefix }}%
\providecommand \urlprefix  [0]{URL }%
\providecommand \Eprint [0]{\href }%
\providecommand \doibase [0]{http://dx.doi.org/}%
\providecommand \selectlanguage [0]{\@gobble}%
\providecommand \bibinfo  [0]{\@secondoftwo}%
\providecommand \bibfield  [0]{\@secondoftwo}%
\providecommand \translation [1]{[#1]}%
\providecommand \BibitemOpen [0]{}%
\providecommand \bibitemStop [0]{}%
\providecommand \bibitemNoStop [0]{.\EOS\space}%
\providecommand \EOS [0]{\spacefactor3000\relax}%
\providecommand \BibitemShut  [1]{\csname bibitem#1\endcsname}%
\let\auto@bib@innerbib\@empty
%</preamble>
\bibitem [{\citenamefont {Maris}\ \emph {et~al.}(2009)\citenamefont {Maris},
  \citenamefont {Vary},\ and\ \citenamefont {Shirokov}}]{Maris:2008ax}%
  \BibitemOpen
  \bibfield  {author} {\bibinfo {author} {\bibfnamefont {P.}~\bibnamefont
  {Maris}}, \bibinfo {author} {\bibfnamefont {J.~P.}\ \bibnamefont {Vary}}, \
  and\ \bibinfo {author} {\bibfnamefont {A.~M.}\ \bibnamefont {Shirokov}},\
  }\href {\doibase 10.1103/PhysRevC.79.014308} {\bibfield  {journal} {\bibinfo
  {journal} {Phys. Rev. C}\ }\textbf {\bibinfo {volume} {79}},\ \bibinfo
  {pages} {014308} (\bibinfo {year} {2009})},\ \Eprint
  {http://arxiv.org/abs/0808.3420} {arXiv:0808.3420 [nucl-th]} \BibitemShut
  {NoStop}%
%%CITATION = 0808.3420;%%
\bibitem [{\citenamefont {Navratil}\ \emph {et~al.}(2009)\citenamefont
  {Navratil}, \citenamefont {Quaglioni}, \citenamefont {Stetcu},\ and\
  \citenamefont {Barrett}}]{Navratil:2009ut}%
  \BibitemOpen
  \bibfield  {author} {\bibinfo {author} {\bibfnamefont {P.}~\bibnamefont
  {Navratil}}, \bibinfo {author} {\bibfnamefont {S.}~\bibnamefont {Quaglioni}},
  \bibinfo {author} {\bibfnamefont {I.}~\bibnamefont {Stetcu}}, \ and\ \bibinfo
  {author} {\bibfnamefont {B.~R.}\ \bibnamefont {Barrett}},\ }\href {\doibase
  10.1088/0954-3899/36/8/083101} {\bibfield  {journal} {\bibinfo  {journal} {J.
  Phys. G}\ }\textbf {\bibinfo {volume} {36}},\ \bibinfo {pages} {083101}
  (\bibinfo {year} {2009})},\ \Eprint {http://arxiv.org/abs/0904.0463}
  {arXiv:0904.0463 [nucl-th]} \BibitemShut {NoStop}%
%%CITATION = 0904.0463;%%
\bibitem [{\citenamefont {Maris}\ \emph
  {et~al.}(2010{\natexlab{a}})\citenamefont {Maris}, \citenamefont {Shirokov},\
  and\ \citenamefont {Vary}}]{Maris:2009bx}%
  \BibitemOpen
  \bibfield  {author} {\bibinfo {author} {\bibfnamefont {P.}~\bibnamefont
  {Maris}}, \bibinfo {author} {\bibfnamefont {A.}~\bibnamefont {Shirokov}}, \
  and\ \bibinfo {author} {\bibfnamefont {J.}~\bibnamefont {Vary}},\ }\href
  {\doibase 10.1103/PhysRevC.81.021301} {\bibfield  {journal} {\bibinfo
  {journal} {Phys. Rev. C}\ }\textbf {\bibinfo {volume} {81}},\ \bibinfo
  {pages} {021301} (\bibinfo {year} {2010}{\natexlab{a}})},\ \Eprint
  {http://arxiv.org/abs/0911.2281} {arXiv:0911.2281 [nucl-th]} \BibitemShut
  {NoStop}%
%%CITATION = ARXIV:0911.2281;%%
\bibitem [{\citenamefont {Maris}\ \emph {et~al.}(2011)\citenamefont {Maris},
  \citenamefont {Vary}, \citenamefont {Navratil}, \citenamefont {Ormand},
  \citenamefont {Nam},\ and\ \citenamefont {Dean}}]{Maris:2011as}%
  \BibitemOpen
  \bibfield  {author} {\bibinfo {author} {\bibfnamefont {P.}~\bibnamefont
  {Maris}}, \bibinfo {author} {\bibfnamefont {J.}~\bibnamefont {Vary}},
  \bibinfo {author} {\bibfnamefont {P.}~\bibnamefont {Navratil}}, \bibinfo
  {author} {\bibfnamefont {W.}~\bibnamefont {Ormand}}, \bibinfo {author}
  {\bibfnamefont {H.}~\bibnamefont {Nam}}, \ and\ \bibinfo {author}
  {\bibfnamefont {D.}~\bibnamefont {Dean}},\ }\href {\doibase
  10.1103/PhysRevLett.106.202502} {\bibfield  {journal} {\bibinfo  {journal}
  {Phys. Rev. Lett.}\ }\textbf {\bibinfo {volume} {106}},\ \bibinfo {pages}
  {202502} (\bibinfo {year} {2011})},\ \Eprint {http://arxiv.org/abs/1101.5124}
  {arXiv:1101.5124 [nucl-th]} \BibitemShut {NoStop}%
%%CITATION = ARXIV:1101.5124;%%
\bibitem [{\citenamefont {Navratil}\ and\ \citenamefont
  {Quaglioni}(2012)}]{Navratil:2011zs}%
  \BibitemOpen
  \bibfield  {author} {\bibinfo {author} {\bibfnamefont {P.}~\bibnamefont
  {Navratil}}\ and\ \bibinfo {author} {\bibfnamefont {S.}~\bibnamefont
  {Quaglioni}},\ }\href {\doibase 10.1103/PhysRevLett.108.042503} {\bibfield
  {journal} {\bibinfo  {journal} {Phys. Rev. Lett.}\ }\textbf {\bibinfo
  {volume} {108}},\ \bibinfo {pages} {042503} (\bibinfo {year} {2012})},\
  \Eprint {http://arxiv.org/abs/1110.0460} {arXiv:1110.0460 [nucl-th]}
  \BibitemShut {NoStop}%
%%CITATION = ARXIV:1110.0460;%%
\bibitem [{\citenamefont {Barrett}\ \emph {et~al.}(2013)\citenamefont
  {Barrett}, \citenamefont {Navratil},\ and\ \citenamefont
  {Vary}}]{Barrett:2013nh}%
  \BibitemOpen
  \bibfield  {author} {\bibinfo {author} {\bibfnamefont {B.~R.}\ \bibnamefont
  {Barrett}}, \bibinfo {author} {\bibfnamefont {P.}~\bibnamefont {Navratil}}, \
  and\ \bibinfo {author} {\bibfnamefont {J.~P.}\ \bibnamefont {Vary}},\ }\href
  {\doibase 10.1016/j.ppnp.2012.10.003} {\bibfield  {journal} {\bibinfo
  {journal} {Prog. Part. Nucl. Phys.}\ }\textbf {\bibinfo {volume} {69}},\
  \bibinfo {pages} {131} (\bibinfo {year} {2013})}\BibitemShut {NoStop}%
%%CITATION = PPNPD,69,131;%%
\bibitem [{\citenamefont {Glazek}\ and\ \citenamefont
  {Wilson}(1993)}]{Glazek:1993rc}%
  \BibitemOpen
  \bibfield  {author} {\bibinfo {author} {\bibfnamefont {S.~D.}\ \bibnamefont
  {Glazek}}\ and\ \bibinfo {author} {\bibfnamefont {K.~G.}\ \bibnamefont
  {Wilson}},\ }\href {\doibase 10.1103/PhysRevD.48.5863} {\bibfield  {journal}
  {\bibinfo  {journal} {Phys. Rev. D}\ }\textbf {\bibinfo {volume} {48}},\
  \bibinfo {pages} {5863} (\bibinfo {year} {1993})}\BibitemShut {NoStop}%
%%CITATION = PHRVA,D48,5863;%%
\bibitem [{\citenamefont {Wegner}(1994)}]{Wegner:1994}%
  \BibitemOpen
  \bibfield  {author} {\bibinfo {author} {\bibfnamefont {F.}~\bibnamefont
  {Wegner}},\ }\href@noop {} {\bibfield  {journal} {\bibinfo  {journal} {Ann.
  Phys.}\ }\textbf {\bibinfo {volume} {506}},\ \bibinfo {pages} {77} (\bibinfo
  {year} {1994})}\BibitemShut {NoStop}%
\bibitem [{\citenamefont {Bogner}\ \emph {et~al.}(2010)\citenamefont {Bogner},
  \citenamefont {Furnstahl},\ and\ \citenamefont {Schwenk}}]{Bogner:2009bt}%
  \BibitemOpen
  \bibfield  {author} {\bibinfo {author} {\bibfnamefont {S.~K.}\ \bibnamefont
  {Bogner}}, \bibinfo {author} {\bibfnamefont {R.~J.}\ \bibnamefont
  {Furnstahl}}, \ and\ \bibinfo {author} {\bibfnamefont {A.}~\bibnamefont
  {Schwenk}},\ }\href {\doibase 10.1016/j.ppnp.2010.03.001} {\bibfield
  {journal} {\bibinfo  {journal} {Prog. Part. Nucl. Phys.}\ }\textbf {\bibinfo
  {volume} {65}},\ \bibinfo {pages} {94} (\bibinfo {year} {2010})},\ \Eprint
  {http://arxiv.org/abs/0912.3688} {arXiv:0912.3688 [nucl-th]} \BibitemShut
  {NoStop}%
%%CITATION = 0912.3688;%%
\bibitem [{\citenamefont {Furnstahl}(2012)}]{Furnstahl:2012fn}%
  \BibitemOpen
  \bibfield  {author} {\bibinfo {author} {\bibfnamefont {R.}~\bibnamefont
  {Furnstahl}},\ }\href {\doibase 10.1016/j.nuclphysbps.2012.06.005} {\bibfield
   {journal} {\bibinfo  {journal} {Nucl.\ Phys.\ Proc.\ Suppl.}\ }\textbf
  {\bibinfo {volume} {228}},\ \bibinfo {pages} {139} (\bibinfo {year}
  {2012})},\ \Eprint {http://arxiv.org/abs/1203.1779} {arXiv:1203.1779
  [nucl-th]} \BibitemShut {NoStop}%
%%CITATION = ARXIV:1203.1779;%%
\bibitem [{\citenamefont {Bogner}\ \emph
  {et~al.}(2007{\natexlab{a}})\citenamefont {Bogner}, \citenamefont
  {Furnstahl},\ and\ \citenamefont {Perry}}]{Bogner:2006pc}%
  \BibitemOpen
  \bibfield  {author} {\bibinfo {author} {\bibfnamefont {S.~K.}\ \bibnamefont
  {Bogner}}, \bibinfo {author} {\bibfnamefont {R.~J.}\ \bibnamefont
  {Furnstahl}}, \ and\ \bibinfo {author} {\bibfnamefont {R.~J.}\ \bibnamefont
  {Perry}},\ }\href@noop {} {\bibfield  {journal} {\bibinfo  {journal} {Phys.
  Rev. C}\ }\textbf {\bibinfo {volume} {75}},\ \bibinfo {pages} {061001}
  (\bibinfo {year} {2007}{\natexlab{a}})},\ \Eprint
  {http://arxiv.org/abs/nucl-th/0611045} {nucl-th/0611045} \BibitemShut
  {NoStop}%
%%CITATION = NUCL-TH/0611045;%%
\bibitem [{\citenamefont {Bogner}\ \emph
  {et~al.}(2007{\natexlab{b}})\citenamefont {Bogner}, \citenamefont
  {Furnstahl}, \citenamefont {Perry},\ and\ \citenamefont
  {Schwenk}}]{Bogner:2007jb}%
  \BibitemOpen
  \bibfield  {author} {\bibinfo {author} {\bibfnamefont {S.~K.}\ \bibnamefont
  {Bogner}}, \bibinfo {author} {\bibfnamefont {R.~J.}\ \bibnamefont
  {Furnstahl}}, \bibinfo {author} {\bibfnamefont {R.~J.}\ \bibnamefont
  {Perry}}, \ and\ \bibinfo {author} {\bibfnamefont {A.}~\bibnamefont
  {Schwenk}},\ }\href {\doibase 10.1016/j.physletb.2007.04.048} {\bibfield
  {journal} {\bibinfo  {journal} {Phys. Lett. B}\ }\textbf {\bibinfo {volume}
  {649}},\ \bibinfo {pages} {488} (\bibinfo {year} {2007}{\natexlab{b}})},\
  \Eprint {http://arxiv.org/abs/nucl-th/0701013} {arXiv:nucl-th/0701013}
  \BibitemShut {NoStop}%
%%CITATION = NUCL-TH/0701013;%%
\bibitem [{\citenamefont {Jurgenson}\ \emph {et~al.}(2008)\citenamefont
  {Jurgenson}, \citenamefont {Bogner}, \citenamefont {Furnstahl},\ and\
  \citenamefont {Perry}}]{Jurgenson:2007td}%
  \BibitemOpen
  \bibfield  {author} {\bibinfo {author} {\bibfnamefont {E.~D.}\ \bibnamefont
  {Jurgenson}}, \bibinfo {author} {\bibfnamefont {S.~K.}\ \bibnamefont
  {Bogner}}, \bibinfo {author} {\bibfnamefont {R.~J.}\ \bibnamefont
  {Furnstahl}}, \ and\ \bibinfo {author} {\bibfnamefont {R.~J.}\ \bibnamefont
  {Perry}},\ }\href {\doibase 10.1103/PhysRevC.78.014003} {\bibfield  {journal}
  {\bibinfo  {journal} {Phys. Rev. C}\ }\textbf {\bibinfo {volume} {78}},\
  \bibinfo {pages} {014003} (\bibinfo {year} {2008})},\ \Eprint
  {http://arxiv.org/abs/0711.4252} {arXiv:0711.4252 [nucl-th]} \BibitemShut
  {NoStop}%
%%CITATION = 0711.4252;%%
\bibitem [{\citenamefont {Bogner}\ \emph {et~al.}(2008)\citenamefont {Bogner},
  \citenamefont {Furnstahl}, \citenamefont {Maris}, \citenamefont {Perry},
  \citenamefont {Schwenk},\ and\ \citenamefont {Vary}}]{Bogner:2007rx}%
  \BibitemOpen
  \bibfield  {author} {\bibinfo {author} {\bibfnamefont {S.~K.}\ \bibnamefont
  {Bogner}}, \bibinfo {author} {\bibfnamefont {R.~J.}\ \bibnamefont
  {Furnstahl}}, \bibinfo {author} {\bibfnamefont {P.}~\bibnamefont {Maris}},
  \bibinfo {author} {\bibfnamefont {R.~J.}\ \bibnamefont {Perry}}, \bibinfo
  {author} {\bibfnamefont {A.}~\bibnamefont {Schwenk}}, \ and\ \bibinfo
  {author} {\bibfnamefont {J.~P.}\ \bibnamefont {Vary}},\ }\href {\doibase
  10.1016/j.nuclphysa.2007.12.008} {\bibfield  {journal} {\bibinfo  {journal}
  {Nucl. Phys. A}\ }\textbf {\bibinfo {volume} {801}},\ \bibinfo {pages} {21}
  (\bibinfo {year} {2008})},\ \Eprint {http://arxiv.org/abs/0708.3754}
  {arXiv:0708.3754 [nucl-th]} \BibitemShut {NoStop}%
%%CITATION = 0708.3754;%%
\bibitem [{\citenamefont {Jurgenson}\ \emph {et~al.}(2009)\citenamefont
  {Jurgenson}, \citenamefont {Navratil},\ and\ \citenamefont
  {Furnstahl}}]{Jurgenson:2009qs}%
  \BibitemOpen
  \bibfield  {author} {\bibinfo {author} {\bibfnamefont {E.~D.}\ \bibnamefont
  {Jurgenson}}, \bibinfo {author} {\bibfnamefont {P.}~\bibnamefont {Navratil}},
  \ and\ \bibinfo {author} {\bibfnamefont {R.~J.}\ \bibnamefont {Furnstahl}},\
  }\href {\doibase 10.1103/PhysRevLett.103.082501} {\bibfield  {journal}
  {\bibinfo  {journal} {Phys. Rev. Lett.}\ }\textbf {\bibinfo {volume} {103}},\
  \bibinfo {pages} {082501} (\bibinfo {year} {2009})},\ \Eprint
  {http://arxiv.org/abs/0905.1873} {arXiv:0905.1873 [nucl-th]} \BibitemShut
  {NoStop}%
%%CITATION = 0905.1873;%%
\bibitem [{\citenamefont {Jurgenson}\ \emph {et~al.}(2011)\citenamefont
  {Jurgenson}, \citenamefont {Navratil},\ and\ \citenamefont
  {Furnstahl}}]{Jurgenson:2010wy}%
  \BibitemOpen
  \bibfield  {author} {\bibinfo {author} {\bibfnamefont {E.}~\bibnamefont
  {Jurgenson}}, \bibinfo {author} {\bibfnamefont {P.}~\bibnamefont {Navratil}},
  \ and\ \bibinfo {author} {\bibfnamefont {R.}~\bibnamefont {Furnstahl}},\
  }\href {\doibase 10.1103/PhysRevC.83.034301} {\bibfield  {journal} {\bibinfo
  {journal} {Phys. Rev. C}\ }\textbf {\bibinfo {volume} {83}},\ \bibinfo
  {pages} {034301} (\bibinfo {year} {2011})},\ \Eprint
  {http://arxiv.org/abs/1011.4085} {arXiv:1011.4085 [nucl-th]} \BibitemShut
  {NoStop}%
\bibitem [{\citenamefont {Roth}(2009)}]{Roth:2009cw}%
  \BibitemOpen
  \bibfield  {author} {\bibinfo {author} {\bibfnamefont {R.}~\bibnamefont
  {Roth}},\ }\href {\doibase 10.1103/PhysRevC.79.064324} {\bibfield  {journal}
  {\bibinfo  {journal} {Phys. Rev. C}\ }\textbf {\bibinfo {volume} {79}},\
  \bibinfo {pages} {064324} (\bibinfo {year} {2009})},\ \Eprint
  {http://arxiv.org/abs/0903.4605} {arXiv:0903.4605 [nucl-th]} \BibitemShut
  {NoStop}%
%%CITATION = ARXIV:0903.4605;%%
\bibitem [{\citenamefont {Roth}\ \emph {et~al.}(2011)\citenamefont {Roth},
  \citenamefont {Langhammer}, \citenamefont {Calci}, \citenamefont {Binder},\
  and\ \citenamefont {Navratil}}]{Roth:2011ar}%
  \BibitemOpen
  \bibfield  {author} {\bibinfo {author} {\bibfnamefont {R.}~\bibnamefont
  {Roth}}, \bibinfo {author} {\bibfnamefont {J.}~\bibnamefont {Langhammer}},
  \bibinfo {author} {\bibfnamefont {A.}~\bibnamefont {Calci}}, \bibinfo
  {author} {\bibfnamefont {S.}~\bibnamefont {Binder}}, \ and\ \bibinfo {author}
  {\bibfnamefont {P.}~\bibnamefont {Navratil}},\ }\href {\doibase
  10.1103/PhysRevLett.107.072501} {\bibfield  {journal} {\bibinfo  {journal}
  {Phys. Rev. Lett.}\ }\textbf {\bibinfo {volume} {107}},\ \bibinfo {pages}
  {072501} (\bibinfo {year} {2011})},\ \Eprint {http://arxiv.org/abs/1105.3173}
  {arXiv:1105.3173 [nucl-th]} \BibitemShut {NoStop}%
\bibitem [{\citenamefont {Roth}\ \emph {et~al.}(2012)\citenamefont {Roth},
  \citenamefont {Binder}, \citenamefont {Vobig}, \citenamefont {Calci},
  \citenamefont {Langhammer},\ and\ \citenamefont {Navratil}}]{Roth:2011vt}%
  \BibitemOpen
  \bibfield  {author} {\bibinfo {author} {\bibfnamefont {R.}~\bibnamefont
  {Roth}}, \bibinfo {author} {\bibfnamefont {S.}~\bibnamefont {Binder}},
  \bibinfo {author} {\bibfnamefont {K.}~\bibnamefont {Vobig}}, \bibinfo
  {author} {\bibfnamefont {A.}~\bibnamefont {Calci}}, \bibinfo {author}
  {\bibfnamefont {J.}~\bibnamefont {Langhammer}}, \ and\ \bibinfo {author}
  {\bibfnamefont {P.}~\bibnamefont {Navratil}},\ }\href {\doibase
  10.1103/PhysRevLett.109.052501} {\bibfield  {journal} {\bibinfo  {journal}
  {Phys. Rev. Lett.}\ }\textbf {\bibinfo {volume} {109}},\ \bibinfo {pages}
  {052501} (\bibinfo {year} {2012})},\ \Eprint {http://arxiv.org/abs/1112.0287}
  {arXiv:1112.0287 [nucl-th]} \BibitemShut {NoStop}%
%%CITATION = ARXIV:1112.0287;%%
\bibitem [{\citenamefont {Hebeler}(2012)}]{Hebeler:2012pr}%
  \BibitemOpen
  \bibfield  {author} {\bibinfo {author} {\bibfnamefont {K.}~\bibnamefont
  {Hebeler}},\ }\href {\doibase 10.1103/PhysRevC.85.021002} {\bibfield
  {journal} {\bibinfo  {journal} {Phys. Rev. C}\ }\textbf {\bibinfo {volume}
  {85}},\ \bibinfo {pages} {021002} (\bibinfo {year} {2012})},\ \Eprint
  {http://arxiv.org/abs/1201.0169} {arXiv:1201.0169 [nucl-th]} \BibitemShut
  {NoStop}%
%%CITATION = ARXIV:1201.0169;%%
\bibitem [{\citenamefont {Entem}\ and\ \citenamefont
  {Machleidt}(2003)}]{Entem:2003ft}%
  \BibitemOpen
  \bibfield  {author} {\bibinfo {author} {\bibfnamefont {D.~R.}\ \bibnamefont
  {Entem}}\ and\ \bibinfo {author} {\bibfnamefont {R.}~\bibnamefont
  {Machleidt}},\ }\href@noop {} {\bibfield  {journal} {\bibinfo  {journal}
  {Phys. Rev. C}\ }\textbf {\bibinfo {volume} {68}},\ \bibinfo {pages} {041001}
  (\bibinfo {year} {2003})},\ \Eprint {http://arxiv.org/abs/nucl-th/0304018}
  {nucl-th/0304018} \BibitemShut {NoStop}%
%%CITATION = NUCL-TH/0304018;%%
\bibitem [{\citenamefont {Epelbaum}\ \emph {et~al.}(2002)\citenamefont
  {Epelbaum}, \citenamefont {Nogga}, \citenamefont {Gloeckle}, \citenamefont
  {Kamada}, \citenamefont {Meissner},\ and\ \citenamefont
  {Witala}}]{Epelbaum:2002vt}%
  \BibitemOpen
  \bibfield  {author} {\bibinfo {author} {\bibfnamefont {E.}~\bibnamefont
  {Epelbaum}}, \bibinfo {author} {\bibfnamefont {A.}~\bibnamefont {Nogga}},
  \bibinfo {author} {\bibfnamefont {W.}~\bibnamefont {Gloeckle}}, \bibinfo
  {author} {\bibfnamefont {H.}~\bibnamefont {Kamada}}, \bibinfo {author}
  {\bibfnamefont {U.-G.}\ \bibnamefont {Meissner}}, \ and\ \bibinfo {author}
  {\bibfnamefont {H.}~\bibnamefont {Witala}},\ }\href {\doibase
  10.1103/PhysRevC.66.064001} {\bibfield  {journal} {\bibinfo  {journal} {Phys.
  Rev. C}\ }\textbf {\bibinfo {volume} {66}},\ \bibinfo {pages} {064001}
  (\bibinfo {year} {2002})},\ \Eprint {http://arxiv.org/abs/nucl-th/0208023}
  {arXiv:nucl-th/0208023} \BibitemShut {NoStop}%
%%CITATION = NUCL-TH/0208023;%%
\bibitem [{\citenamefont {Navratil}(2007)}]{Navratil:2007zn}%
  \BibitemOpen
  \bibfield  {author} {\bibinfo {author} {\bibfnamefont {P.}~\bibnamefont
  {Navratil}},\ }\href {\doibase 10.1007/s00601-007-0193-3} {\bibfield
  {journal} {\bibinfo  {journal} {Few Body Syst.}\ }\textbf {\bibinfo {volume}
  {41}},\ \bibinfo {pages} {117} (\bibinfo {year} {2007})},\ \Eprint
  {http://arxiv.org/abs/0707.4680} {arXiv:0707.4680 [nucl-th]} \BibitemShut
  {NoStop}%
%%CITATION = 0707.4680;%%
\bibitem [{\citenamefont {Navratil}\ \emph {et~al.}(2007)\citenamefont
  {Navratil}, \citenamefont {Gueorguiev}, \citenamefont {Vary}, \citenamefont
  {Ormand},\ and\ \citenamefont {Nogga}}]{Navratil:2007we}%
  \BibitemOpen
  \bibfield  {author} {\bibinfo {author} {\bibfnamefont {P.}~\bibnamefont
  {Navratil}}, \bibinfo {author} {\bibfnamefont {V.~G.}\ \bibnamefont
  {Gueorguiev}}, \bibinfo {author} {\bibfnamefont {J.~P.}\ \bibnamefont
  {Vary}}, \bibinfo {author} {\bibfnamefont {W.~E.}\ \bibnamefont {Ormand}}, \
  and\ \bibinfo {author} {\bibfnamefont {A.}~\bibnamefont {Nogga}},\
  }\href@noop {} {\bibfield  {journal} {\bibinfo  {journal} {Phys. Rev. Lett.}\
  }\textbf {\bibinfo {volume} {99}},\ \bibinfo {pages} {042501} (\bibinfo
  {year} {2007})},\ \Eprint {http://arxiv.org/abs/nucl-th/0701038}
  {nucl-th/0701038} \BibitemShut {NoStop}%
%%CITATION = NUCL-TH/0701038;%%
\bibitem [{\citenamefont {Kehrein}(2006)}]{Kehrein:2006}%
  \BibitemOpen
  \bibfield  {author} {\bibinfo {author} {\bibfnamefont {S.}~\bibnamefont
  {Kehrein}},\ }\href@noop {} {\emph {\bibinfo {title} {The Flow Equation
  Approach to Many-Particle Systems}}}\ (\bibinfo  {publisher} {Springer},\
  \bibinfo {address} {Berlin},\ \bibinfo {year} {2006})\BibitemShut {NoStop}%
\bibitem [{\citenamefont {Anderson}\ \emph {et~al.}(2008)\citenamefont
  {Anderson} \emph {et~al.}}]{Anderson:2008mu}%
  \BibitemOpen
  \bibfield  {author} {\bibinfo {author} {\bibfnamefont {E.}~\bibnamefont
  {Anderson}} \emph {et~al.},\ }\href {\doibase 10.1103/PhysRevC.77.037001}
  {\bibfield  {journal} {\bibinfo  {journal} {Phys. Rev. C}\ }\textbf {\bibinfo
  {volume} {77}},\ \bibinfo {pages} {037001} (\bibinfo {year} {2008})},\
  \Eprint {http://arxiv.org/abs/0801.1098} {arXiv:0801.1098 [nucl-th]}
  \BibitemShut {NoStop}%
%%CITATION = 0801.1098;%%
\bibitem [{\citenamefont {Li}\ \emph {et~al.}(2011)\citenamefont {Li},
  \citenamefont {Anderson},\ and\ \citenamefont {Furnstahl}}]{li:2011sr}%
  \BibitemOpen
  \bibfield  {author} {\bibinfo {author} {\bibfnamefont {W.}~\bibnamefont
  {Li}}, \bibinfo {author} {\bibfnamefont {E.}~\bibnamefont {Anderson}}, \ and\
  \bibinfo {author} {\bibfnamefont {R.}~\bibnamefont {Furnstahl}},\ }\href
  {\doibase 10.1103/PhysRevC.84.054002} {\bibfield  {journal} {\bibinfo
  {journal} {Phys. Rev. C}\ }\textbf {\bibinfo {volume} {84}},\ \bibinfo
  {pages} {054002} (\bibinfo {year} {2011})},\ \Eprint
  {http://arxiv.org/abs/1106.2835} {arXiv:1106.2835 [nucl-th]} \BibitemShut
  {NoStop}%
\bibitem [{\citenamefont {Jurgenson}\ and\ \citenamefont
  {Furnstahl}(2009)}]{Jurgenson:2008jp}%
  \BibitemOpen
  \bibfield  {author} {\bibinfo {author} {\bibfnamefont {E.~D.}\ \bibnamefont
  {Jurgenson}}\ and\ \bibinfo {author} {\bibfnamefont {R.~J.}\ \bibnamefont
  {Furnstahl}},\ }\href {\doibase 10.1016/j.nuclphysa.2008.12.007} {\bibfield
  {journal} {\bibinfo  {journal} {Nucl. Phys. A}\ }\textbf {\bibinfo {volume}
  {818}},\ \bibinfo {pages} {152} (\bibinfo {year} {2009})},\ \Eprint
  {http://arxiv.org/abs/0809.4199} {arXiv:0809.4199 [nucl-th]} \BibitemShut
  {NoStop}%
%%CITATION = 0809.4199;%%
\bibitem [{\citenamefont {Gazit}\ \emph {et~al.}(2009)\citenamefont {Gazit},
  \citenamefont {Quaglioni},\ and\ \citenamefont {Navratil}}]{Gazit:2008ma}%
  \BibitemOpen
  \bibfield  {author} {\bibinfo {author} {\bibfnamefont {D.}~\bibnamefont
  {Gazit}}, \bibinfo {author} {\bibfnamefont {S.}~\bibnamefont {Quaglioni}}, \
  and\ \bibinfo {author} {\bibfnamefont {P.}~\bibnamefont {Navratil}},\ }\href
  {\doibase 10.1103/PhysRevLett.103.102502} {\bibfield  {journal} {\bibinfo
  {journal} {Phys. Rev. Lett.}\ }\textbf {\bibinfo {volume} {103}},\ \bibinfo
  {pages} {102502} (\bibinfo {year} {2009})},\ \Eprint
  {http://arxiv.org/abs/0812.4444} {arXiv:0812.4444 [nucl-th]} \BibitemShut
  {NoStop}%
%%CITATION = 0812.4444;%%
\bibitem [{\citenamefont {Sternberg}\ \emph {et~al.}(2008)\citenamefont
  {Sternberg}, \citenamefont {Ng}, \citenamefont {Yang}, \citenamefont {Maris},
  \citenamefont {Vary}, \citenamefont {Sosonkina},\ and\ \citenamefont
  {Le}}]{Sternberg:2008:ACI:1413370.1413386}%
  \BibitemOpen
  \bibfield  {author} {\bibinfo {author} {\bibfnamefont {P.}~\bibnamefont
  {Sternberg}}, \bibinfo {author} {\bibfnamefont {E.~G.}\ \bibnamefont {Ng}},
  \bibinfo {author} {\bibfnamefont {C.}~\bibnamefont {Yang}}, \bibinfo {author}
  {\bibfnamefont {P.}~\bibnamefont {Maris}}, \bibinfo {author} {\bibfnamefont
  {J.~P.}\ \bibnamefont {Vary}}, \bibinfo {author} {\bibfnamefont
  {M.}~\bibnamefont {Sosonkina}}, \ and\ \bibinfo {author} {\bibfnamefont
  {H.~V.}\ \bibnamefont {Le}},\ }in\ \href
  {http://dl.acm.org/citation.cfm?id=1413370.1413386} {\emph {\bibinfo
  {booktitle} {Proceedings of the 2008 ACM/IEEE conference on
  Supercomputing}}},\ \bibinfo {series and number} {SC '08}\ (\bibinfo
  {publisher} {IEEE Press},\ \bibinfo {address} {Piscataway, NJ, USA},\
  \bibinfo {year} {2008})\ pp.\ \bibinfo {pages} {15:1--15:12}\BibitemShut
  {NoStop}%
\bibitem [{\citenamefont {Maris}\ \emph
  {et~al.}(2010{\natexlab{b}})\citenamefont {Maris}, \citenamefont {Sosonkina},
  \citenamefont {Vary}, \citenamefont {Ng},\ and\ \citenamefont
  {Yang}}]{DBLP:journals/procedia/MarisSVNY10}%
  \BibitemOpen
  \bibfield  {author} {\bibinfo {author} {\bibfnamefont {P.}~\bibnamefont
  {Maris}}, \bibinfo {author} {\bibfnamefont {M.}~\bibnamefont {Sosonkina}},
  \bibinfo {author} {\bibfnamefont {J.~P.}\ \bibnamefont {Vary}}, \bibinfo
  {author} {\bibfnamefont {E.~G.}\ \bibnamefont {Ng}}, \ and\ \bibinfo {author}
  {\bibfnamefont {C.}~\bibnamefont {Yang}},\ }\href@noop {} {\bibfield
  {journal} {\bibinfo  {journal} {Procedia CS}\ }\textbf {\bibinfo {volume}
  {1}},\ \bibinfo {pages} {97} (\bibinfo {year}
  {2010}{\natexlab{b}})}\BibitemShut {NoStop}%
\bibitem [{\citenamefont {Aktulga}\ \emph {et~al.}(2012)\citenamefont
  {Aktulga}, \citenamefont {Yang}, \citenamefont {Ng}, \citenamefont {Maris},\
  and\ \citenamefont {Vary}}]{DBLP:conf/europar/AktulgaYNMV12}%
  \BibitemOpen
  \bibfield  {author} {\bibinfo {author} {\bibfnamefont {H.~M.}\ \bibnamefont
  {Aktulga}}, \bibinfo {author} {\bibfnamefont {C.}~\bibnamefont {Yang}},
  \bibinfo {author} {\bibfnamefont {E.~G.}\ \bibnamefont {Ng}}, \bibinfo
  {author} {\bibfnamefont {P.}~\bibnamefont {Maris}}, \ and\ \bibinfo {author}
  {\bibfnamefont {J.~P.}\ \bibnamefont {Vary}},\ }in\ \href@noop {} {\emph
  {\bibinfo {booktitle} {Euro-Par}}},\ \bibinfo {series} {Lecture Notes in
  Computer Science}, Vol.\ \bibinfo {volume} {7484},\ \bibinfo {editor} {edited
  by\ \bibinfo {editor} {\bibfnamefont {C.}~\bibnamefont {Kaklamanis}},
  \bibinfo {editor} {\bibfnamefont {T.~S.}\ \bibnamefont {Papatheodorou}}, \
  and\ \bibinfo {editor} {\bibfnamefont {P.~G.}\ \bibnamefont {Spirakis}}}\
  (\bibinfo  {publisher} {Springer},\ \bibinfo {year} {2012})\ pp.\ \bibinfo
  {pages} {830--842}\BibitemShut {NoStop}%
\bibitem [{\citenamefont {Maris}\ \emph {et~al.}(2012)\citenamefont {Maris},
  \citenamefont {Aktulga}, \citenamefont {Caprio}, \citenamefont
  {\c{C}ataly\"urek}, \citenamefont {Ng}, \citenamefont {Oryspayev},
  \citenamefont {Potter}, \citenamefont {Saule}, \citenamefont {Sosonkina},
  \citenamefont {Vary}, \citenamefont {Yang},\ and\ \citenamefont
  {Zhou}}]{Maris:2013}%
  \BibitemOpen
  \bibfield  {author} {\bibinfo {author} {\bibfnamefont {P.}~\bibnamefont
  {Maris}}, \bibinfo {author} {\bibfnamefont {H.}~\bibnamefont {Aktulga}},
  \bibinfo {author} {\bibfnamefont {M.}~\bibnamefont {Caprio}}, \bibinfo
  {author} {\bibfnamefont {U.}~\bibnamefont {\c{C}ataly\"urek}}, \bibinfo
  {author} {\bibfnamefont {E.}~\bibnamefont {Ng}}, \bibinfo {author}
  {\bibfnamefont {D.}~\bibnamefont {Oryspayev}}, \bibinfo {author}
  {\bibfnamefont {H.}~\bibnamefont {Potter}}, \bibinfo {author} {\bibfnamefont
  {E.}~\bibnamefont {Saule}}, \bibinfo {author} {\bibfnamefont
  {M.}~\bibnamefont {Sosonkina}}, \bibinfo {author} {\bibfnamefont
  {J.}~\bibnamefont {Vary}}, \bibinfo {author} {\bibfnamefont {C.}~\bibnamefont
  {Yang}}, \ and\ \bibinfo {author} {\bibfnamefont {Z.}~\bibnamefont {Zhou}},\
  }\href {\doibase 10.1088/1742-6596/403/1/012019} {\bibfield  {journal}
  {\bibinfo  {journal} {J. Phys. Conf. Ser.}\ }\textbf {\bibinfo {volume}
  {403}},\ \bibinfo {pages} {012019} (\bibinfo {year} {2012})}\BibitemShut
  {NoStop}%
\bibitem [{\citenamefont {Furnstahl}\ \emph {et~al.}(2012)\citenamefont
  {Furnstahl}, \citenamefont {Hagen},\ and\ \citenamefont
  {Papenbrock}}]{Furnstahl:2012qg}%
  \BibitemOpen
  \bibfield  {author} {\bibinfo {author} {\bibfnamefont {R.}~\bibnamefont
  {Furnstahl}}, \bibinfo {author} {\bibfnamefont {G.}~\bibnamefont {Hagen}}, \
  and\ \bibinfo {author} {\bibfnamefont {T.}~\bibnamefont {Papenbrock}},\
  }\href {\doibase 10.1103/PhysRevC.86.031301} {\bibfield  {journal} {\bibinfo
  {journal} {Phys.\ Rev.\ C}\ }\textbf {\bibinfo {volume} {86}},\ \bibinfo
  {pages} {031301} (\bibinfo {year} {2012})},\ \Eprint
  {http://arxiv.org/abs/1207.6100} {arXiv:1207.6100 [nucl-th]} \BibitemShut
  {NoStop}%
%%CITATION = ARXIV:1207.6100;%%
\bibitem [{\citenamefont {Coon}\ \emph {et~al.}(2012)\citenamefont {Coon},
  \citenamefont {Avetian}, \citenamefont {Kruse}, \citenamefont {van Kolck},
  \citenamefont {Maris},\ and\ \citenamefont {Vary}}]{Coon:2012ab}%
  \BibitemOpen
  \bibfield  {author} {\bibinfo {author} {\bibfnamefont {S.~A.}\ \bibnamefont
  {Coon}}, \bibinfo {author} {\bibfnamefont {M.~I.}\ \bibnamefont {Avetian}},
  \bibinfo {author} {\bibfnamefont {M.~K.}\ \bibnamefont {Kruse}}, \bibinfo
  {author} {\bibfnamefont {U.}~\bibnamefont {van Kolck}}, \bibinfo {author}
  {\bibfnamefont {P.}~\bibnamefont {Maris}}, \ and\ \bibinfo {author}
  {\bibfnamefont {J.~P.}\ \bibnamefont {Vary}},\ }\href {\doibase
  10.1103/PhysRevC.86.054002} {\bibfield  {journal} {\bibinfo  {journal} {Phys.
  Rev. C}\ }\textbf {\bibinfo {volume} {86}},\ \bibinfo {pages} {054002}
  (\bibinfo {year} {2012})},\ \Eprint {http://arxiv.org/abs/1205.3230}
  {arXiv:1205.3230 [nucl-th]} \BibitemShut {NoStop}%
%%CITATION = ARXIV:1205.3230;%%
\bibitem [{\citenamefont {More}\ \emph {et~al.}(2013)\citenamefont {More},
  \citenamefont {{Ekstr\"om}}, \citenamefont {Furnstahl}, \citenamefont
  {Hagen},\ and\ \citenamefont {Papenbrock}}]{More:2013aa}%
  \BibitemOpen
  \bibfield  {author} {\bibinfo {author} {\bibfnamefont {S.}~\bibnamefont
  {More}}, \bibinfo {author} {\bibfnamefont {A.}~\bibnamefont {{Ekstr\"om}}},
  \bibinfo {author} {\bibfnamefont {R.~J.}\ \bibnamefont {Furnstahl}}, \bibinfo
  {author} {\bibfnamefont {G.}~\bibnamefont {Hagen}}, \ and\ \bibinfo {author}
  {\bibfnamefont {T.}~\bibnamefont {Papenbrock}},\ }\href@noop {} {\  (\bibinfo
  {year} {2013})},\ \Eprint {http://arxiv.org/abs/1302.3815} {arXiv:1302.3815
  [nucl-th]} \BibitemShut {NoStop}%
\bibitem [{\citenamefont {Calci}\ \emph {et~al.}(2013)\citenamefont {Calci},
  \citenamefont {Langhammer}, \citenamefont {Maris}, \citenamefont {Roth},\
  and\ \citenamefont {Vary}}]{Calci:2013}%
  \BibitemOpen
  \bibfield  {author} {\bibinfo {author} {\bibfnamefont {A.}~\bibnamefont
  {Calci}}, \bibinfo {author} {\bibfnamefont {J.}~\bibnamefont {Langhammer}},
  \bibinfo {author} {\bibfnamefont {P.}~\bibnamefont {Maris}}, \bibinfo
  {author} {\bibfnamefont {R.}~\bibnamefont {Roth}}, \ and\ \bibinfo {author}
  {\bibfnamefont {J.}~\bibnamefont {Vary}},\ }\href@noop {} {\  (\bibinfo
  {year} {2013})},\ \bibinfo {note} {in preparation}\BibitemShut {NoStop}%
\bibitem [{\citenamefont {Cockrell}\ \emph {et~al.}(2012)\citenamefont
  {Cockrell}, \citenamefont {Vary},\ and\ \citenamefont
  {Maris}}]{Cockrell:2012vd}%
  \BibitemOpen
  \bibfield  {author} {\bibinfo {author} {\bibfnamefont {C.}~\bibnamefont
  {Cockrell}}, \bibinfo {author} {\bibfnamefont {J.~P.}\ \bibnamefont {Vary}},
  \ and\ \bibinfo {author} {\bibfnamefont {P.}~\bibnamefont {Maris}},\ }\href
  {\doibase 10.1103/PhysRevC.86.034325} {\bibfield  {journal} {\bibinfo
  {journal} {Phys. Rev. C}\ }\textbf {\bibinfo {volume} {86}},\ \bibinfo
  {pages} {034325} (\bibinfo {year} {2012})},\ \Eprint
  {http://arxiv.org/abs/1201.0724} {arXiv:1201.0724 [nucl-th]} \BibitemShut
  {NoStop}%
%%CITATION = ARXIV:1201.0724;%%
\bibitem [{\citenamefont {Epelbaum}\ \emph {et~al.}(2009)\citenamefont
  {Epelbaum}, \citenamefont {Hammer},\ and\ \citenamefont
  {Meissner}}]{Epelbaum:2008ga}%
  \BibitemOpen
  \bibfield  {author} {\bibinfo {author} {\bibfnamefont {E.}~\bibnamefont
  {Epelbaum}}, \bibinfo {author} {\bibfnamefont {H.-W.}\ \bibnamefont
  {Hammer}}, \ and\ \bibinfo {author} {\bibfnamefont {U.-G.}\ \bibnamefont
  {Meissner}},\ }\href {\doibase 10.1103/RevModPhys.81.1773} {\bibfield
  {journal} {\bibinfo  {journal} {Rev. Mod. Phys.}\ }\textbf {\bibinfo {volume}
  {81}},\ \bibinfo {pages} {1773} (\bibinfo {year} {2009})},\ \Eprint
  {http://arxiv.org/abs/0811.1338} {arXiv:0811.1338 [nucl-th]} \BibitemShut
  {NoStop}%
%%CITATION = ARXIV:0811.1338;%%
\bibitem [{\citenamefont {Caprio}\ \emph {et~al.}(2013)\citenamefont {Caprio},
  \citenamefont {Maris},\ and\ \citenamefont {Vary}}]{Caprio:2013yp}%
  \BibitemOpen
  \bibfield  {author} {\bibinfo {author} {\bibfnamefont {M.}~\bibnamefont
  {Caprio}}, \bibinfo {author} {\bibfnamefont {P.}~\bibnamefont {Maris}}, \
  and\ \bibinfo {author} {\bibfnamefont {J.}~\bibnamefont {Vary}},\ }\href
  {\doibase 10.1016/j.physletb.2012.12.064} {\bibfield  {journal} {\bibinfo
  {journal} {Phys. Lett.}\ }\textbf {\bibinfo {volume} {B719}},\ \bibinfo
  {pages} {179} (\bibinfo {year} {2013})},\ \Eprint
  {http://arxiv.org/abs/1301.0956} {arXiv:1301.0956 [nucl-th]} \BibitemShut
  {NoStop}%
%%CITATION = ARXIV:1301.0956;%%
\bibitem [{\citenamefont {Marcucci}\ \emph {et~al.}(2008)\citenamefont
  {Marcucci}, \citenamefont {Pervin}, \citenamefont {Pieper}, \citenamefont
  {Schiavilla},\ and\ \citenamefont {Wiringa}}]{PhysRevC.78.065501}%
  \BibitemOpen
  \bibfield  {author} {\bibinfo {author} {\bibfnamefont {L.~E.}\ \bibnamefont
  {Marcucci}}, \bibinfo {author} {\bibfnamefont {M.}~\bibnamefont {Pervin}},
  \bibinfo {author} {\bibfnamefont {S.~C.}\ \bibnamefont {Pieper}}, \bibinfo
  {author} {\bibfnamefont {R.}~\bibnamefont {Schiavilla}}, \ and\ \bibinfo
  {author} {\bibfnamefont {R.~B.}\ \bibnamefont {Wiringa}},\ }\href {\doibase
  10.1103/PhysRevC.78.065501} {\bibfield  {journal} {\bibinfo  {journal} {Phys.
  Rev. C}\ }\textbf {\bibinfo {volume} {78}},\ \bibinfo {pages} {065501}
  (\bibinfo {year} {2008})}\BibitemShut {NoStop}%
\bibitem [{\citenamefont {Pastore}\ \emph {et~al.}(2012)\citenamefont
  {Pastore}, \citenamefont {Pieper}, \citenamefont {Schiavilla},\ and\
  \citenamefont {Wiringa}}]{Pastore:2012rp}%
  \BibitemOpen
  \bibfield  {author} {\bibinfo {author} {\bibfnamefont {S.}~\bibnamefont
  {Pastore}}, \bibinfo {author} {\bibfnamefont {S.~C.}\ \bibnamefont {Pieper}},
  \bibinfo {author} {\bibfnamefont {R.}~\bibnamefont {Schiavilla}}, \ and\
  \bibinfo {author} {\bibfnamefont {R.}~\bibnamefont {Wiringa}},\ }\href@noop
  {} {\  (\bibinfo {year} {2012})},\ \Eprint {http://arxiv.org/abs/1212.3375}
  {arXiv:1212.3375 [nucl-th]} \BibitemShut {NoStop}%
%%CITATION = ARXIV:1212.3375;%%
\bibitem [{\citenamefont {Hagen}\ \emph {et~al.}(2007)\citenamefont {Hagen},
  \citenamefont {Dean}, \citenamefont {Hjorth-Jensen}, \citenamefont
  {Papenbrock},\ and\ \citenamefont {Schwenk}}]{Hagen:2007hi}%
  \BibitemOpen
  \bibfield  {author} {\bibinfo {author} {\bibfnamefont {G.}~\bibnamefont
  {Hagen}}, \bibinfo {author} {\bibfnamefont {D.~J.}\ \bibnamefont {Dean}},
  \bibinfo {author} {\bibfnamefont {M.}~\bibnamefont {Hjorth-Jensen}}, \bibinfo
  {author} {\bibfnamefont {T.}~\bibnamefont {Papenbrock}}, \ and\ \bibinfo
  {author} {\bibfnamefont {A.}~\bibnamefont {Schwenk}},\ }\href {\doibase
  10.1103/PhysRevC.76.044305} {\bibfield  {journal} {\bibinfo  {journal} {Phys.
  Rev. C}\ }\textbf {\bibinfo {volume} {76}},\ \bibinfo {pages} {044305}
  (\bibinfo {year} {2007})},\ \Eprint {http://arxiv.org/abs/0707.1516}
  {arXiv:0707.1516 [nucl-th]} \BibitemShut {NoStop}%
%%CITATION = 0707.1516;%%
\bibitem [{\citenamefont {Drut}\ and\ \citenamefont
  {Platter}(2011)}]{Drut:2011te}%
  \BibitemOpen
  \bibfield  {author} {\bibinfo {author} {\bibfnamefont {J.~E.}\ \bibnamefont
  {Drut}}\ and\ \bibinfo {author} {\bibfnamefont {L.}~\bibnamefont {Platter}},\
  }\href {\doibase 10.1103/PhysRevC.84.014318} {\bibfield  {journal} {\bibinfo
  {journal} {Phys. Rev. C}\ }\textbf {\bibinfo {volume} {84}},\ \bibinfo
  {pages} {014318} (\bibinfo {year} {2011})},\ \Eprint
  {http://arxiv.org/abs/1104.4357} {arXiv:1104.4357 [nucl-th]} \BibitemShut
  {NoStop}%
%%CITATION = ARXIV:1104.4357;%%
\bibitem [{\citenamefont {Quaglioni}\ and\ \citenamefont
  {Navratil}(2008)}]{Quaglioni:2008sm}%
  \BibitemOpen
  \bibfield  {author} {\bibinfo {author} {\bibfnamefont {S.}~\bibnamefont
  {Quaglioni}}\ and\ \bibinfo {author} {\bibfnamefont {P.}~\bibnamefont
  {Navratil}},\ }\href {\doibase 10.1103/PhysRevLett.101.092501} {\bibfield
  {journal} {\bibinfo  {journal} {Phys. Rev. Lett.}\ }\textbf {\bibinfo
  {volume} {101}},\ \bibinfo {pages} {092501} (\bibinfo {year} {2008})},\
  \Eprint {http://arxiv.org/abs/0804.1560} {arXiv:0804.1560 [nucl-th]}
  \BibitemShut {NoStop}%
%%CITATION = 0804.1560;%%
\bibitem [{\citenamefont {Navratil}\ \emph {et~al.}(2010)\citenamefont
  {Navratil}, \citenamefont {Roth},\ and\ \citenamefont
  {Quaglioni}}]{Navratil:2010jn}%
  \BibitemOpen
  \bibfield  {author} {\bibinfo {author} {\bibfnamefont {P.}~\bibnamefont
  {Navratil}}, \bibinfo {author} {\bibfnamefont {R.}~\bibnamefont {Roth}}, \
  and\ \bibinfo {author} {\bibfnamefont {S.}~\bibnamefont {Quaglioni}},\ }\href
  {\doibase 10.1103/PhysRevC.82.034609} {\bibfield  {journal} {\bibinfo
  {journal} {Phys. Rev. C}\ }\textbf {\bibinfo {volume} {82}},\ \bibinfo
  {pages} {034609} (\bibinfo {year} {2010})},\ \Eprint
  {http://arxiv.org/abs/1007.0525} {arXiv:1007.0525 [nucl-th]} \BibitemShut
  {NoStop}%
%%CITATION = ARXIV:1007.0525;%%
\bibitem [{\citenamefont {Navratil}\ \emph {et~al.}(2011)\citenamefont
  {Navratil}, \citenamefont {Roth},\ and\ \citenamefont
  {Quaglioni}}]{Navratil:2011sa}%
  \BibitemOpen
  \bibfield  {author} {\bibinfo {author} {\bibfnamefont {P.}~\bibnamefont
  {Navratil}}, \bibinfo {author} {\bibfnamefont {R.}~\bibnamefont {Roth}}, \
  and\ \bibinfo {author} {\bibfnamefont {S.}~\bibnamefont {Quaglioni}},\ }\href
  {\doibase 10.1016/j.physletb.2011.09.079} {\bibfield  {journal} {\bibinfo
  {journal} {Phys.Lett.}\ }\textbf {\bibinfo {volume} {B704}},\ \bibinfo
  {pages} {379} (\bibinfo {year} {2011})},\ \Eprint
  {http://arxiv.org/abs/1105.5977} {arXiv:1105.5977 [nucl-th]} \BibitemShut
  {NoStop}%
%%CITATION = ARXIV:1105.5977;%%
\bibitem [{\citenamefont {Quaglioni}\ \emph {et~al.}(2012)\citenamefont
  {Quaglioni}, \citenamefont {Navratil}, \citenamefont {Roth},\ and\
  \citenamefont {Horiuchi}}]{Quaglioni:2012bm}%
  \BibitemOpen
  \bibfield  {author} {\bibinfo {author} {\bibfnamefont {S.}~\bibnamefont
  {Quaglioni}}, \bibinfo {author} {\bibfnamefont {P.}~\bibnamefont {Navratil}},
  \bibinfo {author} {\bibfnamefont {R.}~\bibnamefont {Roth}}, \ and\ \bibinfo
  {author} {\bibfnamefont {W.}~\bibnamefont {Horiuchi}},\ }\href@noop {} {\
  (\bibinfo {year} {2012})},\ \Eprint {http://arxiv.org/abs/1203.0268}
  {arXiv:1203.0268 [nucl-th]} \BibitemShut {NoStop}%
%%CITATION = ARXIV:1203.0268;%%
\bibitem [{\citenamefont {Caprio}\ \emph
  {et~al.}(2012{\natexlab{a}})\citenamefont {Caprio}, \citenamefont {Maris},\
  and\ \citenamefont {Vary}}]{Caprio:2012rv}%
  \BibitemOpen
  \bibfield  {author} {\bibinfo {author} {\bibfnamefont {M.}~\bibnamefont
  {Caprio}}, \bibinfo {author} {\bibfnamefont {P.}~\bibnamefont {Maris}}, \
  and\ \bibinfo {author} {\bibfnamefont {J.}~\bibnamefont {Vary}},\ }\href
  {\doibase 10.1103/PhysRevC.86.034312} {\bibfield  {journal} {\bibinfo
  {journal} {Phys. Rev. C}\ }\textbf {\bibinfo {volume} {86}},\ \bibinfo
  {pages} {034312} (\bibinfo {year} {2012}{\natexlab{a}})},\ \Eprint
  {http://arxiv.org/abs/1208.4156} {arXiv:1208.4156 [nucl-th]} \BibitemShut
  {NoStop}%
%%CITATION = ARXIV:1208.4156;%%
\bibitem [{\citenamefont {Caprio}\ \emph
  {et~al.}(2012{\natexlab{b}})\citenamefont {Caprio}, \citenamefont {Maris},\
  and\ \citenamefont {Vary}}]{Caprio:2012as}%
  \BibitemOpen
  \bibfield  {author} {\bibinfo {author} {\bibfnamefont {M.}~\bibnamefont
  {Caprio}}, \bibinfo {author} {\bibfnamefont {P.}~\bibnamefont {Maris}}, \
  and\ \bibinfo {author} {\bibfnamefont {J.}~\bibnamefont {Vary}},\ }\href
  {\doibase 10.1088/1742-6596/403/1/012014} {\bibfield  {journal} {\bibinfo
  {journal} {J. Phys. Conf. Ser.}\ }\textbf {\bibinfo {volume} {403}},\
  \bibinfo {pages} {012014} (\bibinfo {year} {2012}{\natexlab{b}})},\ \Eprint
  {http://arxiv.org/abs/1208.6055} {arXiv:1208.6055 [nucl-th]} \BibitemShut
  {NoStop}%
%%CITATION = ARXIV:1208.6055;%%
\end{thebibliography}%

\end{document}